\documentclass[11pt,a4paper]{article}
\usepackage{jheppub}

\usepackage{amsmath}
\usepackage[dvipsnames]{xcolor}
\usepackage{graphicx}
\usepackage{hyperref}
\usepackage{mathtools}
\usepackage{graphicx}
\usepackage{comment}
\usepackage{cancel}
\usepackage{booktabs}
\usepackage{amssymb}
\usepackage{caption}
\usepackage{float,subcaption}
\usepackage{multirow}
\usepackage{makecell}
\usepackage{dsfont}
\allowdisplaybreaks

\hypersetup{
 bookmarksnumbered=true,
 colorlinks=true,
 linkcolor=[rgb]{0.098,0.098,0.439},
 citecolor=[rgb]{0.098,0.098,0.439},
 urlcolor=[rgb]{0.098,0.098,0.439}
  }

\usepackage[normalem]{ulem}

\newcommand{\mb}[1]{\mathbf{#1}}
\newcommand{\mr}[1]{\mathrm{#1}}
\newcommand{\eps}[0]{\epsilon}

\newcommand{\bY}[1]{\mb{Y}_{#1}}
\newcommand{\bh}[1]{\mb{h}_{#1}}
\newcommand{\bYd}[1]{\mb{Y}^\dagger_{#1}}
\newcommand{\bhd}[1]{\mb{h}^\dagger_{#1}}
\newcommand{\sqm}[1]{\mb{m}^2_{#1}}
\newcommand{\sqmHu}[0]{m^2_{H_u}}
\newcommand{\sqmHd}[0]{m^2_{H_d}}

\title{Dark gauge-mediated supersymmetry breaking with a massless dark photon}

\author[a]{Brian Batell,}
\author[b]{Yechan Kim,}
\author[b]{Hye-Sung Lee,}
\author[b,c]{Jiheon Lee}
\affiliation[a]{Pittsburgh Particle Physics, Astrophysics, and Cosmology Center, Department of Physics and Astronomy, University
of Pittsburgh, Pittsburgh, PA 15217, USA}
\affiliation[b]{Department of Physics, Korea Advanced Institute of Science and Technology, Daejeon 34141, Korea}
\affiliation[c]{Theoretical Physics Department, CERN, 1 Esplanade des Particules, CH-1211 Geneva 23, Switzerland}

\date{\today}

\abstract{
 We study dark gauge-mediated supersymmetry breaking (dark GMSB) in a theory with a new unbroken $U(1)_{D}$ local symmetry and massless dark photon. Messenger fields charged under both Standard Model and dark gauge symmetries produce new soft supersymmetry-breaking terms due to gauge kinetic mixing between $U(1)_Y$ hypercharge and $U(1)_D$.
We show that large kinetic mixing induces significant distortions to the superpartner spectra relative to conventional GMSB. Notably, shifts in the Higgs soft masses impact the conditions for electroweak symmetry breaking, lowering the $\mu$ parameter and yielding a relatively light Higgsino that may be accessible at the LHC. Furthermore, for very simple messenger representations, a very light bino-dark photino mixed state is present in the spectrum, which may be probed through exotic Higgs boson decays at future Higgs factories. We also examine the cosmological and phenomenological consequences of the messengers, the lightest of which is absolutely stable and carries fractional electric charge. }

\begin{document}

\maketitle

\flushbottom

\section{Introduction}
\label{IntroSec}

Supersymmetry (SUSY) is an attractive framework for physics beyond the Standard Model (SM), offering solutions to some of its most pressing mysteries~\cite{Martin:1997ns}. A critical question is how SUSY breaking is communicated to the observable sector. This communication mechanism determines the masses and interactions of the superpartners, profoundly influencing their experimental signatures. Gauge-mediated SUSY breaking (GMSB) offers an appealing answer to this question
\cite{Dine:1981gu,Nappi:1982hm,Alvarez-Gaume:1981abe,Dine:1993yw,Dine:1994vc,Dine:1995ag,Giudice:1998bp,Meade:2008wd}. 
In GMSB, SUSY breaking is transmitted via the SM gauge interactions through messenger fields, ensuring a high degree of flavor universality in the soft SUSY-breaking terms and safeguarding against dangerous flavor-violating effects common in gravity-mediated scenarios.

In this work, we consider a simple extension of GMSB involving an additional unbroken $U(1)_{D}$  
gauge symmetry, which we term {\it dark gauge-mediated SUSY breaking} (dark GMSB). 
The messenger fields are assumed to be charged under both the SM and dark gauge symmetries. Due to the presence of kinetic mixing between the $U(1)_{D}$ and  $U(1)_Y$ hypercharge gauge sectors, with strength parameterized by $\epsilon$, additional contributions to the bino and sfermion soft-SUSY breaking terms are generated in this scenario. The size of these new soft terms depends on the strength of the kinetic mixing and dark gauge coupling. In particular, we explore the implications of large kinetic mixing, $0.1 \lesssim \epsilon \lesssim 1$, on the superpartner spectrum. Such large kinetic mixing is still phenomenologically viable for the case of an unbroken dark gauge symmetry with a massless dark photon provided no additional light matter charged under $U(1)_{D}$ is present, a condition that is satisfied in our setup. 

In this first study, for concreteness we consider two scenarios for the messenger representation: 1) a complete $SU(5)$ GUT representation 
${\bf 5} + {\bf \overline 5}$, and 2) a single vectorlike chiral multiplet in a representation of the SM gauge group (i.e., an incomplete $SU(5)$ multiplet). In both cases, we assume that the messengers are also charged under $U(1)_{D}$. The first scenario corresponds to the messenger content of the minimal GSMB model~\cite{Dine:1993yw,Dine:1994vc,Dine:1995ag} and, as such, provides an interesting point of comparison for dark GSMB. The latter choice of an incomplete GUT multiplet is somewhat non-standard, but has the novel feature of an approximate unbroken SUSY for one linear combination of the $U(1)$ vector multiplets, leading to a very light mixed bino-dark photino neutralino with mass below the weak scale. This feature opens up the possibility of an exotic decay channel of the Higgs boson into the light neutralinos, which can be probed at future Higgs factories. Furthermore, in both scenarios, we show that large values of kinetic mixing impact the conditions for electroweak symmetry breaking and lead to significant modifications to the superpartner spectrum relative to conventional GMSB, with distinctive phenomenological implications. Notably, the $\mu$ parameter required for successful EWSB becomes smaller as the kinetic mixing increases, leading to a relatively light Higgsino that can be searched for at the LHC.  We mainly consider low-scale SUSY breaking in this work, which can lead to distinctive collider signals from the NLSP decay to the gravitino LSP. Finally, we consider cosmological and phenomenological consequences of the messenger fields, which carry fractional electric charge and are absolutely stable.  

The idea of extending the MSSM to include a kinetically mixed $U(1)$ sector has been explored in various contexts. 
The pioneering work of Ref.~\cite{Dienes:1996zr} examined the implications of hidden $U(1)$ $D$-terms in the presence of kinetic mixing and their impact on the MSSM soft masses, and furthermore investigated the expected size of kinetic mixings in UV theories, including string theory. Subsequent works have discussed various implications of SUSY $U(1)$ extensions with kinetic mixing, including for SUSY breaking \cite{Suematsu:2006wh, Chun:2008by, Arkani-Hamed:2008kxc, Baumgart:2009tn, Cheung:2009qd, Morrissey:2009ur, Arvanitaki:2009hb, Cohen:2010kn, Kang:2010mh, Chan:2011aa, Baryakhtar:2012rz, Lee:2017fin}, connections to string theory \cite{Arvanitaki:2009hb, Andreas:2011in}, dark matter and dark sector physics \cite{Kors:2004ri, Feldman:2006wd, Hooper:2008im, Ibarra:2008kn, Arkani-Hamed:2008kxc, Zurek:2008qg, Chun:2008by, Baumgart:2009tn, Cheung:2009qd, Katz:2009qq, Morrissey:2009ur, Feldman:2010wy, Cohen:2010kn, Kang:2010mh, Andreas:2011in, Barnes:2020vsc}, collider searches \cite{Kors:2004ri, Ibarra:2008kn, Arvanitaki:2009hb, Morrissey:2009ur, Barnes:2020vsc, Baryakhtar:2012rz}, and cosmology \cite{Ibarra:2008kn, Pierce:2019ozl}. 
Previous works on massless hidden gauge bosons in SUSY discussed relevant phenomenology in the small kinetic mixing limit~\cite{Ibarra:2008kn, Arvanitaki:2009hb, Baryakhtar:2012rz}. 

The structure of this paper is outlined as follows.
Section~\ref{SUSYmixSec} introduces supersymmetric kinetic mixing between $U(1)_{Y}$ and $U(1)_{D}$.
In Section~\ref{SUSYbreakSec}, we detail the components of dark GMSB and discuss how soft terms depend on kinetic mixing.
The effects of large kinetic mixing on the Higgs scalar and sfermion sectors are examined in Section~\ref{ScalarSec}, while Section~\ref{GauginoSec} explores the neutralino and chargino sectors. Section~\ref{SpectrumSec} discusses the distinct spectra of dark GMSB and phenomenological implications of large kinetic mixing.
Finally, we discuss the cosmology of stable relic particles, including the messengers, in Section~\ref{sec:Cosmology}.
Our conclusions are summarized in Section~\ref{SummarySec}.
Additionally, we provide two appendices.
In Appendix~\ref{appendixRenormaliz}, we present the renormalization group equations for the SUSY parameters at the two-loop level in the presence of a dark $U(1)$ gauge symmetry and kinetic mixing.
Next, Appendix~\ref{appendixRunning} discusses the evolution of the gauge couplings at high scales and explores the impact of large kinetic mixing on gauge coupling unification.

\section{Supersymmetric kinetic mixing}
\label{SUSYmixSec}

The kinetic mixing between $U(1)_Y$ hypercharge and $U(1)_D$ \cite{Holdom:1985ag,Holdom:1986eq} can be extended to a supersymmetric framework. The supersymmetric kinetic terms for the gauge sector are given by \cite{Dienes:1996zr}
\begin{align}
\label{fieldstrength}
\mathcal{L} \supset &
\int d^2 \theta \left( \frac{1}{4} \hat{\mathcal{W}}_\mathbb{B} \hat{\mathcal{W}}_\mathbb{B}
+ \frac{1}{4} \hat{\mathcal{W}}_\mathbb{X} \hat{\mathcal{W}}_\mathbb{X}
+ \frac{\epsilon}{2} \hat{\mathcal{W}}_\mathbb{B} \hat{\mathcal{W}}_\mathbb{X}  \right) +h.c.
\\
\label{kineticMix}
= & -\frac{1}{4} \mathbb{B}_{\mu\nu} \mathbb{B}^{\mu\nu} -\frac{1}{4} \mathbb{X}_{\mu\nu} \mathbb{X}^{\mu\nu} - \frac{\epsilon}{2} \mathbb{B}_{\mu\nu} \mathbb{X}^{\mu\nu}
\nonumber
\\
& + i \tilde{\mathbb{B}}^\dagger \sigma^{\mu} \partial_\mu \tilde{\mathbb{B}}
+ i \tilde{\mathbb{X}}^\dagger \sigma^{\mu} \partial_\mu \tilde{\mathbb{X}}
+ (i \epsilon \tilde{\mathbb{B}}^\dagger \sigma^{\mu} \partial_\mu \tilde{\mathbb{X}} + h.c.)
\\
& +\frac{1}{2} D_\mathbb{B}^2 + \frac{1}{2} D_\mathbb{X}^2 + \epsilon D_\mathbb{B} D_\mathbb{X} ,
\nonumber
\end{align}
where $\hat{\mathcal{W}}_\mathbb{B}$ and $\hat{\mathcal{W}}_\mathbb{X}$ denote the superfield strength associated with the gauge vector superfields $\hat{\mathbb{B}} \supset (\mathbb{B}, \tilde{\mathbb{B}})$ of $U(1)_{Y}$ and
$\hat{\mathbb{X}} \supset (\mathbb{X}, \tilde{\mathbb{X}})$ of $U(1)_D$, respectively.

The component fields $\mathbb{B}$, $\tilde{\mathbb{B}}$, $\mathbb{X}$, and $\tilde{\mathbb{X}}$ correspond to the hypercharge gauge boson, bino, dark photon, and dark photino, respectively. 
The gauge kinetic mixing serves as a portal for the vector bosons ($\mathbb{B}$-$\mathbb{X}$ mixing) and their fermionic superpartners ($\tilde{\mathbb{B}}$-$\tilde{\mathbb{X}}$ mixing).
We adopt the notation $\epsilon = \varepsilon/ \cos \theta_W$, where $\theta_W$ represents the weak mixing angle \cite{Glashow:1961tr}.  The parameter $\varepsilon$ describes the familiar photon-dark photon mixing.

Additionally, there is a cross-term involving the auxiliary fields $D_\mathbb{B}$ and $D_\mathbb{X}$. This interaction can naturally lead to the generation of an effective Fayet–Iliopoulos (FI) $D$-term \cite{Dienes:1996zr, Suematsu:2006wh, Baumgart:2009tn, Cheung:2009qd, Morrissey:2009ur, Cohen:2010kn, Chan:2011aa}, which provides the transfer of SUSY breaking from the dark sector to the visible sector, or vice versa. However, in our scenario, which considers an unbroken $U(1)_D$, there are no additional SUSY breaking contributions from $D$-term mixing.

To explore the physical implications of kinetic mixing it is convenient to move to a basis in which the kinetic terms of the gauge bosons are diagonal. The kinetic terms for each component field can be diagonalized using a $GL(2)$ transformation as outlined in Ref.~\cite{Fabbrichesi:2020wbt}. This process can also be applied at the superfield level as
\begin{align}
\label{diag}
\begin{pmatrix}
\hat{\mathbb{X}} \\ \hat{\mathbb{B}}
\end{pmatrix}
=
\begin{pmatrix}
1 & - \dfrac{\epsilon}{\sqrt{1-\epsilon^2}} \\
0 &  \dfrac{1}{\sqrt{1-\epsilon^2}} 
\end{pmatrix}
\begin{pmatrix}
\cos \omega & - \sin \omega \\
\sin \omega & \cos \omega
\end{pmatrix}
\begin{pmatrix}
\hat{X} \\ \hat{B}
\end{pmatrix}  ,
\end{align}
where $\hat{X}$ and $\hat{B}$ are the physical states after the transformation.\footnote{A requirement that the kinetic terms of the gauge bosons be positive constrains the kinetic mixing to be less than 1~\cite{Burgess:2008ri,Gan:2023jbs}.} While any choice of $\omega$ can diagonalize the kinetic term and is physically equivalent~\cite{Pan:2018dmu}, we adopt the following basis \cite{Holdom:1985ag, Fabbrichesi:2020wbt} for our analysis:\footnote{Although renormalization group running may change the value of $\epsilon$, one can always take the basis of Eq.~\eqref{basis} provided that $U(1)_D$ remains unbroken.}
\begin{align}
\label{basis}
\text{(Our basis)} \quad \sin \omega = 0 , \quad \cos \omega = 1.
\end{align}
Given that the superfield strength is defined as the chiral derivative of an Abelian vector superfield, the $GL(2)$ transformation can be straightforwardly expressed as a linear combination of $\hat{\mathcal{W}}_\mathbb{B}$ and $\hat{\mathcal{W}}_\mathbb{X}$:
\begin{align}
\label{physical}
\hat{\mathcal{W}}_\mathbb{B} = \hat{\mathcal{W}}_B/\sqrt{1-\epsilon^2}  \quad\mathrm{and}\quad \hat{\mathcal{W}}_\mathbb{X} = \hat{\mathcal{W}}_X - \epsilon \hat{\mathcal{W}}_B/\sqrt{1-\epsilon^2}.
\end{align}
This transformation leads to the diagonalized states of the system, effectively illustrating how kinetic mixing impacts the physical properties of the gauge bosons. In this supersymmetric framework, component fields such as gauge bosons, gauginos, and auxiliary fields can be coherently rotated by a unified basis transformation. We note that one can take different rotation angle $\omega$ for each component, without changing any physical results. However, this alignment facilitates a clear delineation of interaction terms in the Lagrangian, even after diagonalizing the kinetic mixing terms.

In this basis, the gauge boson $B$ can be interpreted as the physical $U(1)_Y$ gauge boson since SM particles interact exclusively with $B$. The interaction terms for the gauge boson in this basis are expressed as
\begin{align}
\label{Int}
\mathcal{L} \supset &\; g_Y^{\prime} Y J_Y^{\mu} \mathbb{B}_{\mu} + g_D D J_D^{\mu} \mathbb{X}_{\mu} \\
\label{effInt}
=&\; \left[ - \frac{g_D \epsilon}{\sqrt{1-\epsilon^2}} D J_D^{\mu}
+ g_Y Y J_Y^{\mu} \right] B_{\mu}
+ g_D D J_D^{\mu} X_{\mu}, 
\end{align}
where $g_Y = g_Y^{\prime} / \sqrt{1-\epsilon^2}$ represents the effective $U(1)_Y$ coupling after the diagonalization, and $g_D$ is $U(1)_{D}$ coupling.
$Y$ denotes the hypercharge associated with the visible current $J^{\mu}_Y$, and $D$ represents the $U(1)_{D}$ charge of the dark current $J^{\mu}_D$. Therefore, this basis isolates the specific linear combination of the gauge fields that is invisible to SM particles as the $X$ field, simplifying the interpretation of experimental results.

If $J_Y^{\mu}$ and $J^{\mu}_{D}$ take the same form, the hypercharge is effectively modified by the contribution from the dark gauge symmetry as
\begin{equation}
Y_{\mathrm{eff}} = Y -\frac{g_D }{g_Y} \frac{\epsilon}{\sqrt{1-\epsilon^2}} D.
\label{eq:effCharge}
\end{equation}
This suggests that within the GMSB framework, messengers that carry dark charge could influence the generation of mass for visible superpartners. 
A quantitative analysis will be provided in Section~\ref{ScalarSec}. In general, Eq.~\eqref{effInt} indicates that any particles with a non-zero dark charge will have a fractional hypercharge \cite{Holdom:1985ag,Holdom:1986eq}. As a result, constraints on a massless dark photon primarily arise from the presence of milli-charged exotic particles \cite{Prinz:1998ua, Davidson:2000hf, PhysRevD.75.032004, Jaeckel:2012yz,Vogel:2013raa, Chang:2018rso,Magill:2018tbb, Stebbins:2019xjr,Fiorillo:2024upk}. Therefore, we assume that there are no light states with a non-zero dark charge, allowing the kinetic mixing parameter $\epsilon$ to remain largely unconstrained under these constraints. 

Similar to the treatment of gauge bosons, the gaugino kinetic terms can be diagonalized using the transformation specified in Eq.~\eqref{diag}, following the basis in Eq.~\eqref{basis}. The transformations are given by
\begin{equation}
     \tilde{\mathbb{B}} = \tilde{B}/\sqrt{1-\epsilon^2}  \quad\mathrm{and}\quad \tilde{\mathbb{X}} = \tilde{X}-\epsilon\tilde{B}/\sqrt{1-\epsilon^2}.
     \label{eq:gauginoKinDiag}
\end{equation}
When SUSY is broken, gauginos can acquire masses. 
Kinetic mixing may then generate mixing between the bino and dark photino, leading to observable effects. 
 The extent and nature of these effects depend heavily on the details of the SUSY breaking scenario employed. In the following sections, we will explore the observable impacts of kinetic mixing on the particle mass spectra, particularly in regimes of large kinetic mixing.

The presence of large kinetic mixing leads to novel phenomena not present in the small $\epsilon$ limit. Specifically, with significant kinetic mixing, the $B$ boson, traditionally a gauge boson of the visible sector, can exhibit a stronger coupling to the dark current $J_D$ than to the visible current $J_Y$. This effect is quantitatively expressed by Eq.~\eqref{effInt}, and occurs under the condition
\begin{align}
\label{crit1}
\left|g_Y Y \right| < \left| \frac{g_D D \epsilon}{\sqrt{1-\epsilon^2}} \right|   \quad \Leftrightarrow \quad  \epsilon_{\ast} \equiv \frac{g_Y Y}{\sqrt{(g_Y Y)^2+(g_D D)^2}} < \epsilon.
\end{align}
For example, one obtains $\epsilon_{\ast} = 0.15$, by taking $Y=1/6$, with $g_D = 0.4$ and $D=1$. This clearly shows that the effect of dark current could be significant in the large $\epsilon$ limit. Also, the $J_D$-$B$ and $J_Y$-$B$ couplings may cancel each other at $\epsilon \approx \epsilon_{\ast}$.

Similarly, it is instructive to consider which gauge boson, $B$ or $X$, interacts more strongly with a matter field characterized by hypercharge $Y$ and dark charge $D$, and how this interaction varies with the value of $\epsilon$. This analysis involves comparing the $J_{D,Y}$-$B$ coupling to the $J_D$-$X$ coupling. With significant kinetic mixing, the $X$ boson can interact more strongly with such a matter field than the $B$ boson. This condition can be expressed as \begin{equation} \left| g_Y Y - \frac{g_D D \epsilon}{\sqrt{1-\epsilon^2}} \right| < | g_D D |. \end{equation} Under this scenario, the dark gauge boson $X$ exhibits a stronger interaction with the matter current than the $B$ boson. This criterion becomes particularly relevant when both visible and dark currents are present, highlighting the influence of $\epsilon$ on the interaction dynamics.

The key insight here is that large kinetic mixing induces exotic behaviors. Specifically, with substantial kinetic mixing, the $B$ boson exhibits a stronger coupling with the dark current than with the visible current; similarly, the dark current may couple more strongly with the $B$ boson than with the $X$ boson. These dynamics underscore the significant role of large $\epsilon$ in the interactions between gauge bosons and currents.

Kinetic mixing can affect electroweak symmetry breaking (EWSB), as the value of the $\mu$ parameter is determined by a combination of Higgs mass parameters to ensure EWSB. So, at large $\epsilon$ values, the Higgs mass parameter at the low-energy scale may not satisfy the EWSB condition. This detail is discussed in Section~\ref{ScalarSec} and Appendix~\ref{appendixRenormaliz}. 
These aspects will be further explored in subsequent sections to assess the impact of large $\epsilon$. Caution is necessary when considering large values of $\epsilon$ and $g_D$, and, in particular, the couplings may diverge in the $\epsilon \rightarrow 1$ limit (see Eq.~\eqref{effInt}) leading to a non-perturbative regime and a breakdown of perturbative unitarity. Moreover, from the perspective of UV physics, elevated coupling strengths could lead to the emergence of Landau poles at energy scales close to those of IR physics, an issue addressed in Appendix~\ref{appendixRunning}. Consequently, kinetic mixing cannot be arbitrarily large, although we demonstrate that substantial kinetic mixing remains viable.

Lastly, the large kinetic mixing can impact gauge coupling unification \cite{Redondo:2008zf,Takahashi:2016iph,Daido:2016kez}. While typical gauge coupling unification in the MSSM relies on complete messenger multiplets, substantial kinetic mixing can enable unification even with incomplete messengers, as we argue in Appendix~\ref{App.Gauge coupling unification}. In the following analysis, however, we do not restrict ourselves to the parameter space where gauge coupling unification is achieved.

\section{Dark gauge-mediated supersymmetry breaking (Dark GMSB)}
\label{SUSYbreakSec}

In the gauge-mediated SUSY breaking scenario \cite{Dine:1993yw, Dine:1994vc, Dine:1995ag, Martin:1997ns, Giudice:1998bp,Meade:2008wd} (see Refs.~\cite{Nappi:1982hm,Dine:1981gu,Alvarez-Gaume:1981abe} for earlier pioneering studies), SUSY breaking is communicated from a SM gauge-singlet chiral superfield $\hat{S} \supset (S, \tilde{S})$, which acquires  scalar and $F$-term vacuum expectation values, to the MSSM superpartners through vector-like messenger chiral superfields 
($\hat{\Psi}_i$, $\hat{\bar{\Psi}}_i$) that are charged under the SM gauge symmetries. Note that $\hat S$ is also a singlet under the dark gauge symmetry, so it does not contribute to the mass of the dark photon. In this framework, the masses of the superpartners are generated by the gauge interactions involving these messenger fields. When this model is extended to include a dark sector, the messengers may also carry dark charges, allowing SUSY breaking effects to be transmitted from the dark sector to the visible sector via kinetic mixing (dark GMSB). 

We consider two scenarios for the representations of the messenger fields, as detailed in Table~\ref{tab:represen}. To ensure gauge anomaly cancellation, we assume that the messenger fields are in a vector-like representation.

\begin{table}[tb]
\centering
\begin{tabular}{c|c|c|c}
\hline\hline Scenario & Superfield & Component fields & Representation 
\\ \hline
I &\makecell{$\hat{\Psi}_1$ \\$\hat{\bar{\Psi}}_1$\\$\hat{\Psi}_2$\\ $\hat{\bar{\Psi}}_2$ } &\makecell{$ \psi_1, \tilde{\psi}_1$\\$ \bar{\psi}_1, \tilde{\bar{\psi}}_1$\\ $ \psi_2, \tilde{\psi}_2$\\ $ \bar{\psi}_2, \tilde{\bar{\psi}}_2$} &\makecell{$(\textbf{3}, 1, -1/3, D_{\Psi})$\\$(\bar{\textbf{3}}, 1, 1/3 , -D_{\Psi})$ \\$(1, \textbf{2}, 1/2, D_{\Psi})$\\$(1, \textbf{2}, -1/2, -D_{\Psi})$  }
\\ 
\hline 
\multirow{3}{*}{II} & \multirow{3}{*}{\makecell{$\hat{\Psi}$ \\ $\hat{\bar{\Psi}}$ }}  & \multirow{3}{*}{\makecell{ $ \psi, \tilde{\psi}$\\$ \bar{\psi}, \tilde{\bar{\psi}}$}}  & \multirow{3}{*}{\makecell{$(\textbf{3}, \textbf{2}, 1/6, D_{\Psi})$\\ $(\bar{\textbf{3}}, \textbf{2}, -1/6, -D_{\Psi})$ }} \\  
  & & &  
  \\ & & & 
  \\
\hline\hline
\end{tabular}
\caption{Messenger representations for two distinct scenarios, which we will discuss in this paper. We denote the representation as $(SU(3)_C,SU(2)_L,U(1)_Y,U(1)_{D})$.
We assume the messenger fields are in a vector-like representation to avoid the gauge anomaly. Scenario I employs a $SU(5)$ complete representation (the fundamental ${\bf 5}+\bar{\bf 5}$), whereas Scenario II utilizes a $SU(5)$ incomplete representation. For concreteness, we fix $D_{\Psi}=1$.} 
\label{tab:represen}
\end{table}

In this paper, we consider only $R$-parity conserving terms in the superpotential $W$, and the relevant terms for mediating SUSY breaking are expressed as 
\begin{equation}
W \supset y \hat{S}  \hat{\bar{\Psi}}_i \hat{\Psi}_i,
\label{superpotential}
\end{equation}
where the messenger mass scale is given by $M_{\text{mess}} = y \langle S \rangle$, with $\langle S \rangle$ denoting the vacuum expectation value (VEV) of $S$.

In the GMSB framework, the scale of the soft mass terms is determined by \cite{Dine:1994vc}
\begin{equation}
\label{GMSB}
m_{\text{soft}} \simeq \frac{g^2}{16\pi^2} \frac{F}{M_{\text{mess}}},
\end{equation}
where $g$ is the coupling constant for the gauge interactions involving the messenger fields, and $F$ is the VEV of the auxiliary field of $\hat{S}$, representing the SUSY breaking scale.
The soft mass scale aligns with the electroweak (EW) scale when the ratio $F / M_{\text{mess}} \simeq \mathcal{O}(100)$ TeV. The mass of the messenger fermions $\psi$ and $\bar{\psi}$ is given by $M_{\text{mess}}$, while the masses of the messenger scalars $\tilde{\psi}$ and $\tilde{\bar{\psi}}$ are split by $\sqrt{F}$. The requirement $\sqrt{F} \lesssim M_{\text{mess}}$ ensures positive squared masses for these scalars, as discussed in Ref.~\cite{Giudice:1998bp}.
Given that the soft mass scale should at least match the EW scale, the condition $F / M_{\text{mess}} \gtrsim \mathcal{O}(100)$ TeV is necessary. Consequently, the permissible range for the messenger mass scale is $M_\text{mess} \gtrsim 100 \, \text{TeV}$. 
For subsequent discussions, we will often set 
$F/M^2_{\rm mess} = 2/3$ for concreteness, corresponding to a low messenger scale.

The gravitino mass $m_{3/2}$ is given by 
\begin{equation}
m_{3/2} \simeq \frac{F_0}{\sqrt{3} M_{\text{Pl}}},
\end{equation}
where $F_0$ represents the fundamental scale of SUSY breaking, related to $F$ by $F = k F_0$, with $k$ depending on the mechanism through which SUSY breaking is transmitted to the messenger fields \cite{Giudice:1998bp}; we assume $k \simeq 1$. 
Generally, $m_{3/2}$ is much smaller than the soft mass $m_{\text{soft}}$ from GMSB when $\sqrt{F}$ is considerably lower than $M_{\text{Pl}}$.

Consequently, in a standard GMSB scenario, the gravitino emerges as the lightest supersymmetric particle (LSP). If $R$-parity is conserved, the LSP is stable, potentially leading to a significant relic density. The implications of such an LSP overabundance are further explored in Section~\ref{sec:Cosmology}. Interestingly, the lightest messenger state can also be stable, as it could be the lightest particle carrying a dark charge. Due to the fractional contribution of the dark charge to its $U(1)$ hypercharge [Eq.~\eqref{eq:effCharge}], this state would manifest as a massive charged particle. If the reheating temperature is not too high, these messenger particles could exist in the present universe without conflicting with observations. We provide the viable parameter space where such a messenger could exist in Section~\ref{sec:Cosmology}.

\begin{figure}[tb]
    \centering
    \begin{subfigure}{0.49 \textwidth}  
        \includegraphics[width=0.99\textwidth]{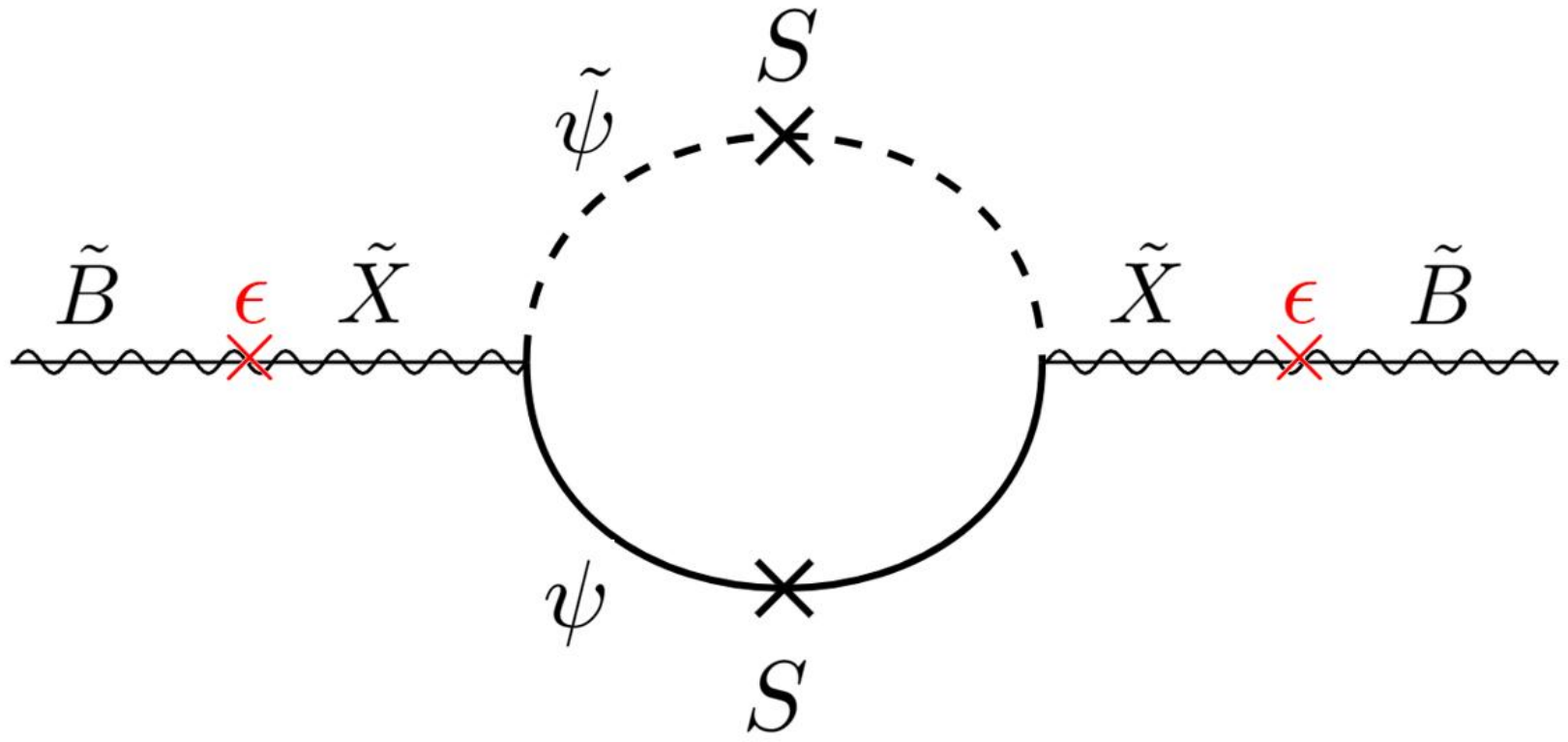}
        \subcaption{}
        \label{fig:Diagram1}
    \end{subfigure}
    \begin{subfigure}{0.49 \textwidth}  
       \includegraphics[width=0.99\textwidth]{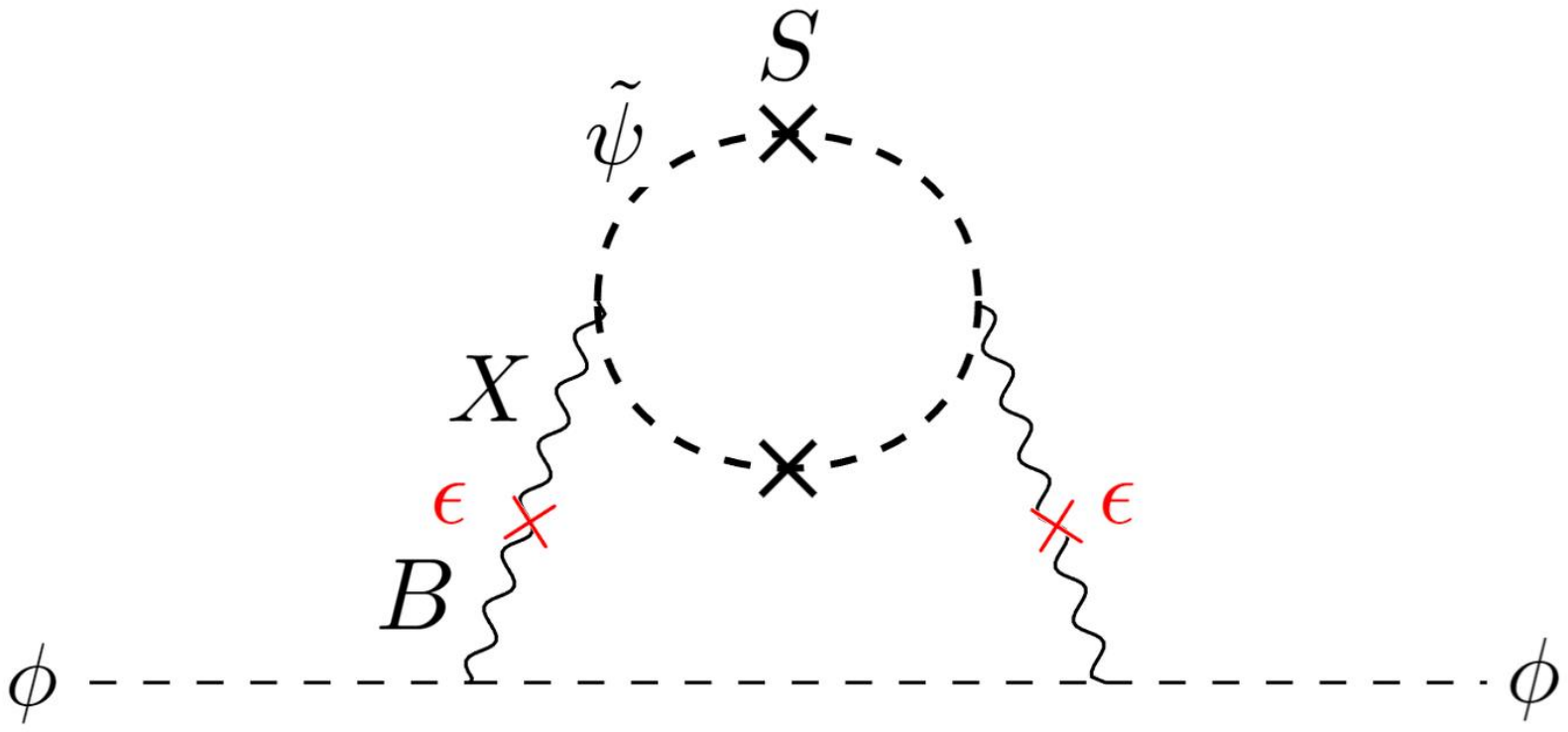}
       \subcaption{}
        \label{fig:Diagram2}
    \end{subfigure}
    \caption{(a) One of the diagrams for the bino mass generated by dark GMSB at the one-loop level is described here. In this diagram, the messengers $\psi$ and $\tilde{\psi}$ act as a dark current. SUSY breaking from the singlet $\hat{S}$ is initially transferred to the dark sector (dark photino $\tilde{X}$), and subsequently to the hypercharge sector (bino $\tilde{B}$) through kinetic mixing. As a result, the bino mass acquires $\epsilon$ dependence.
    (b) One of the diagrams that gives the scalar mass. The scalar squared soft mass term is affected by kinetic mixing. $\phi$ represents a scalar charged under $U(1)_Y$, such as the Higgs scalar or a sfermion. 
    }
    \label{Diagram1}
\end{figure}

Now, let us discuss the generation of soft terms when the messenger fields possess a dark charge. A naive expectation is the generation of soft terms in the dark sector, i.e., the dark photino mass term. However, the effect of dark GMSB is not limited to the dark sector. If kinetic mixing between the visible and dark sectors is present, this configuration allows the SUSY breaking effect to permeate into the visible sector through the following steps.

Initially, SUSY breaking is transmitted from the singlet $\hat{S}$ to the messenger fields $\hat{\Psi}$ and $\hat{\bar{\Psi}}$ via the superpotential term described in Eq.~\eqref{superpotential}. Subsequently, the messenger fields $\hat{\Psi}$ and $\hat{\bar{\Psi}}$ communicate this SUSY breaking to $\hat{X}$ in the dark sector via the $U(1)_{D}$ gauge interaction. Finally, the SUSY breaking effect is transferred from $\hat{X}$ in the dark sector to $\hat{B}$ in the hypercharge sector via kinetic mixing. This description is based on the original basis with Eq.~\eqref{kineticMix}.

\begin{figure}[t]
    \centering
    \includegraphics[width=0.7\textwidth]{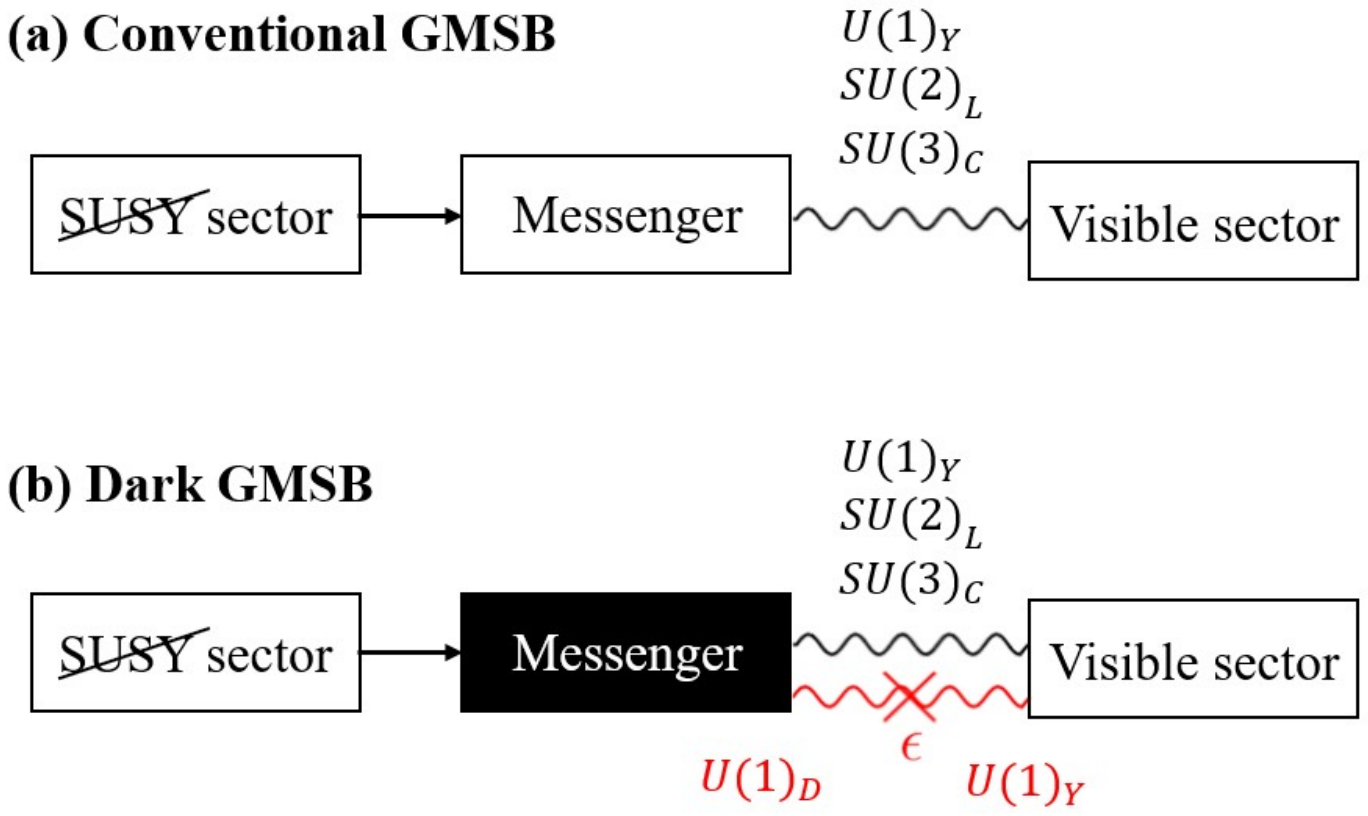}
    \caption{
    Schematic diagram for conventional GMSB and dark GMSB models. SUSY breaking is first transferred to the messenger and then to the visible sector. In the dark GMSB model, the messenger has a dark charge, which can provide additional contributions to the hypercharge part $U(1)_Y$ via the kinetic mixing $\epsilon$.
    }
    \label{schematic}
\end{figure}

This sequence effectively generates soft masses for the bino and hypercharged scalar superpartners
through kinetic mixing, as depicted in Figure~\ref{Diagram1}. We term this mechanism of SUSY breaking transfer via kinetic mixing  {\it dark gauge-mediated supersymmetry breaking}. 
Dark GMSB generates additional contributions  to the soft terms for the bino and hypercharge fields only, as represented in Figure~\ref{schematic}.

A distinctive feature of dark GMSB is the $\epsilon$-dependence of the soft mass terms, which can be quantitatively described as 
\begin{equation}
\label{soft} 
m_{\text{soft}}(g_D, \epsilon) \simeq \frac{g_\text{eff}^2(g_D, \epsilon)}{16\pi^2} \frac{F}{M_{\text{mess}}}. 
\end{equation} Here, $g_{\text{eff}}(g_D, \epsilon)$ represents the effective coupling, which acquires $\epsilon$-dependence following diagonalization, as shown in Eq.~\eqref{effInt}. In the upcoming sections, we will explore how this kinetic mixing influences the mass scales of various SUSY particles, specifically within the Higgs, sfermion, neutralino, and chargino sectors.
We will first discuss the scalar sectors, and demonstrate how the $\mu$ term also depends on kinetic mixing to satisfy the EWSB condition and then move on to the other sectors.

\section{Scalar soft masses in the dark GMSB}
\label{ScalarSec}

The scalar squared soft mass terms, $m^2_{\phi_i}$, are generated at the two-loop level by gauge-mediated SUSY breaking mechanisms. The expression for the mass term arising from the non-Abelian gauge group, $G_a$, is given by 
\begin{align}
\label{m2phi}
m^2_{\phi_i} = 2 \sum_{\Psi} N_{G_a}   \left( \frac{g_a^2}{16\pi^2} \right)^2 \left( \frac{F}{M_\text{mess}} \right)^2 2 S_a C_a \mathcal{F}(x),
\end{align}
where $\mathrm{Tr}[T_a^{\alpha}T_a^{\beta}]=S_a \delta^{\alpha\beta}$ is defined as the trace of the generators ($T^{\alpha}_a$) of the corresponding gauge group representation, $C_a \mathds{1}= \sum_\alpha T_a^{\alpha}T_a^{\alpha}$ represents the quadratic Casimir of the scalar under the gauge group representation associated with the coupling $g_a$, and $\sum_{\Psi} N_{G_a}$ is the summation over vector-like messengers with the number of $G_a$ multiplets. Specifically for $SU(N)$, $S_a = 1/2$, and $C_a = \frac{N^2-1}{2N}$ for fundamental representation.
The loop function $\mathcal{F}(x)$ in Eq.~\eqref{m2phi} is \cite{Dimopoulos:1996gy}
\begin{align}
\mathcal{F}(x) = \frac{1+x}{x^2} \left[ \ln (1+x) - 2 \text{Li}_2\left( \frac{x}{1+x} \right) + \frac{1}{2} \text{Li}_2\left( \frac{2x}{1+x} \right) \right] + (x \rightarrow -x),
\end{align}
where $\text{Li}_2(x)$ is the dilogarithm function. Here, $x = F/M_{\text{mess}}^2$, and $\mathcal{F} \simeq 1$ unless $x \gtrsim 0.95$.

We calculate the expression for the soft mass from the Abelian gauge symmetry by taking  $S_a= Y_{\mathrm{eff},\Psi}^2$, $C_a=Y_{\phi_i}^2$:
\begin{equation}
    m^2_{\phi_i} = 2 \sum_{\Psi} N_{U(1)}  \left( \frac{g_Y^2}{16\pi^2} \right)^2\left( \frac{F}{M_\text{mess}} \right)^2 2 Y_{\mathrm{eff},\Psi}^2 Y_{\phi_i}^2 \mathcal{F}(x).
\end{equation}  
One can interpret the result that the messengers acquire fractional charges proportional to their dark charges due to the kinetic mixing \cite{Holdom:1986eq}.

Kinetic mixing facilitates the mediation of SUSY breaking to the Higgs or sfermion sector, as depicted in Figure~\ref{Diagram1}. 
The squared soft mass term of the Higgs, incorporating the effective charge detailed in Eq.~\eqref{eq:effCharge}, can be expressed as
\begin{equation}
m_{H_i}^2(g_D, \epsilon)  = \sum_{\Psi} \tilde{M}_{\text{mess,1}}^2 
\left[ N_{SU(2)}\frac{3g_2^4}{8}
+  N_{U(1)}g_Y^2Y_{H_i}^2 \left( g_Y Y_\Psi - \frac{g_D \epsilon D_\Psi}{\sqrt{1-\epsilon^2}}  \right)^2  \right].
\label{epsilonHiggs2}
\end{equation}
Here, $\tilde{M}_{\text{mess,1}}^2 \equiv 4 \mathcal{F}(x) \left[\frac{ x M_{\text{mess}}}{16\pi^2}\right]^2$. 
The squared soft mass can increase with the degree of kinetic mixing. Similarly, the soft masses of the sfermions are given as
\begin{align}
\label{mqL}
m_{\tilde{q}_L}^2 (g_D, \epsilon)
& = \sum_{\Psi} \tilde{M}_{\text{mess,1}}^2
\left[ 
N_{SU(3)}\frac{2 g_3^4 }{3}
+N_{SU(2)}\frac{3g_2^4 }{8}
+ N_{U(1)} g_Y^2 Y^2_{\tilde{q}_L}\left( g_Y Y_\Psi -\frac{g_D \epsilon D_\Psi}{\sqrt{1-\epsilon^2}}  \right)^2 
\right],
\\
\label{muR}
m_{\tilde{u}_R}^2 (g_D, \epsilon)
& = \sum_{\Psi} \tilde{M}_{\text{mess,1}}^2
\left[ 
N_{SU(3)}\frac{2g_3^4}{3}
+ N_{U(1)}g_Y^2 Y^2_{\tilde{u}_R} \left( g_Y Y_\Psi - \frac{g_D \epsilon D_\Psi}{\sqrt{1-\epsilon^2}}  \right)^2
\right],
\\
m_{\tilde{d}_R}^2 (g_D, \epsilon)
& = \sum_{\Psi} \tilde{M}_{\text{mess,1}}^2
\left[ 
 N_{SU(3)}\frac{2 g_3^4}{3}
+ N_{U(1)} g_Y^2Y^2_{\tilde{d}_R}\left( g_Y Y_\Psi - \frac{g_D \epsilon D_\Psi}{\sqrt{1-\epsilon^2}}  \right)^2
\right],
\\
m_{\tilde{\ell}_L}^2 (g_D, \epsilon)
& = \sum_{\Psi} \tilde{M}_{\text{mess,1}}^2
\left[ 
 N_{SU(2)}\frac{3 g_2^4}{8}
+ N_{U(1)} g_Y^2Y^2_{\tilde{\ell}_L} \left( g_Y Y_\Psi - \frac{g_D \epsilon D_\Psi}{\sqrt{1-\epsilon^2}}  \right)^2
\right],
\\
m_{\tilde{e}_R}^2 (g_D, \epsilon)
& = \sum_{\Psi} \tilde{M}_{\text{mess,1}}^2 \, 
\left[N_{U(1)}g_Y^2Y^2_{\tilde{e}_R}\left( g_Y Y_\Psi - \frac{g_D \epsilon D_\Psi}{\sqrt{1-\epsilon^2}}  \right)^2\right].
\label{eR}
\end{align}
 The dark GMSB effect is represented by the combination of $g_D$ and $\epsilon$ as $g_{\mathrm{eff}} \equiv g_D \epsilon/\sqrt{1-\eps^2}$.

In addition to the given soft masses, there is a mixing between the left- and right-handed sfermions due to the trilinear $A$ terms, $D$ terms, and $F$ terms. 
Typically, these contributions are insignificant because the size of the mixing terms are governed by the masses of SM particles.

\subsection{Higgs mass and electroweak symmetry breaking}

The scalar potential of the neutral Higgs fields is given by 
\begin{align}
V(H_u^0, H_d^0) = & \sum_{i=u,d} \left( |\mu|^2 + m_{H_i}^2 \right) |H_i^0|^2 - \left(b_{\mu} H_u^0 H_d^0 + \text{h.c.}\right) \nonumber \\
& + \frac{1}{8} (g_Y^2 + g_2^2) \left( |H_u^0|^2 - |H_d^0|^2 \right)^2.
\end{align}
At the potential minimum, where $\partial V / \partial H_u^0 = \partial V / \partial H_d^0 = 0$, EWSB requires 
\begin{align}
\label{EWSBcond}
b_{\mu} &= \frac{\sin (2\beta) }{2} \left[2 |\mu|^2 + m_{H_u}^2 + m_{H_d}^2 \right],\\
\label{eq:EWSBcond2}
|\mu|^2 &= - \frac{m_Z^2}{2} - \frac{m_{H_u}^2 + m_{H_d}^2}{2} + \frac{m_{H_u}^2 - m_{H_d}^2}{2 \cos (2\beta)}.
\end{align}
The $\mu$-term is a supersymmetric mass parameter appearing in the superpotential that couples the two Higgs doublets, while the $b_\mu$-term is a soft supersymmetry-breaking bilinear term in the Higgs potential.
Generating $\mu$ and $b_\mu$ terms required by the above relations for EWSB without fine tuning is an acute problem in the GMSB~\cite{Giudice:1998bp}. In this paper, we do not address this issue directly; instead, as commonly done in typical studies of GMSB, we assume that $\mu$ and $b_\mu$ can be generated to satisfy the relations in Eqs.~\eqref{EWSBcond} and \eqref{eq:EWSBcond2}.
In the conventional GMSB, EWSB can occur through the renormalization group (RG) running of 
$m_{H_u}^2$ and $m_{H_d}^2$, which generates the hierarchy $m_{H_u}^2<m_{H_d}^2$ at the EW scale. 
This is the case in dark GMSB as well.\footnote{In both conventional GMSB and dark GMSB, $m_{H_u}^2 = m_{H_d}^2$ 
at the messenger scale, and thus cannot satisfy Eq.~\eqref{eq:EWSBcond2}.} A key difference in dark GMSB, however, is that the bino and mixed bino-dark photino soft masses can contribute significantly to the running of $m_{H_u}^2$ and $m_{H_d}^2$. Large values of $g_D$ and $\epsilon$ significantly enhance these soft masses parameters, eventually preventing $m_{H_u}^2$ from being driven to negative values.
(See Eqs.~\eqref{MXB}, \eqref{MB}, and Appendix~\ref{appendixRenormaliz}.)
The one-loop beta functions of $m_{H_u}^2$ and $m_{H_d}^2$ receive identical negative contributions from $g_D$ and $\epsilon$. Thus, regardless of the value of $\beta$, large $g_D$ and $\epsilon$ drive the right-handed side of Eq.~\eqref{eq:EWSBcond2} to become negative and must therefore be constrained. On the other hand, this also implies that $|\mu|$ could be significantly smaller than in conventional GMSB for moderate values of $g_D$ and $\epsilon$, which brings down the mass of the higgsino. 

\begin{figure}[tb]
    \centering
    \begin{subfigure}{0.477 \textwidth}    \includegraphics[width=0.99\textwidth]{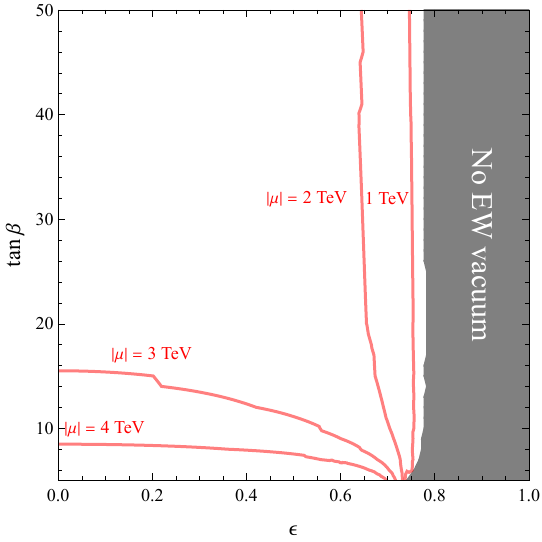}
        \subcaption{}
        \label{fig:muconstA}
    \end{subfigure}
        \begin{subfigure}{0.49 \textwidth}
        \includegraphics[width=0.99\textwidth]{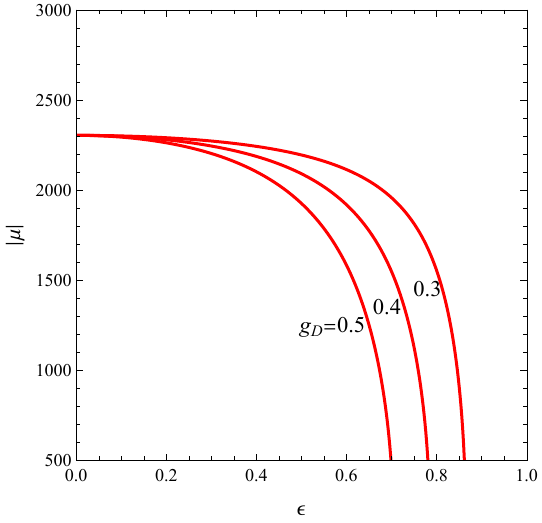}
        \subcaption{}
        \label{fig:muconstB}
    \end{subfigure}
    \caption{(a) 
    Isocontours of $|\mu|$ 
    in the $\tan\beta - \epsilon$ plane for a complete $SU(5)$ messenger multiplet (Scenario I). Here, $F/M$ is adjusted to consistently achieve $m_{h^0} = 125$ GeV, with $F/M_{\mathrm{mess}}^2 = 2/3$ fixed. The EW vacuum cannot be achieved in the gray region, which is therefore ruled out.
    (b) The variation of $|\mu|$ with respect to $\epsilon$ is shown for $g_D=0.3$, $0.4$, and $0.5$. This shows that $\mu$ sharply turns imaginary as $\epsilon$ increases, 
    obstructing successful EWSB. We take $F/M_{\mathrm{mess}} = 800$ TeV, $M_{\mathrm{mess}} = 1200$ TeV, and $\tan \beta = 15$ — values that match our benchmark point except for $g_D$ (see Table~\ref{input}). }
    \label{muconst}
\end{figure}

We use the public software {\tt SOFTSUSY 4.1.20} \cite{Allanach:2001kg} to include 2-loop renormalization group equations for evaluating EWSB conditions and calculating MSSM parameters near the EW scale. In order to implement the dark GMSB model, we modify the relevant parts of the code, incorporating additional degrees of freedom from dark gauge coupling, kinetic mixing, and gaugino mass matrix. Additionally, we adjust the corresponding $\beta$-functions to account for these parameters. The $\beta$-functions are listed in Appendix~\ref{appendixRenormaliz}. 

Figure~\ref{muconst} illustrates the effect of dark GMSB on $|\mu|$ in detail. Figure~\ref{fig:muconstA} shows the 
values of $|\mu|$ as a function of $\tan\beta$ and $\epsilon$ needed to achieve EWSB and the correct Higgs mass. Since $1/\cos(2\beta)$ diverges when $\tan \beta \rightarrow 1$, the contour of $|\mu|$ becomes parallel to the constant $\tan \beta$ line. This figure also shows one crucial 
feature of dark GMSB, which is that $|\mu|$ is affected by $\epsilon$. The value of $|\mu|$ decreases as $\epsilon$ increases, with the variation becoming more sensitive at larger $\epsilon$, as shown in Figure~\ref{fig:muconstB}.
The boundary of the gray region is where $|\mu|\simeq 0$. Higher-order loop corrections slightly shift the precise location of the boundary of the gray region; however, the qualitative behavior, where $|\mu|$ steeply decreases near the boundary is expected to remain similar. If the values of $g_D$ and $\epsilon$ lie closer to the gray region, it makes $|\mu|$ to be small. Therefore, the spectrum of dark GMSB can contain a light higgsino, with mass determined by $|\mu|$, which will be demonstrated in Section~\ref{GauginoSec}.

A crucial requirement of any supersymmetric model is the reproduction of successful  EWSB and observed Higgs mass of  $m_{h^0} = 125$ GeV \cite{ATLAS:2012yve,CMS:2012qbp}. It is known that the tree-level Higgs potential in the MSSM cannot reproduce the correct SM-like Higgs mass
and higher order corrections, especially from the top and stop quark, are required to this end. For instance, the corresponding one-loop correction to SM-like light Higgs mass is given by \cite{Casas:1994us,Carena:1995bx,Haber:1996fp,Draper:2011aa}
\begin{equation}
\begin{split}
\label{HiggsFormula}
m_{h^0}^2 & = m_Z^2 \cos^2 2\beta + \frac{3 m_t^4}{2\pi^2 v^2} \left[\ln \left( \frac{M_s^2}{m_t^2} \right)+\frac{X_t^2}{M_s^2}\left(1-\frac{X_t^2}{12M_s^2}\right)\right]\\
& \simeq m_Z^2 \cos^2 2\beta + \frac{3 m_t^4}{2\pi^2 v^2} \ln \left( \frac{M_s^2}{m_t^2} \right),
\end{split}
\end{equation}
where $M_s = \sqrt{m_{\tilde{t}_1}m_{\tilde{t}_2}}$, and $X_t=A_t -\mu \cot \beta$.
The first term corresponds to the tree-level mass. Due to the minimal contributions of the trilinear $A$-terms in the GMSB, $X_t^2/M_s^2 \ll 1$ and $M_s$ must be in the range of $\mathcal{O}(1-10)$ TeV, depending on $\beta$, to accommodate the observed Higgs mass of $m_{h^0} = 125$ GeV \cite{Arbey:2011ab, Draper:2011aa}. 

The dark gauge mediation effect can induce a small fractional change in the Higgs mass through the change of soft stop mass, Eqs.~\eqref{mqL}, \eqref{muR}. The effect can be explicitly written by 
\begin{equation}
    \delta m^2_{h^0}(g_{\rm eff}) \simeq \frac{17m_t^4}{32\pi^2v^2} \dfrac{
    \sum\limits_{\Psi} N_{U(1)} g_Y^2 g_{\mathrm{eff}} D_{\Psi} \left(-2 g_Y Y_{\Psi}+g_{\mathrm{eff}} D_{\Psi} \right) }{\sum\limits_{\Psi} N_{SU(3)} g_3^4} .
\end{equation} 
We find that Higgs mass can only be modified by at most $\mathcal{O}(0.1)$ GeV when considering the constraints on $g_D$ and $\epsilon$ from the EW symmetry breaking discussed in the following paragraphs of this section. Therefore, once $\tan\beta$ is fixed, the ratio $F/M_{\text{mess}}$ is determined to ensure the correct Higgs mass.

\subsection{Scalar particle spectrum}

\begin{table}[tb]
    \centering
    \begin{tabular}{c| c| c| c|  c}
\hline\hline
     $F/M_{\text{mess}}$ & $M_{\text{mess}}$ & $g_D$  & $\tan \beta$ &  $F/M_{\mathrm{mess}}^2$
    \\ \hline 
   800 TeV &  1200 TeV & 0.4  & 15 & 2/3 \\ \hline \hline
    \end{tabular}
    \caption{The parameters listed in this table are utilized for illustrations unless specifically stated otherwise.
    This parameters setup is fit to
    obtain the observed SM-like Higgs mass of $m_{{h^0}}=125$ GeV.
    } 
    \label{input}
\end{table}

\begin{figure}[tb]
    \centering
    \begin{subfigure}{0.49\textwidth}
    \includegraphics[width=0.99\textwidth]{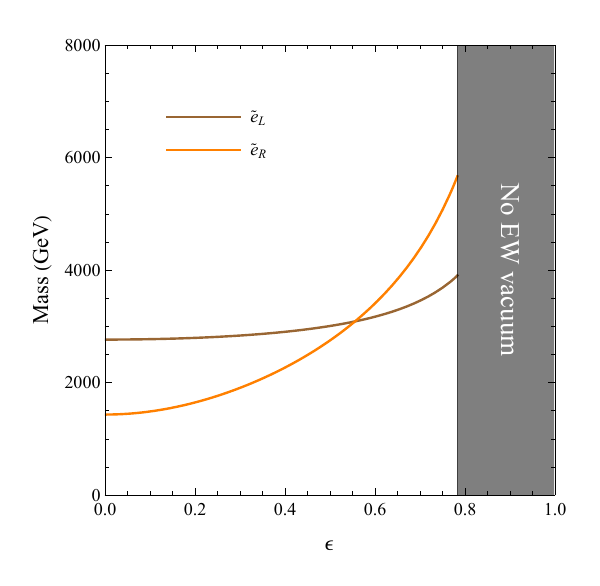}    \subcaption{Scenario I}
    \end{subfigure}
    \begin{subfigure}{0.49\textwidth}
    \includegraphics[width=0.99\textwidth]{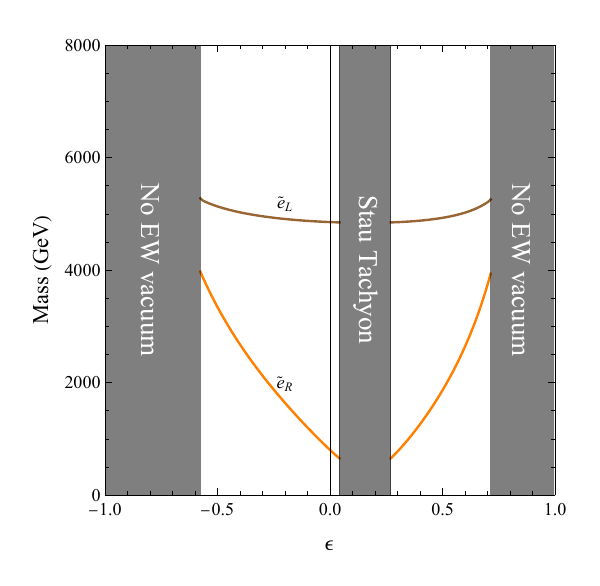} 
    \subcaption{Scenario II}
    \end{subfigure}
    \caption{
    The mass spectrum of $\tilde{e}_R$ and $\tilde{e}_L$ 
    as a function of the kinetic mixing $\epsilon$ in both scenarios.
    The parameters are set as in Table~\ref{input}.
    Since the hypercharge of $\tilde{e}_R$ is larger, it is more sensitive to the change of $\epsilon$. 
    There is no EW vacuum 
    for large enough $|\epsilon|$ (gray regions).
    (a): The mass spectrum is symmetric under $\epsilon \rightarrow -\epsilon$ since the soft mass terms only contain quadratic terms of $\epsilon$. The minimum (maximum) of the masses appears at $|\epsilon| = 0.0 \; (0.783)$.
    (b): The stau can become tachyonic in the specific interval of $\epsilon$ where the effective coupling part in Eq.~\eqref{eR} nearly vanishes. The minimum (maximum) of the sfermion masses appears at $\epsilon = 0.042 \; (-0.576)$.
    }
    \label{eLeR}
\end{figure}

\begin{figure}[tb]
    \centering
    \includegraphics[width=0.56\textwidth]{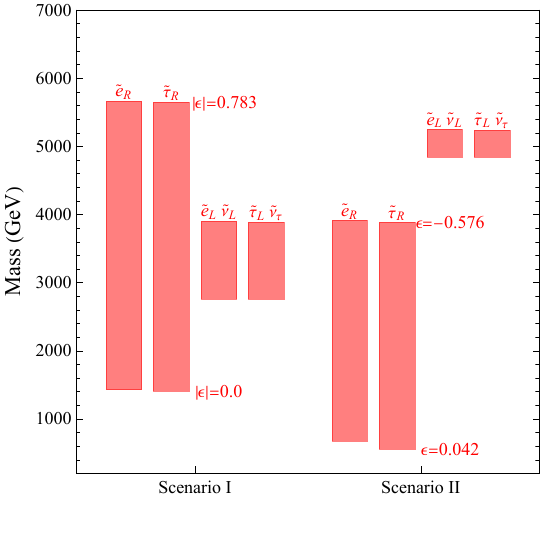} 
    \caption{
    The mass spectrum of the sleptons for dark GMSB scenarios. 
    The parameters are set as in Table~\ref{input}. 
    In Scenario I, the lower (upper) boundary of the mass bands correspond to $|\epsilon| = 0.0$ (0.783). The masses of $\tilde{e}_R$ and $\tilde{\tau}_R$ are particularly sensitive to kinetic mixing due to the large hypercharge of the sfermions.
    In Scenario II, the bands represent the full possible range of the mass spectrum.
    The minimum (maximum) point corresponds to $\epsilon = 0.042 \; (-0.576)$, representing the full mass range. 
    }
    \label{slepton}
\end{figure}

\begin{figure}[tb]
    \centering
    \begin{subfigure}{0.49\textwidth}
    \includegraphics[width=0.99\textwidth]{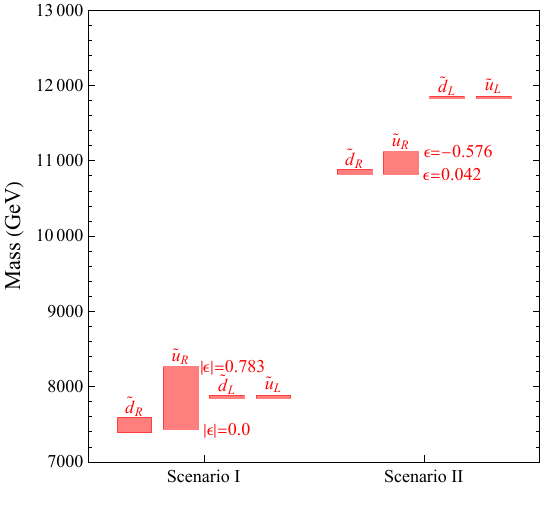}  
    \subcaption{First and second generations}
    \end{subfigure}
    \begin{subfigure}{0.49\textwidth}
    \includegraphics[width=0.99\textwidth]{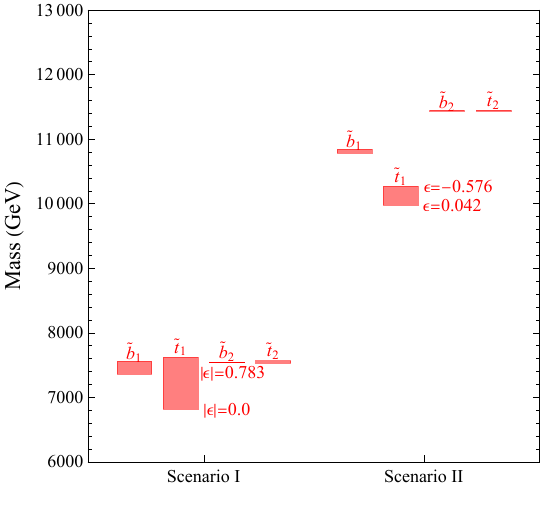}
    \subcaption{Third generation}
    \end{subfigure}
    \caption{
    The mass spectrum of (a) the first/second generation squarks and (b) the third generation squarks for dark GMSB scenarios. The parameter setup and notations are the same as in Figure~\ref{slepton}. 
    }
    \label{squark}
\end{figure}

The resulting mass spectrum of the scalar particles shows dependence on the kinetic mixing $\epsilon$, but it differs for the two scenarios we consider. 
The parameters are set as follows: $g_D = 0.4$, $F/M_{\text{mess}}^2 = 2/3$, $\tan\beta = 15$, and $F/M_{\text{mess}} = 800$ TeV.
These values are chosen to align with the SM-like Higgs mass and the EWSB conditions discussed in the previous subsection. The benchmark parameter values are highlighted in Table~\ref{input}.
As an example, in Figure~\ref{eLeR} we show the masses of $\tilde{e}_R$ and $\tilde{e}_L$ as a function of the kinetic mixing $\epsilon$ for both scenarios.
In Scenario I, the terms linear 
in $\epsilon$ are canceled out in Eq.~\eqref{epsilonHiggs2}, leaving only terms quadratic in $\epsilon$. As a result, the mass spectrum increases monotonically with the magnitude of $|\epsilon|$ and is symmetric under $\epsilon \rightarrow -\epsilon$. 
In Scenario II, however, the soft mass contains a term linear in $\epsilon$, and the soft mass at the messenger scale vanishes when the effective coupling term, $g_Y Y_\Psi - g_D \epsilon D_\Psi / \sqrt{1-\epsilon^2}$, becomes zero for a non-zero value of $\epsilon$. 
As a result, the stau could turn tachyonic around this choice of $\epsilon$ during RG evolution, as shown in Figure~\ref{eLeR}. Such points are excluded, as they indicate the breakdown of $U(1)_{\mathrm{em}}$~\cite{Ibe:2007km}.

The mass spectrum of the sfermions is illustrated in Figs.~\ref{slepton} and \ref{squark} for sleptons and squarks, respectively. 
The minimum and maximum points of the boxes for Scenario II correspond to $\epsilon = 0.042$, and $-0.576$, respectively.
While the dark GMSB effects modify the mass of all sfermions, the most drastic changes occur in the sleptons, especially right-handed sleptons. This pronounced effect arises because the slepton masses lack contributions from the strong interaction, and right-handed sleptons do not receive the 
$SU(2)_L$ interaction contribution either. 
Additionally, 
the right-handed sleptons possess the largest hypercharge and thus experience the largest effects. In the conventional GMSB scenarios, the left-handed sleptons are heavier than the right-handed ones. However, this hierarchy can be reversed when there is a large $g_{\mathrm{eff}}$, which could be a distinct characteristic of the dark GMSB model. In particular, we see that in Scenario II the sleptons can be relatively light, even below the TeV scale.

Inside two Higgs doublets, besides the SM-like Higgs, there exist another neutral scalar, $H^0$, one neutral pseudo-scalar, $A^0$, and a 
charged scalar, $H^{\pm}$. (While many extensions of the SUSY model include an extended Higgs sector \cite{Barger:2006dh}, the dark GMSB model with a massless dark photon does not introduce additional Higgs scalars.)
The tree-level mass spectrum for these Higgs sector components is expressed as
\begin{align}
m_{A^0}^2 & = 2|\mu|^2 + m_{H_u}^2 + m_{H_d}^2, \\
m_{H^0}^2 & = \frac{1}{2} \left[ m_{A^0}^2 + m_Z^2  +\sqrt{ \left(m_{A^0}^2 - m_Z^2 \right)^2 + 4 m_Z^2 m_{A^0}^2 \sin^2 (2\beta) } \right], \\
m_{H^\pm}^2 & = m_{A^0}^2 + m_W^2,
\end{align}
where $m_W^{}$ is the mass of the $W$ boson. 

\begin{figure}[tb]
    \centering
    \includegraphics[width=0.49\textwidth]{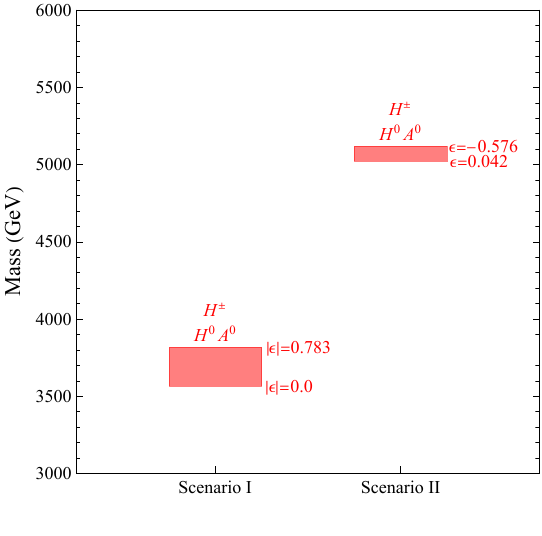}
    \caption{
    The heavy Higgs mass spectrum for dark GMSB scenarios is depicted.
    The parameter setup and notations are the same as in Figure~\ref{slepton}.
    }
    \label{Higgs}
\end{figure}

Figure~\ref{Higgs} presents the heavy Higgs mass spectrum for Scenarios I and II as a function of $\epsilon$. The heavy Higgs spectrum depends on $g_{\mathrm{eff}}$ through the mild change in $m_{\tilde{t}}$. An increase in $g_{\mathrm{eff}}$ results in higher masses for $A^0$, $H^0$, and $H^\pm$. These masses are nearly degenerate because the contributions from $m_Z$ and $m_W$ are substantially smaller than those from the soft terms.
Given the range of $\mu$, the condition $m_{A^0} \gg m_Z$ consistently holds. In this regime, the tree-level mass of $h^0$ remains relatively insensitive to changes in kinetic mixing. 
Specifically, variations in the mass of $h^0$ are limited to less than $\mathcal{O}(0.1)$ GeV over the entire allowed range of kinetic mixing $\epsilon$.

In summary, once the basic model parameters, $\tan\beta$, $g_D$, $\epsilon$, and $M_{\mathrm{mess}}$ are chosen, other parameters such as $F/M_{\mathrm{mess}}$, $\mu$ and $b_{\mu}$ are determined via the conditions on the observed Higgs mass and EW symmetry breaking.
Furthermore, we have seen that very large values of $g_{\rm eff}$  are incompatible with these conditions.
The mass spectrum of the Higgs and sfermion sectors is influenced by kinetic mixing. A distinctive feature is the sensitivity of sleptons to variations in $g_{\mathrm{eff}}$. Especially right-handed sleptons have a significant dependence on $g_{\mathrm{eff}}$, while squarks and exotic Higgs states are less sensitive to $g_{\mathrm{eff}}$ due to the contribution from the strong interaction.

\section{Gaugino soft masses in the dark GMSB}
\label{GauginoSec}

The inclusion of kinetic mixing affects the neutralino mass spectrum, particularly due to effects related to the dark photino. Furthermore, as previously discussed, even though it preserves supersymmetry the $\mu$ term obtains kinetic mixing dependence via the requirements of successful EWSB.
This causes the higgsino-dominant neutralino and chargino masses to change with $\epsilon$. In contrast, the wino and gluino remain unaffected, as the couplings of messengers to these sectors do not involve kinetic mixing.

\subsection{Neutralino sector}
\label{subSec:Neutralino sector}

To derive explicit formulas for these components, we start from the basis where the gaugino kinetic terms are already diagonalized, as achieved by the transformation given in Eq.~\eqref{eq:gauginoKinDiag}. Diagonal gaugino mass terms, denoted as $M_a$, are generated at the one-loop level in gauge mediation scenarios,
\begin{align}
M_a = \sum_{\Psi} N_{G(a)} \frac{g_a^2}{16\pi^2} \frac{F}{M_{\text{mess}}} 2 S_a \mathcal{G}(x).
\end{align}
Here, the loop function $\mathcal{G}(x)$ is
\begin{align}
\mathcal{G}(x) = \frac{1}{x^2} \left[ (1+x) \ln(1+x) + (1-x) \ln(1-x) \right],
\end{align}
where $x = F / M_{\text{mess}}^2$. This function approximates to $\mathcal{G} \simeq 1$ unless $x \gtrsim 0.7$.

For Abelian gauginos, $M_a$ is expressed as
\begin{align}
M_a = \sum_{\Psi} N_{U(1)} \frac{g_a^2}{16\pi^2} \frac{F}{M_{\text{mess}}} 2 Q_a^2 \mathcal{G}(x),
\end{align}
where $Q_a$ is the charge of the messenger corresponding to the gauge group, i.e., $Q_a = Y_{\mathrm{eff},\psi}$ for $U(1)_Y$, and $Q_a = D_\psi$ for $U(1)_D$. 
There is also an off-diagonal mass mixing term between the bino and dark photino, which is obtained by replacing $g_a^2 Q_a^2$ with $g_Y g_D Y_{\mathrm{eff},\psi} D_\psi$ in the above formula.

The mass sub-matrix for the dark photino and bino, considering the effects of kinetic mixing, is given by
\begin{align}
\label{submatrix}
\mathbf{M}_{\tilde{N}}^{2 \times 2}
& =\begin{pmatrix} 
M_D & M_K \\
M_K & M_1
\end{pmatrix}.
\end{align}
The components of $\mathbf{M}_{\tilde{N}}^{2 \times 2}$ are
\begin{align}
\label{MX}
M_D(g_D) & = \sum_{\Psi} N_{U(1)} g_D^2 D_\Psi^2 \tilde{M}_{\text{mess,2}}, \\
\label{MXB}
M_K(g_D,\epsilon) & = \sum_{\Psi} N_{U(1)} g_D D_\Psi \left( g_Y Y_\Psi - \frac{g_D \epsilon D_\Psi}{\sqrt{1-\epsilon^2}} \right) \tilde{M}_{\text{mess,2}}, \\
\label{MB}
M_1(g_D,\epsilon) & = \sum_{\Psi} N_{U(1)} \left( g_Y Y_\Psi - \frac{g_D \epsilon D_\Psi}{\sqrt{1-\epsilon^2}} \right)^2 \tilde{M}_{\text{mess,2}},
\end{align}
where $\tilde{M}_{\text{mess,2}} \equiv 2 \mathcal{G}(x) x M_{\text{mess}} / (16\pi^2)$. Here, $M_D$ represents the dark photino soft mass, $M_K$ is the dark photino-bino mass mixing term, and $M_1$ is the bino soft mass. This expression reflects modifications to the bino vertex due to kinetic mixing (The structure is the same as [Eq.~\eqref{effInt}]).  The kinetic mixing effects are encapsulated in $M_K$ and $M_1$, indicating their $\epsilon$-dependence.

\begin{figure}[tb]
    \centering
    \begin{subfigure}{0.49\textwidth}
    \includegraphics[width=0.99\textwidth]{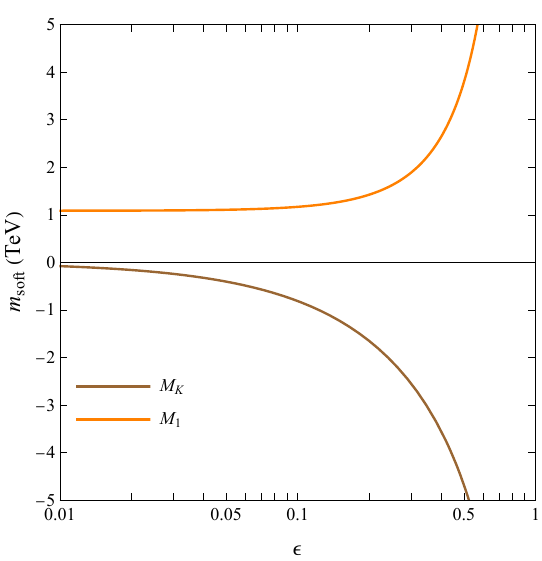} 
    \caption{Scenario I}
    \end{subfigure}
    \begin{subfigure}{0.49\textwidth}
    \includegraphics[width=0.99\textwidth]{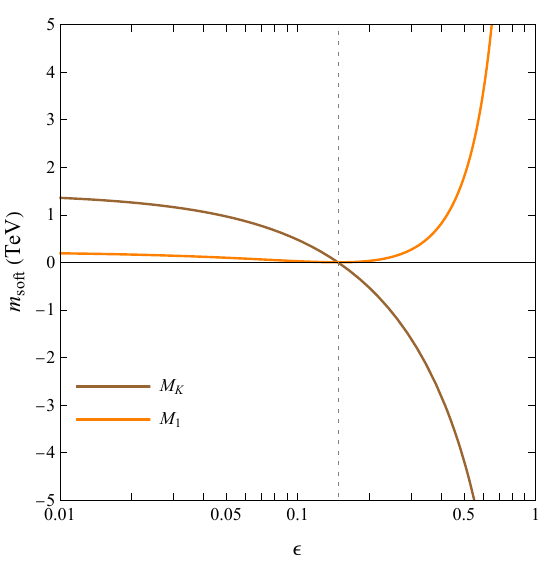} 
    \caption{Scenario II}
    \end{subfigure}
    \caption{ The gaugino soft terms $M_K$ and $M_1$ as a function of kinetic mixing $\epsilon$.
We show both Scenario I and Scenario II with parameters listed in Table~\ref{input}. (There are no dependencies on $\tan\beta$ or $\mu$ in the plots.) The soft terms undergo substantial changes with large kinetic mixing. In Scenario II, $M_K$ and $M_1$ vanish at $\epsilon_{\ast} = (g_Y/6) /\sqrt{(g_Y/6)^2 + g_D^2} \simeq 0.15$ (vertical dashed line) from Eq.~\eqref{crit1}. Beyond this value, $M_K$ becomes negative, while $M_1$ continues to increase.}
    \label{softPlot}
\end{figure}

Figure~\ref{softPlot} shows the dependence of these terms on kinetic mixing for Scenarios I and II. The figure illustrates that the soft terms $M_1$ and $M_K$ can be significantly affected by kinetic mixing. 
In Scenario I, $M_1$ is even in $\epsilon$, while $M_K$ is odd. Although $M_K$ depends on the sign of $\epsilon$, we have numerically verified that the neutralino mass spectrum depends only on the absolute value of $\epsilon$.   
In Scenario II, on the other hand, $M_K$ and $M_1$ is not symmetric in $\epsilon$ but vanish at $\epsilon_{\ast}$ [Eq.~\eqref{crit1}] due to a cancellation between $U(1)_Y$ and $U(1)_D$ contributions. This accidental cancellation decouples the dark photino even with large kinetic mixing.

In Scenario I, the determinant of $\mathbf{M}_{\tilde{N}}^{2 \times 2}$ does not vanish. As a result, the lightest neutralino  
is generally heavier than the EW scale. In Scenario II, however, the determinant of mass matrix is ${\rm det} \, \mathbf{M}_{\tilde{N}}^{2 \times 2}= M_D M_1 - M_K^2$ always vanishes, a consequence of the simplicity of the messenger representation (i.e., the messenger being in a single representation of the SM gauge group),
regardless of the values of $g_D$, $\epsilon$, and $D_{\Psi}$.\footnote{This can be understood very simply by moving to a basis defined in Eq.~(\ref{diag}) by $\sin\omega = -g_D D_\Psi/\sqrt{(g_D D_\psi)^2 + (g_Y Y_\Psi - g_{\rm eff} D_\Psi)^2}$, $\cos\omega =(g_Y Y_\Psi - g_{\rm eff} D_\Psi)/\sqrt{(g_D D_\psi)^2 + (g_Y Y_\Psi - g_{\rm eff} D_\Psi)^2}$. In this basis, the single messenger field in Scenario II only couples to one of the vector superfields, whose gaugino component acquires mass at one-loop via GMSB. The other gaugino remains massless at one loop, i.e., there is an approximate supersymmetry preserved in that sector. The same transformation diagonalizes the $2\times 2$ bino-dark photino mass matrix, Eq.~(\ref{submatrix}). In scenarios with messengers in multiple SM representations, it is not possible in general to choose a basis which decouples all messengers from one linear combination of vector multiplets.} This leads to the nearly massless bino-dark photino mixture neutralino state 
at the messenger scale, before accounting for RG running to the EW scale. Additionally, in Scenario II the ratio of each component in the lighter neutralino within $\mathbf{M}_{\tilde{N}}^{2 \times 2}$, denoted $|{N}^{2\times2}_{0i}|$ ($i = X, B$), takes the simple form 
\begin{equation}
\begin{split}
|{N}^{2\times2}_{0X}| : |{N}^{2\times2}_{0B}| & = |M_K(\epsilon)| : |M_D| = |M_1(\epsilon)| : |M_K(\epsilon)|
\end{split}
\label{mixture}
\end{equation}
This equation indicates that ${N}^{2\times2}_0$ will be dark photino-dominant if $|M_K| > |M_D|$ and bino-dominant if $|M_K| < |M_D|$. The dominance of the dark photino or bino in the composition thus depends on the kinetic mixing. This relationship generally holds even when considering additional contributions from the wino and higgsino unless there is a degeneracy between these components.

A complete analysis of the neutralino spectrum requires including higgsinos and wino in the mass matrix. There are three types of terms contributing to the neutralino mass: gaugino soft mass, gaugino-higgsino-Higgs coupling with VEV, and the $\mu$ term in the superpotential. 
(For a comprehensive discussion on the neutralino sector across various SUSY model extensions, see Ref.~\cite{Barger:2005hb}.) 
In the basis $\tilde \psi^0 = \begin{pmatrix}
\tilde{X}, & \tilde{B}, & \tilde{W}^3, & \tilde{H}_d^0, & \tilde{H}_u^0 \end{pmatrix}^T$, the neutralino mass terms are 
\begin{align}
\label{neutralino}
\mathcal{L}_{\tilde{N}} &= -\frac{1}{2} (\tilde \psi^0)^T \, \mathbf{M}_{\tilde{N}} \, \tilde \psi^0 + \text{h.c.}, \\
\label{MN}
\mathbf{M}_{\tilde{N}} &= \begin{pmatrix} 
M_D & M_K & 0 & 0 & 0 \\
M_K & M_1 & 0 & -c_\beta s_W m_Z & s_\beta s_W m_Z \\
0 & 0 & M_2 & c_\beta c_W m_Z & -s_\beta c_W m_Z \\
0 & -c_\beta s_W m_Z & c_\beta c_W m_Z & 0 & -\mu \\
0 & s_\beta s_W m_Z & -s_\beta c_W m_Z & -\mu & 0 
\end{pmatrix},
\end{align}
where $\mathbf{M}_{\tilde{N}}$ is the neutralino mass matrix. Here, $M_2 \simeq \frac{g_2^2}{2} \tilde{M}_{\text{mess,2}}$, $c_{\beta} = \cos\beta$, $s_{\beta} = \sin\beta$ with $\tan\beta = v_u/v_d$, $c_W = \cos\theta_W$, $s_W = \sin\theta_W$ with the weak mixing angle $\tan\theta_W = g_Y/g_2$, and the $Z$ boson mass is $m_Z = v \sqrt{g_Y^2 + g_2^2}/2$ where $v = \sqrt{v_u^2 + v_d^2}$.

The off-diagonal term $M_K$ links the visible and dark sectors. Off-diagonal terms between the bino and the wino/higgsino blocks are typically small compared to the soft mass terms, resulting in negligible mixing between the wino/higgsino components and the bino or dark photino-like states.

\begin{figure}[t]
    \centering
    \includegraphics[width=0.56\textwidth]{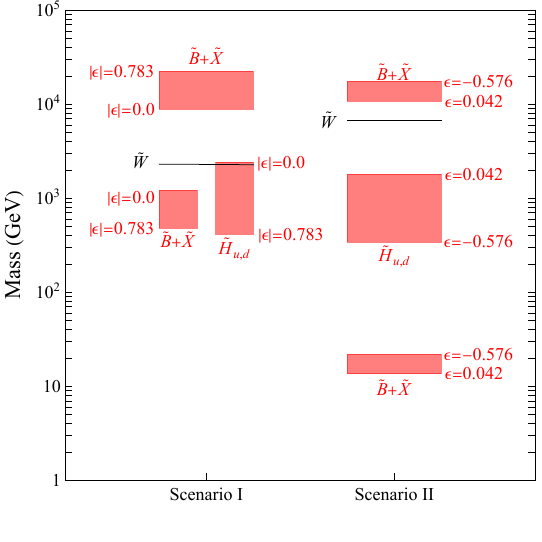}
    \caption{
    The dark GMSB neutralino mass spectrum for both Scenario I and Scenario II. The parameter values are listed in Table~\ref{input}. The neutralino states $\tilde{N}_i$ denote the dominant or mixed components of each neutralino mass eigenstate. 
    In Scenario I, the lightest neutralino could be predominantly a higgsino or a mixture of bino, dark photino, and higgsino, depending on the value of $\epsilon$.
    In Scenario II, a photino-bino mixture emerges as the lightest neutralino ($\tilde{N}_0$) with a suppressed mass around $\mathcal{O}(10)$ GeV, while another becomes the heaviest ($\tilde{N}_4$), showing significant sensitivity to changes in $\epsilon$. 
    }
    \label{Neutralino}
\end{figure}

The mass matrix can be diagonalized by a suitable unitary 
transformation $\mathbf{N}$ as $\mathbf{N}^{\ast} \mathbf{M}_{\tilde{N}} \mathbf{N}^{-1} = \text{diag}(m_{\tilde{N}_0}, m_{\tilde{N}_1}, m_{\tilde{N}_2}, m_{\tilde{N}_3}, m_{\tilde{N}_4})$, which orders the masses. The relation between the neutralinos in the gauge and mass basis is then $\psi^0_i = \mathbf{N}_{ij} \tilde N_j$. The determinant of $\mathbf{M}_{\tilde{N}}$ is calculated as
\begin{equation}
\label{TotalDet}
\det \mathbf{M}_{\tilde{N}} = (M_1 M_D - M_K^2) \left[ -M_2 \mu^2 + m_Z^2 \mu s_{2\beta} c_W^2 \right] + M_D M_2 m_Z^2 \mu s_{2\beta} s_W^2.
\end{equation}
The first term vanishes under the condition that $\mathbf{M}_{\tilde{N}}^{2 \times 2}$ has zero determinant (Scenario II), and the second term is maximized when $\beta = \pi/4$.

\begin{figure}[tb]
    \centering
    \begin{subfigure}{0.49 \textwidth}
        \includegraphics[width=0.99\textwidth]{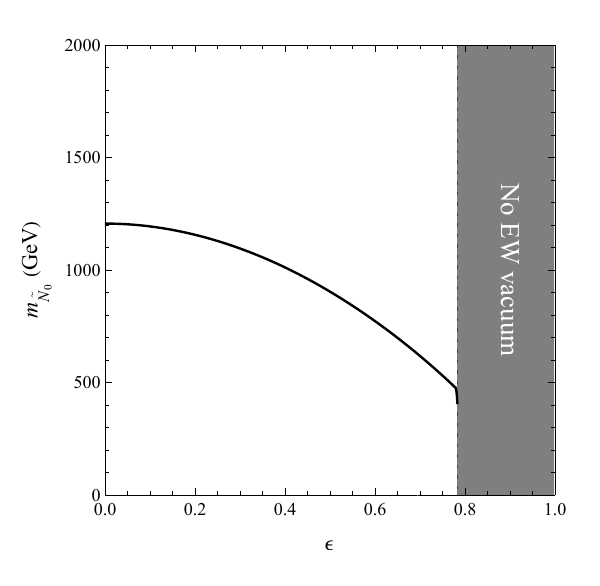}
        \subcaption{Scenario I}
    \end{subfigure}
        \begin{subfigure}{0.49 \textwidth}
        \includegraphics[width=0.99\textwidth]{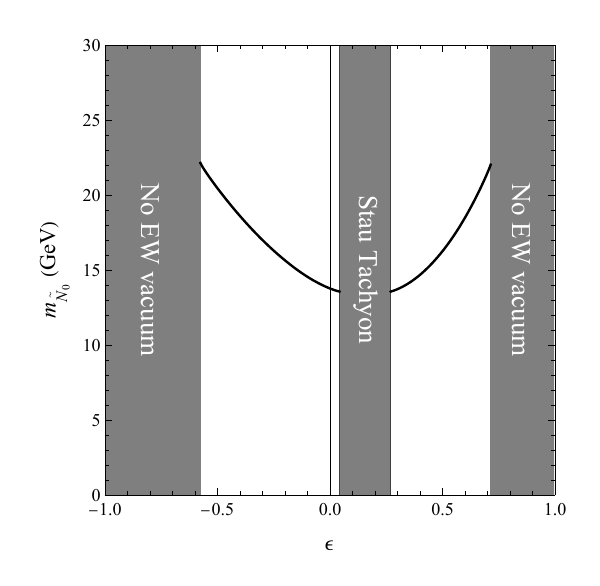}
        \subcaption{Scenario II}
    \end{subfigure}
    \caption{
    The mass of the lightest neutralino as a function of the kinetic mixing $\epsilon$.
    The parameter values are listed in Table~\ref{input}.
    The lightest neutralino has a suppressed mass in Scenario II.
    }
    \label{Nmass}
\end{figure}

The mass spectrum in the dark GMSB model largely depends on the representation of the messengers chosen in our two scenarios, as outlined in Table~\ref{tab:represen}. We summarize the key features of the two scenarios in the following.

\begin{itemize}
\item Scenario I: The messengers belong to the $(\textbf{3},1, -1/3, D_{\Psi})$ and $(1, \textbf{2}, 1/2, D_{\Psi})$ representations and their conjugates (a complete ${\bf 5}+{\bf \bar 5}$ of $SU(5)$). This configuration leads to a non-vanishing determinant for $\mathbf{M}_{\tilde{N}}^{2 \times 2}$, meaning the mass of the bino/dark photino-like neutralino is not suppressed.

\item Scenario II: The messengers belong to the $(\textbf{3}, \textbf{2}, 1/6, D_{\Psi})$ representation and its conjugate. Here, the lightest neutralino is a mixture of the bino and dark photino. Since $M_D M_1 = M_K^2$ at the messenger scale, the bino–dark photino mixture has a substantially suppressed mass. However, the RG running of $M_1$ and $M_K$ at the two-loop order breaks the accidental cancellation of the determinant of the bino–dark photino submatrix, determining the mass of the lightest neutralino.

\end{itemize}

\begin{figure}[tb]
    \centering
    \begin{subfigure}{0.49 \textwidth}
        \includegraphics[width=0.99\textwidth]{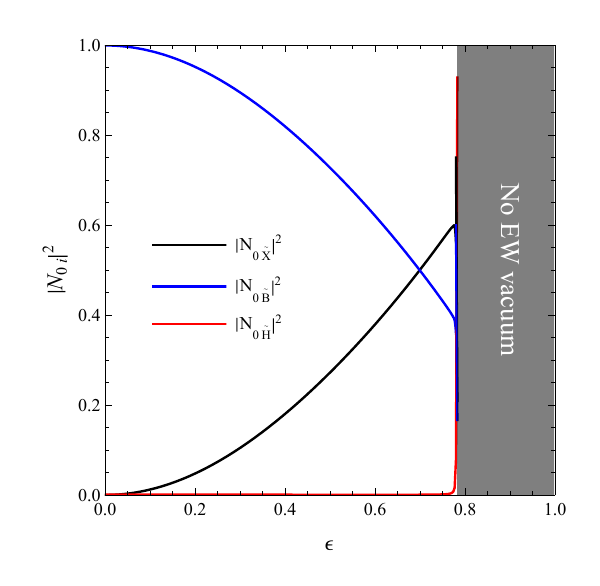}
        \subcaption{Scenario I}
    \end{subfigure}
        \begin{subfigure}{0.49 \textwidth}
        \includegraphics[width=0.99\textwidth]{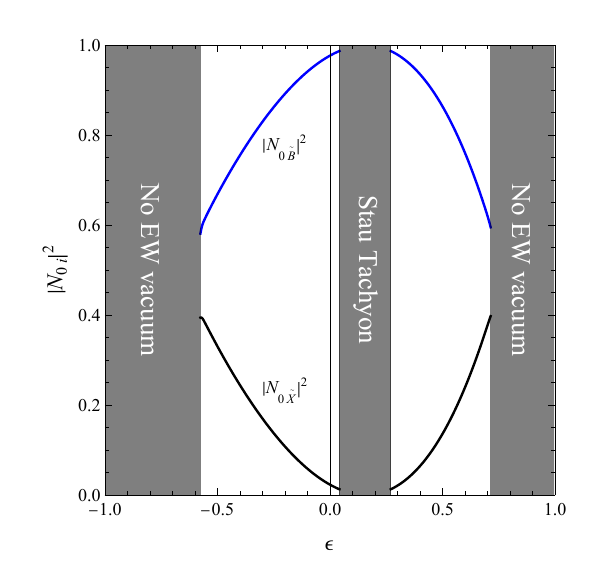}
        \subcaption{Scenario II}
    \end{subfigure}
    \caption{
    Variation of the lightest neutralino components $|N_{0i}|^2$ ($i = X, B$) with kinetic mixing $\epsilon$ in both dark GMSB Scenario I and Scenario II. The parameter values are listed in Table~\ref{input}. 
    }
    \label{case2}
\end{figure}

Figure~\ref{Neutralino} displays the neutralino mass spectra for our benchmark scenarios with varying $\epsilon$.
These results are obtained using 
{\tt SOFTSUSY 4.1.20} with suitable modifications to implement our dark GMSB model. The parameter $\epsilon$ affects the mass of the bino–dark photino mixture by modifying $M_1$ and $M_K$, while the higgsino masses are influenced by changes in $|\mu|$. Large values of $\epsilon$ or $g_D$ can significantly decrease $|\mu|$, thereby lowering the higgsino masses. Figure~\ref{Nmass} shows the mass spectrum of the lightest neutralino states as a function of the kinetic mixing $\epsilon$ in both scenarios.

The composition of the lightest neutralino, $\tilde{N}_0$, in both scenarios is illustrated in Figure~\ref{case2}. The value of the kinetic mixing can change the dominant component of the lightest neutralino mass eigenstate.

A crucial distinction between the two scenarios lies in the characteristics of the bino–dark photino mixture states. In Scenario I, the mass of the lightest neutralino bino–dark photino state is considerably heavy, approximately $M_{\tilde{N}_0} \simeq 500$–$1000$ GeV. On the other hand, in Scenario II, the mass of the light bino–dark photino state remains low, approximately $M_{\tilde{N}_0} \simeq \mathcal{O}(10)$ GeV, due to the vanishing determinant of $\mathbf{M}_{\tilde{N}}^{2 \times 2}$ at the messenger scale. 

The neutralino sector within the dark GMSB framework exhibits significant dependence on kinetic mixing. The mass of the dark photino–bino mixture is affected by changes in $\epsilon$. Notably, their dominant compositions can be swapped around the specific large value of the kinetic mixing.

\subsection{Chargino sector}
\label{subSec:Chargino sector}

\begin{figure}[tb]
    \centering
        \includegraphics[width=0.56\textwidth]{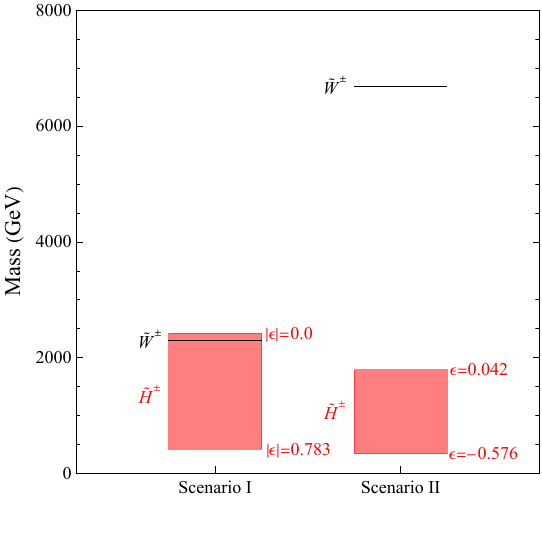}
    \caption{The chargino mass spectrum for both Scenario I and Scenario II. The parameter values are listed in Table~\ref{input}. The mass of each chargino state can be approximated as $m_{\tilde{H}^{\pm}} \simeq |\mu|$ and $m_{\tilde{W}^{\pm}} \simeq M_2 $. Thus, the lightest, mostly higgsino, chargino mass depends on $\epsilon$ through $\mu$. 
    }
    \label{chargino}
\end{figure}

The chargino mass spectrum is controlled by the wino mass $M_2$ and higgsino mass $\mu$, and the latter depends on $g_D$ and $\epsilon$. The chargino masses are given by 
\begin{equation}
\begin{split}
    m^2_{\tilde{C}_i} = &\frac{1}{2} \Big[ M_2^2+\mu^2+2m_W^2 \pm\sqrt{(M_2^2+\mu^2+2m_W^2)^2 -4(\mu M_2-m_W^2\sin(2\beta)  )^2} \;\Big].
\end{split}
\end{equation}
In the common situation where $|\mu|$ and $M_2$ are much larger than $m_W$, the chargino masses are approximately 
given by $|\mu|$ or $M_2$.

Therefore, the chargino state that is mostly higgsino is influenced by $g_D$ and $\epsilon$, while the one that is mostly wino is largely unaffected. The chargino mass spectrum is illustrated in Figure~\ref{chargino}, 
which shows that the higgsino-dominant chargino mass can be significantly reduced for large $\epsilon$.

\section{Particle spectra and implications for phenomenology}
\label{SpectrumSec}

\begin{figure}[t]
    \centering
    \includegraphics[width=0.99\textwidth]{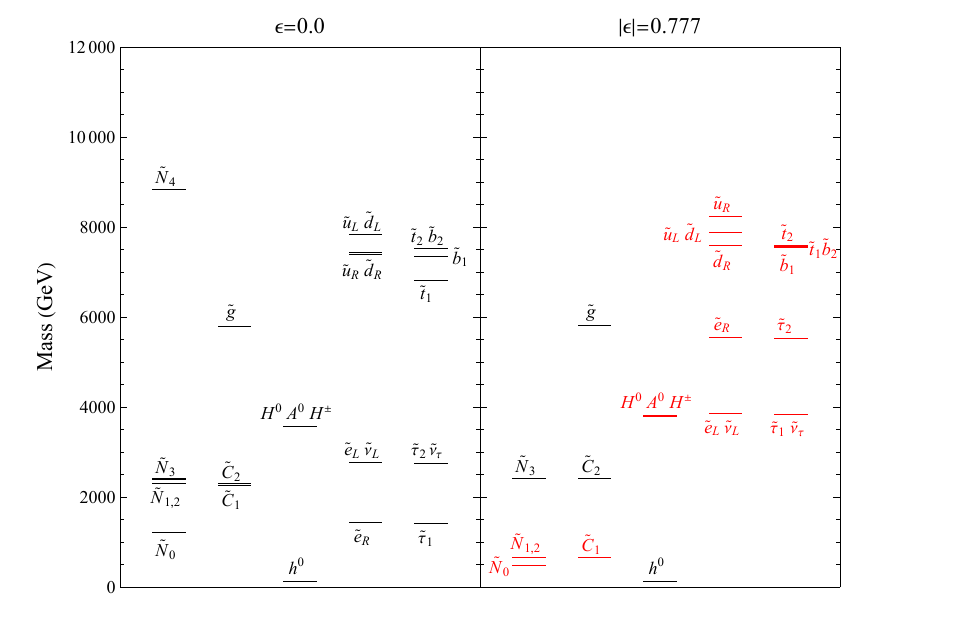}   
    \caption{
    The mass spectrum in dark GMSB for Scenario I with a complete $SU(5)$ ${{\bf 5}+ \bar {\bf 5}}$ messenger representation. The parameters are set as in Table~\ref{input}.  States indicated in red exhibit changes to their masses in comparison to conventional GMSB.  
    The distinctive features include (1) the lightest neutralino as a bino-dark photino mixture, (2) a relatively light Higgsino, and (3) heavier sleptons with a reversed mass hierarchy between the left-handed and right-handed sleptons.  
    The SM Higgs mass of 125 GeV is reproduced for our parameter choices. 
    The gravitino LSP $\tilde G$, with sub-keV mass, and the heaviest neutralino $\tilde{N}_4$, with mass around 9 to 22 TeV, depending on $\epsilon$, are not represented here. 
      }
    \label{FinalComplete}
\end{figure}

\begin{figure}[t]
    \centering
    \includegraphics[width=0.99\textwidth]{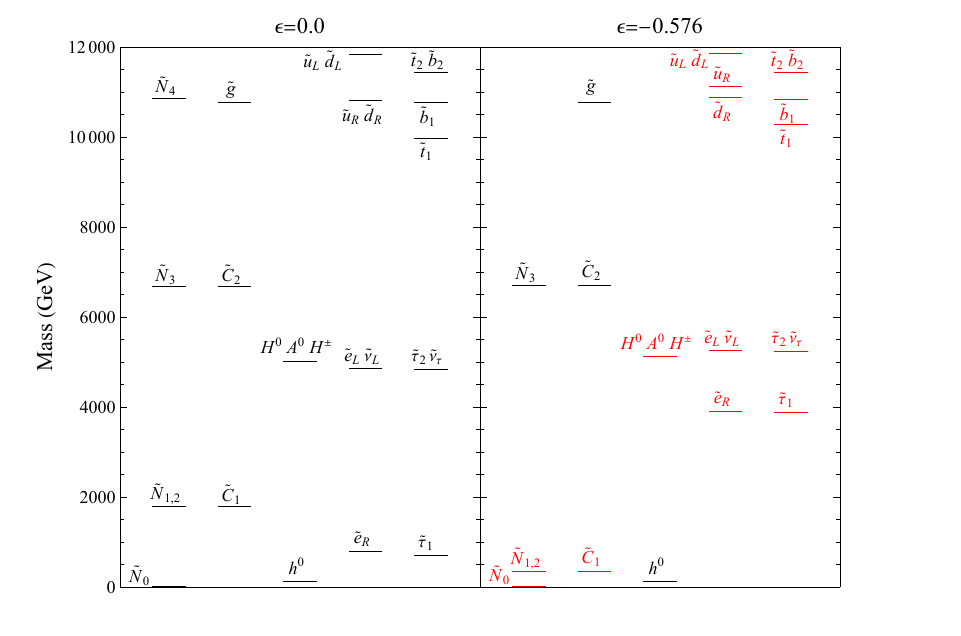}   
    \caption{
    The mass spectrum in dark GMSB with Scenario II for an incomplete messenger representation. The parameters are set as in Table~\ref{input}. States indicated in red exhibit changes to their masses in comparison to conventional GMSB.
The distinctive features include
(1) a very light bino–dark photino mixture as the neutralino NLSP,
(2) a relatively light Higgsino, and
(3) significantly heavier right-handed sleptons.
The SM Higgs mass of 125 GeV is reproduced for our parameter choices. 
The gravitino LSP $\tilde G$, with sub-keV mass is not represented here, while on the right panel the heaviest neutralino $\tilde{N}_4$ as a bino-dark photino mixture has a mass around 17 TeV and is not shown in the plot. 
    }
    \label{FinalPlot}
\end{figure}

As discussed in the previous sections, dark GMSB models with large kinetic mixing can lead to pronounced changes in the spectra of the neutralino, sfermion, and Higgs sectors when compared to conventional GMSB models. These effects could be drastic for some particles, while others are less impacted. This section examines the overall spectrum in dark GMSB and highlights some of the phenomenological implications of large kinetic mixing. 

The spectra are shown in Figure~\ref{FinalComplete} and Figure~\ref{FinalPlot} for Scenarios I and II, respectively. We note that the gravitino LSP, with mass $ m_{3/2} \lesssim \mathcal{O}({\rm keV})$, is not shown in these figures. We see that the gluino $\tilde{g}$ and wino $\tilde{W}$-dominant neutralino/chargino masses are unaffected by kinetic mixing, as only the $SU(3)_C$ and $SU(2)_L$ gauge interactions are relevant to these states, respectively. Furthermore, the colored sfermion spectra are only modestly impacted by kinetic mixing as the dominant contribution to their masses comes from the $SU(3)_C$ strong interaction.
Instead, the color neutral sfermions, particularly the right-handed sleptons, the bino-dark photino system, and the higgsino all experience significant changes to their masses at large kinetic mixing. Concerning the higgsino, even though $\mu$ is a SUSY-preserving parameter, a dependence on the kinetic mixing is introduced to $\mu$ 
through the EWSB condition. 
Although the left panels in both figures consider a non-zero $g_D$, the spectra approximate those of the corresponding conventional GMSB scenarios,
except for the extremely light bino-like neutralino in Scenario II and the presence of a heavy dark photino-like neutralino in both scenarios.\footnote{For conventional GMSB with an incomplete messenger as in Scenario II, the mass of the bino-like lightest neutralino is about 230 GeV, assuming the same parameters except for $g_D$, $\epsilon \rightarrow 0$.} 

We first discuss Scenario I with a complete $SU(5)$ ${{\bf 5}+ \bar {\bf 5}}$ messenger representation. For $\epsilon = 0$ we see that all superpartners other than the gravitino LSP are above the TeV scale, with the NLSP bino in the TeV range followed by right-handed sleptons in the 1.5 TeV range and all other superpartners in the multi-TeV range. This spectrum, which is essentially that of minimal GMSB except for the addition of the heavy dark photino, is inaccessible to the LHC but could be probed at future high energy colliders. 

\begin{figure}[tb]
    \centering
    \begin{subfigure}{0.49 \textwidth}
        \includegraphics[width=0.99\textwidth]{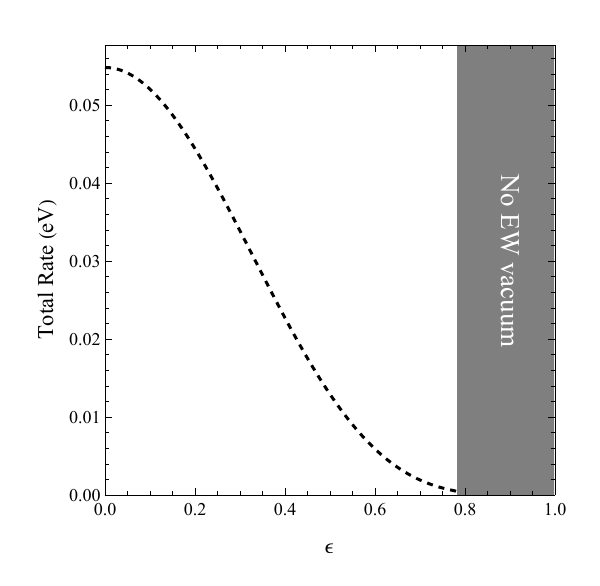}
        \subcaption{Total ${\tilde N}_0$ decay rate in Scenario I}
    \end{subfigure}
        \begin{subfigure}{0.49 \textwidth}
        \includegraphics[width=0.99\textwidth]{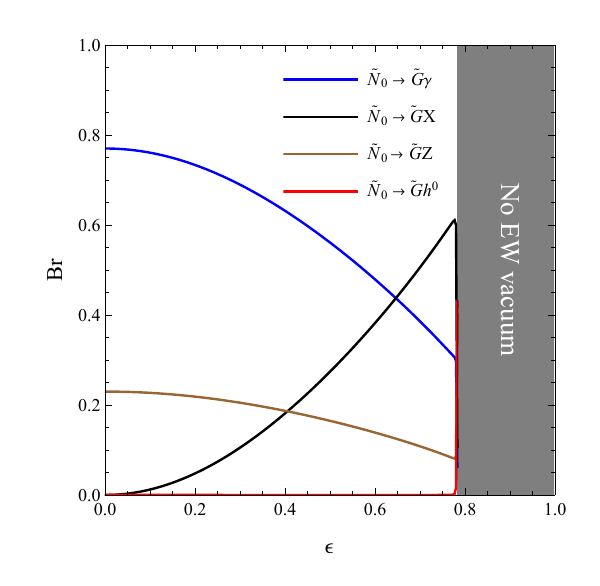}
        \subcaption{${\tilde N}_0$ branching ratios in Scenario I}
    \end{subfigure}
    \\
    \begin{subfigure}{0.49 \textwidth}
        \includegraphics[width=0.99\textwidth]{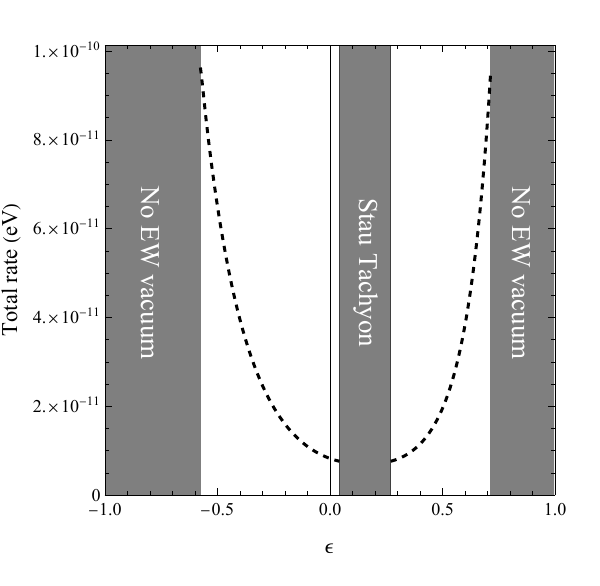}
        \subcaption{Total ${\tilde N}_0$ decay rate in Scenario II}
    \end{subfigure}
        \begin{subfigure}{0.49 \textwidth}
        \includegraphics[width=0.99\textwidth]{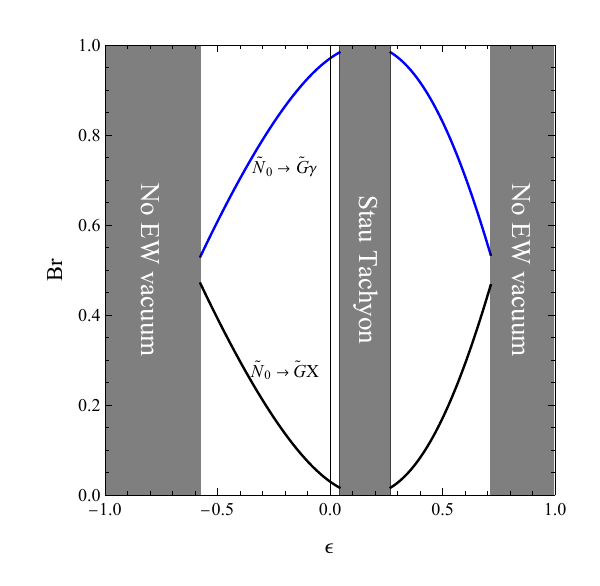}
        \subcaption{${\tilde N}_0$ branching ratios in Scenario II}
    \end{subfigure}
    \caption{
    Total decay width (left) and branching ratios (right) of the NLSP $\tilde{N}_0$ decay to the gravitino $\tilde{G}$ for Scenario I (top) and Scenario II (bottom). The parameter choices are given in Table~\ref{input}.
    The total decay rate exhibits a steep $m_{{\tilde N}_0}^5$ dependence on the neutralino mass, leading to a fast (slow) decay for Scenario I (Scenario II).
    The branching ratio is governed by the composition of $\tilde{N}_0$.
    For Scenario I, the photon and $Z$ boson decay modes dominate for small $\epsilon$, while the dark photon decay mode becomes more important as $\epsilon$ is increased. For large $\epsilon$ near the boundary of viable EWSB, 
    the Higgs boson mode becomes dominant due to the substantial 
    higgsino component in $\tilde{N}_0$.
    In Scenario II, the $Z$ boson and Higgs boson decay modes are kinematically forbidden because the neutralino NLSP has the suppressed $\tilde{N}_0$ mass.
    }
    \label{Grav}
\end{figure}

The situation changes in the presence of kinetic mixing,  $\epsilon \neq 0$. The higgsino becomes lighter and can be below the TeV scale for large enough kinetic mixing (Figure~\ref{muconst}). Furthermore, the bino and dark photino undergo substantial mass mixing, causing one linear combination to become lighter (Figure~\ref{Nmass}). Thus the typical spectrum of the lightest states at large $\epsilon$ consists of a bino/dark photino NLSP and a somewhat heavier higgsino. If light enough, these states could be within reach of the LHC. The dominant production channel is the pair production of a chargino and a heavy neutralino. The produced chargino can then decay to a neutralino NLSP and $W$ boson, while the heavy neutralino can decay to a neutralino NLSP and a $Z$ or Higgs boson. Finally, the neutralino NLSP will subsequently decay to the gravitino and either a dark photon, photon, $Z$, or $h^0$~\cite{Ambrosanio:1996jn, Dimopoulos:1996yq}, with respective partial decay widths 
\begin{align}
\label{gravitino-X}
\Gamma (\tilde{N}_0  \rightarrow  \tilde{G} X) & = \frac{m_{\tilde{N}_0}^5}{16\pi F^2} |N_{0 \tilde{X}} |^2 ,
\\
\label{gravitino-gamma}
\Gamma (\tilde{N}_0 \rightarrow \tilde{G} \gamma) & = \frac{m_{\tilde{N}_0}^5}{16\pi F^2} |N_{0 \tilde{B}} \cos \theta_W  + N_{0 \tilde{W}} \sin \theta_W|^2 ,
\\
\label{gravitino-Z}
\Gamma (\tilde{N}_0  \rightarrow \tilde{G} Z) & = \frac{m_{\tilde{N}_0}^5}{16\pi F^2} \left( 1 - \frac{m_Z^2}{m_{\tilde{N_0}}^2} \right)^4  
\\
& \quad \times \left( |N_{0 \tilde{B}} \sin \theta_W  - N_{0 \tilde{W}} \cos \theta_W|^2 + \frac{1}{2} | N_{0 \tilde{H}_d} \cos \beta - N_{0 \tilde{H}_u} \sin \beta |^2 \right) \nonumber, 
\\
\label{gravitino-h}
\Gamma (\tilde{N}_0  \rightarrow  \tilde{G}h^0) & = \frac{m_{\tilde{N}_0}^5}{32\pi F^2} \left( 1 - \frac{m_{h^0}^2}{m_{\tilde{N_0}}^2} \right)^4 |N_{0 \tilde{H}_d}  \sin \alpha - N_{0 \tilde{H}_u} \cos \alpha |^2 ,
\end{align} 
where $\alpha$ is the mixing angle for the neutral $CP$-even Higgs sector, satisfying the relations $\sin 2\alpha = - (m_{H^0}^2 + m_{h^0}^2) / (m_{H^0}^2 - m_{h^0}^2) \sin 2\beta$ and $\tan 2\alpha =(m_{A^0}^2 +m_Z^2)/(m_{A^0}^2 -m_Z^2) \tan 2\beta$. 
Figure~\ref{Grav} exhibits the $\epsilon$-dependence of the total rate and branching ratios of the NLSP $\tilde{N}_0$ decay to the gravitino $\tilde{G}$.
For Scenario I, with $m_{\tilde{N}_0}$ in the 500 GeV -- 1 TeV range   
and $\sqrt{F} \sim \mathcal{O}(10^3\, {\rm TeV})$, the total decay width is $\mathcal{O}(10^{-2})$ eV, corresponding to a decay length $c \, \tau_{\tilde{N}_0}$ of order 10 microns. Thus, the additional visible particles from the NLSP decays $\tilde G \gamma$ and $\tilde G Z$ may provide an extra handle in the experimental searches at the LHC. We also see from Figure~\ref{Grav} that for large $\epsilon$ there is typically a sizable branching ratio for the $\tilde G X$ channel.  As both the gravitino and dark photon would escape the detector, the NLSP  neutralino in this case would manifest as missing energy. Ultimately, discerning the NLSP branching ratios, along with information about the spectrum, could provide a means of determining $\epsilon$ and other parameters in dark GMSB.     

For Scenario I, it is worth remarking that for very large $\epsilon$, near the boundary of viable EWSB, the $\mu$ parameter may be driven to small enough values so that the higgsino becomes the NLSP. The typical splitting between chargino and neutralino states in this case can be quite small, leading to a compressed spectrum with relatively softer visible particles produced in the chargino decay to the neutralino NLSP. Still, in this case it is expected that visible particles are produced in the NLSP decay at the last step, where now the $\tilde G h^0$ mode becomes important, see Figure~\ref{Grav}.

Turning now to Scenario II with an incomplete $SU(5)$ messenger, Figure~\ref{FinalPlot} illustrates the corresponding dark GMSB spectrum.
 As in Scenario I, the higgsino can become relatively light as the kinetic mixing strength increases, enhancing the prospects for its production and detection at the LHC. 
 The neutralino sector contains dark photino–bino mixtures, whose masses and composition depend on $\epsilon$ (Figure~\ref{softPlot}). In particular, one of the bino–dark photino mixed states has a suppressed mass on the order of 10 GeV,
 making it the lightest neutralino and NLSP. This occurs because the determinant of mass matrix for the bino–dark photino system vanishes at the messenger scale, although it is revived through RG running to the weak scale. 

The presence of such a light neutralino NLSP opens up the interesting possibility of producing it in a rare decay of the 125 GeV Higgs boson, $h^0 \rightarrow \tilde N_0 \tilde N_0$. 
The corresponding decay rate is \cite{Griest:1987qv, Djouadi:1996mj, Dreiner:2012ex} 
\begin{align}
\Gamma(h^0 \rightarrow \tilde{N}_0 \tilde{N}_0 )
= \frac{g_2^2 m_{h^0}}{16 \pi} \left( 1 - \frac{4 m_{\tilde{N}_0}^2}{m_{h^0}^2} \right)^{3/2}
\big| (N_{0 \tilde{W} } - N_{0 \tilde{B} } \tan \theta_W  )
( N_{0 \tilde{H}_u} \cos \alpha + N_{0 \tilde{H}_d} \sin \alpha  )
\big|^2 .
\label{HiggsDecayEq}
\end{align}
Once produced, the light neutralino NLSP will decay either via $\tilde{N}_0 \rightarrow \tilde{G} \gamma$ or $\tilde{N}_0 \rightarrow \tilde{G} X$, see Figure~\ref{Grav}.
Due to the strong $m_{\tilde N_0}^5$ dependence of $\Gamma_{\tilde{N}_0}$, Eqs.~(\ref{gravitino-X} -- \ref{gravitino-h}), for $m_{\tilde{N}_0} \lesssim 20$ GeV and $\sqrt{F} \simeq \mathcal{O}(10^3\, {\rm TeV})$  as appropriate here, the lifetime of $\tilde{N}_0$ is at least $\mathcal{O}(10^{-6})$ seconds, which is long enough for it to propagate outside the detector. 
The signature of the $h^0 \rightarrow \tilde N_0 \tilde N_0$ is therefore an invisible Higgs decay. 

\begin{figure}[tb]
    \centering
    \begin{subfigure}{0.49 \textwidth}
        \includegraphics[width=0.99\textwidth]{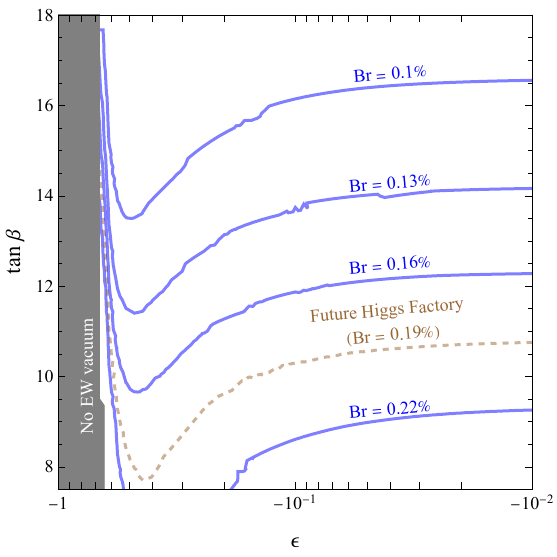}
        \subcaption{Negative $\epsilon$}
    \end{subfigure}
        \begin{subfigure}{0.49 \textwidth}
        \includegraphics[width=0.99\textwidth]{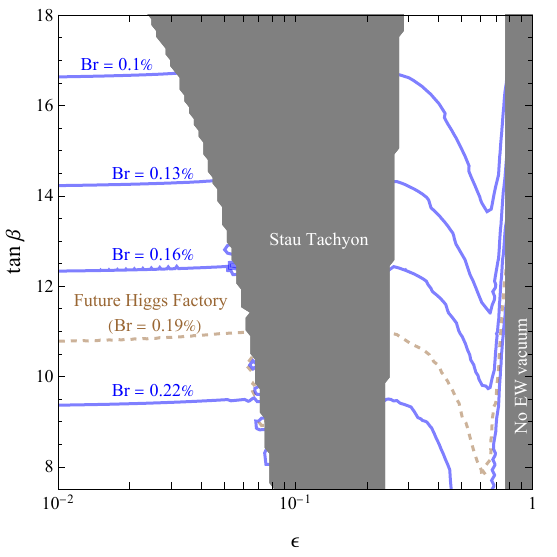}
        \subcaption{Positive $\epsilon$}
    \end{subfigure}
    \caption{
    The branching ratio for the decay $h^0 \rightarrow \tilde{N}_0 \tilde{N}_0$ 
    in the parameter space of $\tan \beta$ and $\epsilon$ in Scenario II, which predicts a very light neutralino NLSP.
    Here, the value of $F/M$ is adjusted at each point to fit $m_{h^0} = 125$ GeV, while other parameters are chosen as in Table~\ref{input}.
    The brown dashed curves with $\mathrm{Br}=0.19\%$ correspond to the expected sensitivities of the future Higgs factories~\cite{deBlas:2019rxi}.
    }
    \label{fig:HiggsDecay}
\end{figure}    

Figure~\ref{fig:HiggsDecay} shows isocontours of the branching ratio 
of the Higgs to the lightest neutralinos for negative and positive $\epsilon$ (left and right panels, respectively).
A sharp rise in the branching ratio is observed near the boundary of the EW vacuum constraint, where $\mu$ decreases rapidly and the Higgsino components of $\tilde N^0$ correspondingly increase.
Furthermore, it is evident that the branching ratio increases as $\tan \beta$ becomes smaller. This is because the NLSP contains a relatively larger $\tilde{H}_d$ component, and since $\sin\alpha \simeq \cos \beta \simeq 1/\tan\beta$ in the $m_{H^0} \gg m_{h^0}$ and large $\tan\beta$ limits, the $N_{0\tilde H_d} \sin \alpha$ term Eq.~(\ref{HiggsDecayEq}) grows as $\tan \beta$ decreases.
For small $\tan\beta$, a large stop mass, i.e., large $F/M$, is required to achieve the observed Higgs mass of $125$ GeV. This also sufficiently raises the $N_0$ mass as $\tan \beta$ becomes smaller than about 5-6 so that the exotic Higgs decays may not occur because $m_{\tilde{N}_0} > m_{h_0}/2$. Furthermore, the numerical evaluation of the spectrum becomes unstable due to convergence issues below $\tan\beta \lesssim 7$.
Interestingly, some of the parameter space of $\tan \beta$ and $\epsilon$ is accessible to the future Higgs factories such as FCC-ee~\cite{FCC:2018evy}, ILC~\cite{ILC:2007oiw}, or CEPC~\cite{CEPCStudyGroup:2023quu} where we have taken $\text{Br}_{\text{inv}} \simeq 0.19\%$ from $\Gamma_{h^0}^{\mathrm{tot}} = 4.07 \times 10^{-3}\,\mathrm{GeV}$ as a projected upper bound on the invisible Higgs branching ratio~\cite{deBlas:2019rxi}.

The invisible decay of $Z$ boson to the $\tilde{N}_0$ is also viable with the following decay rate~\cite{Dreiner:2012ex}
\begin{align}
\Gamma(Z \rightarrow \tilde{N}_0 \tilde{N}_0) = \frac{g_2^2 m_Z}{12 \pi} \left( 1 - \frac{4 m_{\tilde{N}_0}^2}{m_Z^2} \right)^{3/2} \, \left| \frac{1}{2 \cos\theta_W \sin\theta_W} \left(N_{0 \tilde{H}_u}^2 - N_{0 \tilde{H}_d}^2 \right) \right|^2
\end{align}
However, due to the suppressed Higgsino composition in $\tilde{N}_0$, the LEP constraint $\Gamma(Z \rightarrow \tilde{N}_0 \tilde{N}_0) < 3\,\mathrm{MeV}$ from the $Z$ boson invisible decay width~\cite{ALEPH:2005ab} is not relevant.

As discussed above, there are still potentially interesting prospects at the LHC for large $\epsilon$, where the Higgsino becomes relatively light. Nevertheless, as in conventional GMSB models, to fully probe the heavier superpartners in the spectrum would only be possible with a future high energy collider. A detailed determination of the spectrum would allow for a test of the dark GMSB scenarios relative to the corresponding conventional GMSB ones.

\section{Cosmology}
\label{sec:Cosmology}

In this section, we discuss the cosmology and relic densities of the stable particles in dark GMSB. The stable states include the gravitino as the LSP, the lightest messenger state, which is the lightest of the dark-charged states and is stable due to the unbroken $U(1)_D$ symmetry, and the massless dark photon.

\subsection{Relic Gravitinos}

In typical GMSB models, the gravitino problem arises when the gravitino mass exceeds approximately 1 keV, leading to an overclosure of the universe~\cite{Pagels:1981ke,Weinberg:1982zq}. Recent studies have imposed even stricter upper limits on the gravitino mass; Lyman-alpha forest data constrain it to around 16 eV~\cite{Viel:2005qj}, while observations of the CMB and large-scale structure further tighten the bound to approximately 4 eV~\cite{Osato:2016ixc}. Typically, to evade these bounds one must invoke  
a low reheating temperature or 
additional entropy production to dilute the gravitino abundance.\footnote{If $R$-parity is violated, the gravitino LSP can decay into SM particles. However, for a light gravitino ($m_{3/2} \lesssim 1$ GeV), its lifetime is still larger than the age of the universe \cite{Borgani:1996ag,Takayama:2000uz,Moreau:2001sr}.} 
Although we do not provide a direct solution to the gravitino problem and must also appeal to a non-standard cosmology, in the following we briefly argue our scenario does not significantly exacerbate the existing constraints.

The NLSP can contribute to the relic gravitino density~\cite{Borgani:1996ag}. If the NLSP decouples from the thermal plasma while still relativistic, its subsequent decay produces gravitinos whose relic abundance can be comparable to that generated directly through thermal freeze-out. This arises because the freeze-out densities of both the gravitino and the NLSP scale with their masses. Such a scenario occurs when the NLSP’s thermal cross section is relatively small, as in Scenario II, due to a light neutralino NLSP of approximately 10 GeV. By contrast, in Scenario I, where the neutralino NLSP has a mass on the order of 1 TeV, the NLSP decouples nonrelativistically, and its contribution to the gravitino relic density is exponentially suppressed by the Boltzmann factor.

Consequently, in both scenarios, the additional contribution from NLSP decays amounts to only a marginal correction to the total gravitino relic density.

\subsection{Relic Messengers}

Due to the unbroken $U(1)_{D}$ gauge symmetry, the lightest messenger particle charged under $U(1)_{D}$ is absolutely stable. Depending on its representation, this lightest messenger is naturally populated through strong or EW gauge interactions within the SM thermal bath. Messengers could easily reach thermal equilibrium with the SM thermal bath, potentially leading to an overabundance \cite{Giudice:1998bp}. Moreover, the messenger fields charged under $U(1)_{D}$ would acquire an electric charge through kinetic mixing proportional to $\epsilon$, making them fractionally charged particles (FCPs). 

The existence of stable FCPs is severely constrained by a variety experimental searches. Underground detectors are capable of identifying FCPs among cosmic rays carrying energies at least on the order of hundreds of GeV~\cite{Kamiokande,Aglietta:1994iv,MACRO:2000bht,MACRO:2004iiu,CDMS:2014ane,Majorana:2018gib}, while on-orbit experiments can probe FCPs as light as a few GeV~\cite{Sbarra:2003ur,Fuke:2008zza}. Beyond these direct detection approaches, searches have been carried out for FCPs in bulk matter by using the levitometer techniques~\cite{Larue:1981jc,Marinelli:1983nd,Smith:1986ik,Smith:1987mj,Jones:1989cq,Homer:1992cz} or liquid drop methods~\cite{Joyce:1983tj,Savage:1986vg,Halyo:1999wq,Lee:2002sa,Kim:2007zzs}.

For order one fractional charge, the most stringent constraint comes from a search for FCPs accumulated in sea water~\cite{Kudo:2001ie} from an experiment employing 
the magnetic levitation technique: \cite{Homer:1992cz}
\begin{equation}
\frac{n_{\Psi}}{n_H}\bigg|_{\mathrm{earth}} \simeq 5 \times 10^{-5} \left(\frac{ \mathrm{GeV} }{m_{\Psi}} \right)\Omega_{\Psi} h^2 \lesssim 10^{-27},
\end{equation}
by taking $\Omega_{\mathrm{CDM}}=0.265$ and $H_0=68$ km/s/Mpc.

However, the relic density of the messengers could be substantially suppressed if the reheating temperature $T_{\mathrm{reh}}$ is significantly lower than $m_{\Psi}$. Then, the production of the messengers is inhibited by the large Boltzmann factor, $e^{-m_{\Psi}/T_{\mathrm{max}}}$.\footnote{The maximum temperature during reheating, $T_{\mathrm{max}}\sim (H_{\mathrm{inf}} M_{\mathrm{Pl}})^{1/4}T_{\mathrm{reh}}^{1/2}$~\cite{Giudice:2000ex}, can be much higher than $T_{\mathrm{reh}}\sim(\Gamma_{\mathrm{inf}} M_{\mathrm{Pl}})^{1/2}$, which is the temperature at the end of reheating. Here, $H_{\mathrm{inf}}$ is the Hubble scale during the inflation, and $\Gamma_{\mathrm{inf}}$ is the decay rate of the inflaton. Once the temperature during reheating reaches $T_{\mathrm{max}}$, it decreases proportionally to $T \propto T_{\mathrm{max}} a^{-3/8}$, where $a$ is the scale factor. }
Even if the messengers achieve thermal equilibrium with the thermal bath, they would decouple during reheating, and their number density would be diluted by subsequent entropy production \cite{Giudice:2000ex}. If $T_{\mathrm{max}} \simeq m_{\Psi}$, the relic density of the messengers is given by \cite{Giudice:2000ex}
\begin{equation}
\Omega_{\Psi} h^2 \simeq\begin{dcases}
  \frac{3 \sqrt{5}(17/2e)^{17/2}}{512 \pi^{11/2}} \frac{g^2 g_{\ast}^{3/2}(T_{\mathrm{reh}})}{g_{\ast}^{3}(T_{{\ast}})} \frac{M_{\mathrm{Pl}}T_{\mathrm{reh}}^7}{m_{\Psi}^7 T_0}  A_s  \Omega_R h^2 &\quad (\text{freeze-in}),\\
  \frac{5\sqrt{5}}{8\sqrt{2}\pi} \frac{g_{\ast}^{1/2}(T_{\mathrm{reh}})}{g_{\ast}(T_{F})}  \frac{T_{\mathrm{reh}}^{3}}{T_0m_{\Psi}M_{\mathrm{Pl}}} \frac{\Omega_R h^2}{A_s x_F^{-4}} &\quad (\text{freeze-out}),
 \end{dcases}
 \label{eq:relicM}
\end{equation}
where $g$ is the number of degrees of freedom of the lightest messenger, $g_{\ast}(T)$ is the number of degrees of freedom of the thermal bath at $T$, $\Omega_R h^2 = 2.5 \times 10^{-5}$ \cite{ParticleDataGroup:2022pth}, $T_0 = 2.7$ K, $T_{\ast} = 4m_{\Psi}/17$, 
and the normalized $s$-wave coefficient $A_s$ is defined as $\langle\sigma v\rangle \simeq A_s/m_{\Psi}^2$. In the freeze-in case, messenger production predominantly occurs at $T_{\ast}$, so the relic density shows no dependence on $T_{\mathrm{max}}$. Similarly, the freeze-out relic density is also independent of $T_{\mathrm{max}}$, as the number density is determined at the freeze-out temperature. This temperature is calculated using the following relation \cite{Giudice:2000ex}:
\begin{equation}
    x_F= \ln \left[\frac{3}{\sqrt{5}\pi^{5/2}} \frac{g g_{\ast}^{1/2}(T_{\mathrm{reh}})}{g_{\ast}(T_F)}\frac{M_{\mathrm{Pl}} T^2_{\mathrm{reh}}}{m_{\Psi}^3}A_s x_F^{5/2} \right].
\end{equation}

\begin{figure}[tb]
    \centering
    \begin{subfigure}{0.49 \textwidth}
    \includegraphics[width=0.99\textwidth]{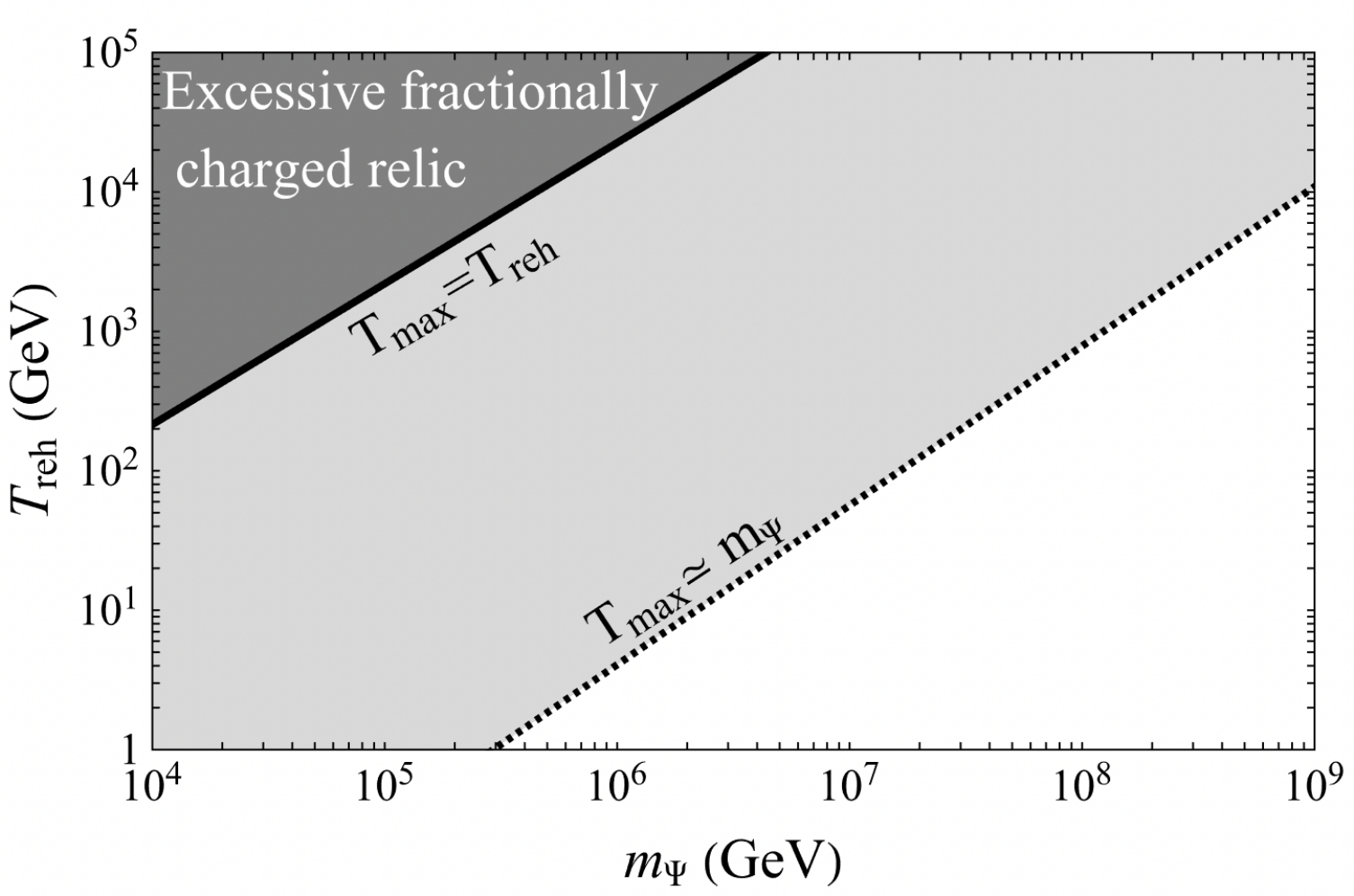}
    \subcaption{}
    \label{fig:relic:a}
    \end{subfigure}
    \begin{subfigure}{0.49 \textwidth}
    \includegraphics[width=0.99\textwidth]{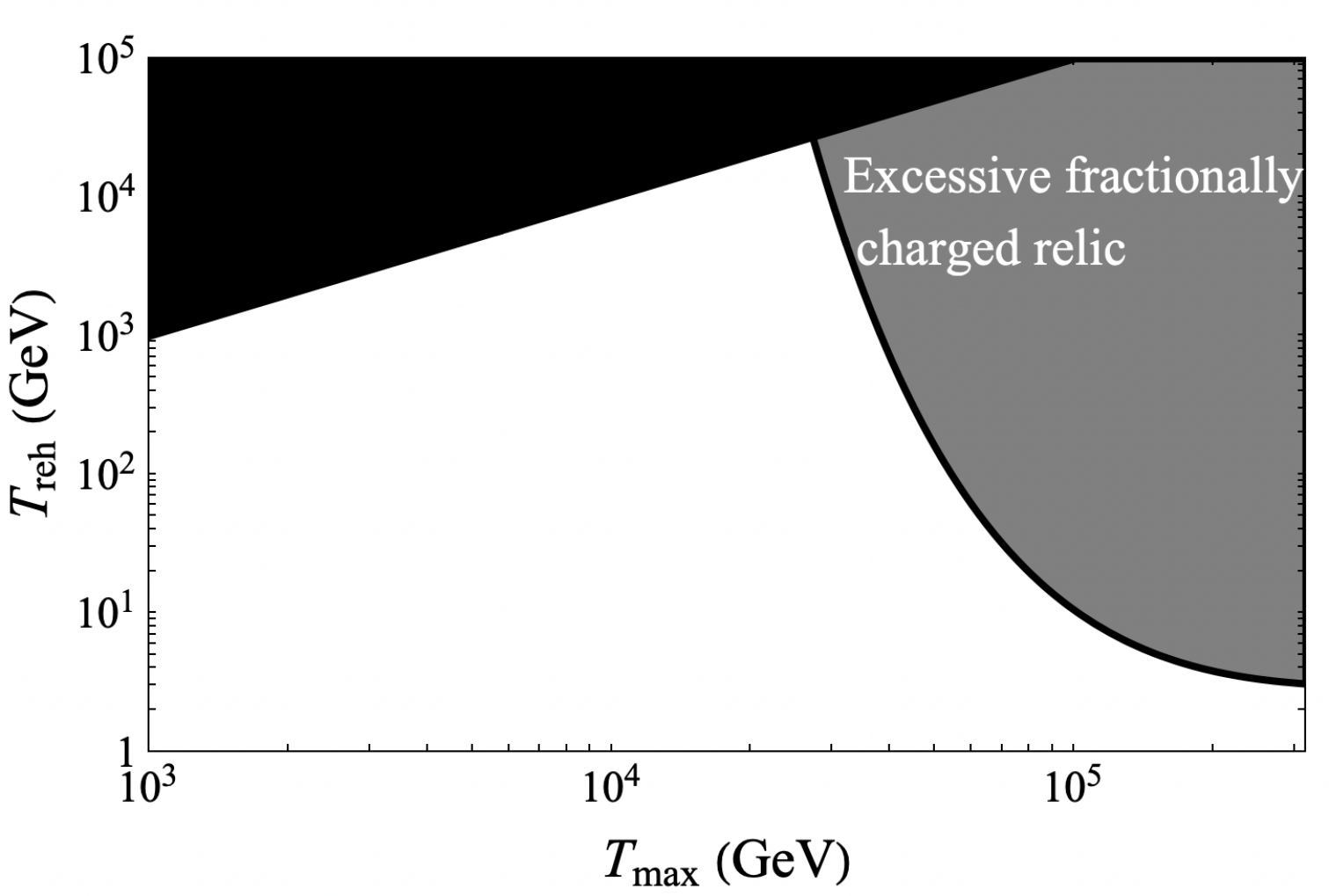}        
    \subcaption{}
    \label{fig:relic:b}
    \end{subfigure}
    \caption{
    Constraints on the reheating temperature due to fractionally charged relic messenger particles. (a): The solid curve represents the upper bound on $T_{\mathrm{reh}}$ when $T_{\mathrm{max}} = T_{\mathrm{reh}}$, and the dotted curve represents the case when $T_{\mathrm{max}} \simeq m_{\Psi}$. Any contour representing the upper bound by $T_{\mathrm{max}}$ between $T_{\mathrm{reh}}$ and $m_{\Psi}$ is located between these solid and dotted curves. (b): Upper bound on $T_{\mathrm{reh}}$ and $T_{\mathrm{max}}$, for a fixed messenger mass, $m_{\Psi}=1200$ TeV.
    }
    \label{fig:relic}
\end{figure}

On the other hand, if $T_{\mathrm{max}} \ll m_{\Psi}$, the relic density is given by
\begin{equation}
\Omega_{\mathrm{\Psi}} h^2 \simeq
 \frac{9 \sqrt{5}}{2^{25/2} \pi^{6}} \frac{g^2 g_{\ast}^{3/2}(T_{\mathrm{reh}})}{g_{\ast}^{3}(T_{\mathrm{\ast}})}\mathcal{B}\left(\frac{2 m_{\Psi}}{1.3 T_{\mathrm{max}}}\right) \left(\frac{2 m_{\Psi}}{1.3 T_{\mathrm{max}}}\right)^9 \frac{M_{\mathrm{Pl}} T_{\mathrm{reh}}^7}{m_{\Psi}^7 T_0}  A_s  \Omega_R h^2 \quad (T_{\mathrm{max}} \ll m_{\Psi}),
  \label{eq:relicMLowTmax}
\end{equation}
where 
\begin{equation}
    \mathcal{B}(x) = \int^{\infty}_{A_i} A^{19/8}  e^{-x A^{3/8}}\;dA,
\end{equation}
    with $A_{i} = 2.07$.\footnote{The variable $A$ is the rescaled scale factor such that the reheating process begins at $A=1$~\cite{Giudice:2000ex}. $A_i$ is set to ensure that the temperature scaling, $T\propto A^{-3/8}$, remains a good approximation. We checked that this choice agrees with the numerical solutions of the Boltzmann equations presented in Ref.~\cite{Giudice:2000ex}.}  Since $\mathcal{B}(x)\propto  e^{-xA_{i}^{3/8}}/x$ for $x\gg 1$, 
the relic density is suppressed by the large exponential factor $e^{-2m_{\Psi}/T_{\mathrm{max}}}$.

Figure~\ref{fig:relic} displays the upper bound on $T_{\mathrm{reh}}$ derived from constraints on FCPs.  We assume that the cross-section is governed by the EW process, $A_s = \pi\alpha_2^2 (1 + 1/(2c_W^4))/4$ \cite{Dimopoulos:1996gy}. Results from the strong interaction process, characterized by $A_s = \pi\alpha_3^2$, provide a similar constraint within an $\mathcal{O}(1)$ factor. 
 In Figure~\ref{fig:relic:a}, the solid curve shows the upper bound on $T_{\mathrm{reh}}$ when $T_{\mathrm{max}} = T_{\mathrm{reh}}$ (instant reheating). The constraint gets stronger for the slower reheating scenarios, and when $4m_{\Psi}/17 < T_{\text{max}} \lesssim m_{\Psi}$, the strongest upper bound is obtained by the dotted curve. The constraint on our benchmark scenario is given in Figure~\ref{fig:relic:b}. The result indicates that the constraint on the relic messenger can be evaded when $T_{\mathrm{reh}} \ll m_{\Psi}$. 
 
The constraint can be further relaxed if the electric charge of the messenger is either close to zero or $|e|$, depending on its representation and the value of kinetic mixing. If the electric charge of the messenger is much smaller than $|e|$, constraints on millicharged heavy particles \cite{Stebbins:2019xjr} should be considered. Alternatively, constraints on the charged stable particles \cite{SMITH1979525,SMITH1982333,PhysRevD.41.2074,PhysRevLett.68.1116,PhysRevD.47.1231} can be applied when the electric charge of the messenger is close to $|e|$. Moreover, FCPs may be depleted from the galactic disk due to supernova explosions~\cite{Chuzhoy:2008zy}. These considerations may alleviate the upper bounds shown in Figure~\ref{fig:relic}.

\subsection{Relic Dark Photons}

We note that the dark photon can be isolated from the MSSM sector and interacts only with the messengers. Therefore, its primary production channels are Compton-like scattering and pair annihilation processes involving the messengers. Under the assumption of a low reheating temperature, as shown in Figure~\ref{fig:relic}, the messenger population, and thus dark photon production, is suppressed. As a result, the dark photon does not substantially contribute to the effective number of neutrino species, $N_{\text{eff}}$.

\section{Summary and outlook}
\label{SummarySec}

In this paper, we have explored an extension of conventional gauge-mediated SUSY breaking by including an additional unbroken $U(1)_{D}$ gauge symmetry with large kinetic mixing. For messengers charged under both SM and dark gauge symmetries, 
extra contributions to the bino and sfermion soft-SUSY breaking terms appear with sizes that are governed by the strength of the kinetic mixing and dark gauge coupling. Significant distortions to the sparticle masses relative to conventional GMSB scenarios are possible when the kinetic mixing is sizable, most notably for the neutralino, Higgsino, and slepton sectors, with interesting phenomenological implications. As the kinetic mixing increases, the $\mu$ parameter required for successful EWSB decreases, implying the presence of a light Higgsino in the spectrum. Furthermore, for certain simple messenger representations, a light bino-dark photino is present in the spectrum, which could be probed via exotic Higgs decays. 
While a full exploration of the various phenomena in dark GMSB would only be possible at future colliders, there may still be room for interesting signatures to show up at the LHC. 

Looking ahead, it would be valuable to carry out a more detailed study of the collider signatures and detection prospects for our scenarios, both for the LHC and for future colliders. It would also be worthwhile to explore how different choices for the messenger representations and messenger scales impact the spectrum of superpartners and their phenomenology, as well as other issues such as gauge coupling unification and cosmology.  
Altogether, these considerations highlight the rich phenomenological landscape opened up by dark GMSB scenarios with large kinetic mixing, providing motivation for more detailed future studies and dedicated searches at present and upcoming collider experiments.

\begin{acknowledgments}
HL thanks Tianjun Li and Min-Seok Seo for the helpful discussions. JL thanks CERN-CKC Ph.D. student visiting program for the support.
This work was supported in part by the U.S. Department of Energy (Grant No. DE–SC0007914) and the National Research Foundation of Korea (Grant No. RS-2024-00352537). 
\end{acknowledgments}

\appendix

\section{Renormalization group beta functions}
\label{appendixRenormaliz}

In this appendix, we compile the beta functions for various quantities in the $\overline{\rm DR}$ scheme below the messenger scales. The two-loop RG equations are taken from Ref.~\cite{Martin:1993zk}, adopting the substitution rules in Ref.~\cite{Fonseca:2011vn}, which account for an additional $U(1)$ gauge group with kinetic mixing. 
For a generic coupling $X$, the renormalization group (RG) equation
takes the general form
\begin{equation}
    Q\frac{d}{dQ}X = \frac{1}{16\pi^2} \beta^{(1)}_X + \frac{1}{(16\pi^2)^2} \beta^{(2)}_X,
\end{equation}
where $Q$ is the renormalization scale, and $\beta_X^{(1)}$, $\beta_X^{(2)}$ correspond to the one- and two-loop beta functions, respectively. The beta functions for the MSSM can be found in Ref.~\cite{Martin:1993zk}. Here we only present the beta functions for the newly added parameters and those that have undergone any changes.

The beta functions for the gauge couplings remain unchanged if one uses the physical basis for the massless dark photon [Eq.~\eqref{effInt}]. 
The beta functions for the kinetic mixing and bino-dark photino mass mixing terms are as follows:
\begin{align}
\beta_{\eps}^{(1)} =&\frac{33\eps(1-\eps^2) g_1^2}{5} , &\\ 
    \beta_{\eps}^{(2)} =& \frac{\eps(1-\eps^2) g_1^2}{5}  \bigg\{ -\mathrm{Tr}[26\mathbf{Y}_u^{\dagger}\mathbf{Y}_u+14 \mathbf{Y}_d^{\dagger}\mathbf{Y}_d+18\mathbf{Y}_e^{\dagger}\mathbf{Y}_e]+88 g_3^2  +27g_2^2 +\frac{199 g_1^2}{5} \bigg\} ,\hspace{14.5 em}
\end{align}

\begin{align}
  \beta_{M_K}^{(1)} =& \frac{33 g_1^2M_K}{5}, \\
    \beta_{M_K}^{(2)} =&  \frac{2g_1^2M_K}{5}\left\{ 44g_3^2+\frac{27}{2} g_2^2+\frac{199g_1^2}{10 }-\mathrm{Tr}[13\mathbf{Y}_u^{\dagger}\mathbf{Y}_u+7\mathbf{Y}_d^{\dagger}\mathbf{Y}_d+9\mathbf{Y}_e^{\dagger}\mathbf{Y}_e]  \right\},\hspace{30 em}
\end{align}
where $g_1 \equiv \sqrt{5/3} g_Y$.
The beta functions of the dark gauge coupling $g_D$ and the dark photino mass $M_D$ vanish below the messenger scale, as there are no dark charged states.

Among the MSSM parameters, modifications occur only to the beta functions for the sfermion and Higgs mass terms. Some useful factors in the beta functions for these parameters are as follows \cite{Martin:1993zk}:
\begin{equation}
    \mathcal{S}\equiv m_{H_u}^2 -m_{H_d}^2 +\mathrm{Tr}[\mathbf{m}^2_Q-2\mathbf{m}^2_u+\mathbf{m}^2_d-\mathbf{m}_L^2+\mathbf{m}^2_e],
\end{equation}
\begin{align}
    \mathcal{S}^{\prime} \equiv& \mathrm{Tr}[-(3m_{H_u}^2+\mathbf{m}_Q^2)\mathbf{Y}_u^\dagger\mathbf{Y}_u +4\mathbf{Y}_u^\dagger \mathbf{m}_u^2\mathbf{Y}_u +(3m_{H_d}^2-\mathbf{m}_Q^2)\mathbf{Y}_d^\dagger\mathbf{Y}_d -2 \mathbf{Y}_d^\dagger \mathbf{m}_d^2\mathbf{Y}_d  \nonumber\\
    &\hspace{3 em}+(m_{H_d}^2+\mathbf{m}_L^2)\mathbf{Y}_e^\dagger\mathbf{Y}_e -2 \mathbf{Y}_e^\dagger\mathbf{m}_e^2\mathbf{Y}_e] + \left(\frac{3}{2}g_2^2 +\frac{3g_1^2}{10 }\right)\left\{ m_{H_u}^2 -m_{H_d}^2 -\mathrm{Tr}[\mathbf{m}_L^2]  \right\} \nonumber\\ 
    &+\left(\frac{8}{3}g_3^2+\frac{3}{2}g_2^2 +\frac{g_1^2}{30 }\right)\mathrm{Tr}[\mathbf{m}_Q^2]-\left(\frac{16}{3}g_3^2+\frac{16g_1^2}{15 }\right)\mathrm{Tr}[\mathbf{m}_u^2]\nonumber\\ &+\left(\frac{8}{3}g_3^2+\frac{2g_1^2}{15 }\right)\mathrm{Tr}[\mathbf{m}_d^2]+\frac{6g_1^2}{5 }\mathrm{Tr}[\mathbf{m}_e^2],
\end{align}
\begin{align}
 \sigma_1 &= \frac{g_1^2}{5 } \left\{ 3(m_{H_u}^2+m_{H_d}^2) +\mathrm{Tr}[\mathbf{m}^2_Q+3\mathbf{m}^2_L+8\mathbf{m}^2_u+2\mathbf{m}^2_d+6\mathbf{m}^2_e] \right\},  \\ 
 \sigma_2 &=  g_2^2 \left\{ m_{H_u}^2+m_{H_d}^2 +\mathrm{Tr}[3\mathbf{m}^2_Q+\mathbf{m}^2_L] \right\}, \\
 \sigma_3 &= g_3^2\left\{\mathrm{Tr}[2\mathbf{m}^2_Q+\mathbf{m}^2_u+\mathbf{m}^2_d] \right\} ,
\end{align}
Then, the beta functions of the scalar soft mass terms receive additional $M_K$ contributions compared to 
conventional GMSB,
\begin{align}
  \beta_{m^{2}_{H_u}}^{(1)}   =&    6\mathrm{Tr}[(m_{H_u}^2+\mathbf{m}_Q^2)\mathbf{Y}_u^{\dagger}\mathbf{Y}_u+ \mathbf{Y}_u^{\dagger}\mathbf{m}_u^2 \mathbf{Y}_u +\mathbf{h}_u^{\dagger}\mathbf{h}_u] -6g_2^2|M_2|^2-\frac{6g_1^2}{5 }(|M_1|^2+|M_K|^2)\nonumber\\ 
  &  +\frac{3g_1^2 \mathcal{S}}{5 },\\ 
  \beta_{m^{2}_{H_u}}^{(2)}   =&   -6\mathrm{Tr}[6(m_{H_{u}}^2 +\mathbf{m}_Q^2 )\mathbf{Y}^{\dagger}_u\mathbf{Y}_u\mathbf{Y}^{\dagger}_u\mathbf{Y}_u +6\mathbf{Y}_u^{\dagger}\mathbf{m}_u^2\mathbf{Y}_u\mathbf{Y}_u^{\dagger}\mathbf{Y}_u  \nonumber\\
  &\hspace{3 em } +(m_{H_{u}}^2 +m_{H_{d}}^2 +\mathbf{m}_Q^2 )\mathbf{Y}^{\dagger}_u\mathbf{Y}_u\mathbf{Y}^{\dagger}_d\mathbf{Y}_d +\mathbf{Y}_u^{\dagger}\mathbf{m}_u^2\mathbf{Y}_u\mathbf{Y}_d^{\dagger}\mathbf{Y}_d + \mathbf{Y}_u^{\dagger}\mathbf{Y}_u\mathbf{m}_Q^2\mathbf{Y}_d^{\dagger}\mathbf{Y}_d\nonumber\\
  &\hspace{3 em}+\mathbf{Y}_u^{\dagger}\mathbf{Y}_u\mathbf{Y}_d^{\dagger}\mathbf{m}_d^2\mathbf{Y}_d + 6\mathbf{h}_u^{\dagger}\mathbf{h}_u\mathbf{Y}_u^{\dagger}\mathbf{Y}_u+6\mathbf{h}_u^{\dagger}\mathbf{Y}_u\mathbf{Y}_u^{\dagger}\mathbf{h}_u+\mathbf{h}_d^{\dagger}\mathbf{h}_d\mathbf{Y}_u^{\dagger}\mathbf{Y}_u \nonumber
  \\&\hspace{3 em}+\mathbf{Y}_d^{\dagger}\mathbf{Y}_d\mathbf{h}_u^{\dagger}\mathbf{h}_u +\mathbf{h}_d^{\dagger}\mathbf{Y}_d\mathbf{Y}_u^{\dagger}\mathbf{h}_u +\mathbf{Y}_d^{\dagger}\mathbf{h}_d\mathbf{h}_u^{\dagger}\mathbf{Y}_u] \nonumber
  \\& + \left( 32 g_3^2+\frac{8g_1^2}{5 } \right) \mathrm{Tr}[(m_{H_{u}}^2 +\mathbf{m}_Q^2 )\mathbf{Y}^{\dagger}_u\mathbf{Y}_u+\mathbf{Y}^{\dagger}_u\mathbf{m}_u^2\mathbf{Y}_u+\mathbf{h}^{\dagger}_u\mathbf{h}_u] \nonumber\\
  &+ 32 g_3^2 \left\{2 |M_3|^2 \mathrm{Tr}[\mathbf{Y}^{\dagger}_u\mathbf{Y}_u] - M_3^\ast \mathrm{Tr}[\mathbf{Y}^{\dagger}_u\mathbf{h}_u]-  M_3 \mathrm{Tr}[\mathbf{h}^{\dagger}_u\mathbf{Y}_u]   \right\} \nonumber\\
& + \frac{8 g_1^2}{5 } \left\{2 (|M_1|^2+|M_K|^2) \mathrm{Tr}[\mathbf{Y}^{\dagger}_u\mathbf{Y}_u] - M_1^\ast \mathrm{Tr}[\mathbf{Y}^{\dagger}_u\mathbf{h}_u]-  M_1 \mathrm{Tr}[\mathbf{h}^{\dagger}_u\mathbf{Y}_u]   \right\}  \nonumber\\ 
&+\frac{6g_1^2}{5 } \mathcal{S}^{\prime} +33g_2^4|M_2|^2 +\frac{18g_2^2g_1^2}{5 } (|M_2|^2+|M_1|^2+|M_K|^2 +\mathrm{Re}[M_1M_2^\ast]) \nonumber\\  
&+\frac{621g_1^4}{25}|M_1|^2 +\frac{414g_1^4}{25}|M_K|^2 + 3g_2^2 \sigma_2 +\frac{3g_1^2\sigma_1}{5 } ,
\end{align}

\begin{align}
  \beta_{m^{2}_{H_d}}^{(1)}   =&     \mathrm{Tr}[ 6(m_{H_d}^2+\mathbf{m}_Q^2)\mathbf{Y}_d^{\dagger}\mathbf{Y}_d+ 6\mathbf{Y}_d^{\dagger}\mathbf{m}_d^2 \mathbf{Y}_d +2(m_{H_d}^2+\mathbf{m}_L^2)\mathbf{Y}_e^{\dagger}\mathbf{Y}_e+2\mathbf{Y}_e^{\dagger}\mathbf{m}_e^2\mathbf{Y}_e\nonumber\\
  +& 6\mathbf{h}_d^{\dagger}\mathbf{h}_d +2 \mathbf{h}_e^{\dagger}\mathbf{h}_e]- 6g_2^2|M_2|^2  -\frac{6g_1^2}{5 }(|M_1|^2+|M_K|^2)-\frac{3g_1^2 \mathcal{S}}{5 },  \\
  \beta_{m^{2}_{H_d}}^{(2)}   =&  -6\mathrm{Tr}[6(m_{H_{d}}^2 +\mathbf{m}_Q^2 )\mathbf{Y}^{\dagger}_d\mathbf{Y}_d\mathbf{Y}^{\dagger}_d\mathbf{Y}_d +6\mathbf{Y}_d^{\dagger}\mathbf{m}_d^2\mathbf{Y}_d\mathbf{Y}_d^{\dagger}\mathbf{Y}_d  \nonumber\\
  &\hspace{3 em } +(m_{H_{u}}^2 +m_{H_{d}}^2 +\mathbf{m}_Q^2 )\mathbf{Y}^{\dagger}_u\mathbf{Y}_u\mathbf{Y}^{\dagger}_d\mathbf{Y}_d +\mathbf{Y}_u^{\dagger}\mathbf{m}_u^2\mathbf{Y}_u\mathbf{Y}_d^{\dagger}\mathbf{Y}_d+\mathbf{Y}_u^{\dagger}\mathbf{Y}_u\mathbf{m}_Q^2\mathbf{Y}_d^{\dagger}\mathbf{Y}_d\nonumber \\
  &\hspace{3 em}+\mathbf{Y}_u^{\dagger}\mathbf{Y}_u\mathbf{Y}_d^{\dagger}\mathbf{m}_d^2\mathbf{Y}_d + 2(m_{H_{d}}^2 +\mathbf{m}_L^2 )\mathbf{Y}^{\dagger}_e\mathbf{Y}_e\mathbf{Y}^{\dagger}_e\mathbf{Y}_e +2\mathbf{Y}_e^{\dagger}\mathbf{m}_e^2\mathbf{Y}_e\mathbf{Y}_e^{\dagger}\mathbf{Y}_e \nonumber\\
  &\hspace{3 em}+6\mathbf{h}_d^{\dagger}\mathbf{h}_d\mathbf{Y}_d^{\dagger}\mathbf{Y}_d+6\mathbf{h}_d^{\dagger}\mathbf{Y}_d\mathbf{Y}_d^{\dagger}\mathbf{h}_d+\mathbf{h}_u^{\dagger}\mathbf{h}_u\mathbf{Y}_d^{\dagger}\mathbf{Y}_d+\mathbf{Y}_u^{\dagger}\mathbf{Y}_u\mathbf{h}_d^{\dagger}\mathbf{h}_d +\mathbf{h}_u^{\dagger}\mathbf{Y}_u\mathbf{Y}_d^{\dagger}\mathbf{h}_d  \nonumber \\&\hspace{3 em} +\mathbf{Y}_u^{\dagger}\mathbf{h}_u\mathbf{h}_d^{\dagger}\mathbf{Y}_d+2\mathbf{h}_e^{\dagger}\mathbf{h}_e\mathbf{Y}_e^{\dagger}\mathbf{Y}_e+2\mathbf{h}_e^{\dagger}\mathbf{Y}_e\mathbf{Y}_e^{\dagger}\mathbf{h}_e ] \nonumber \\
  &+ \left( 32 g_3^2- \frac{4g_1^2}{5 } \right) \mathrm{Tr}[(m_{H_{d}}^2 +\mathbf{m}_Q^2 )\mathbf{Y}^{\dagger}_d\mathbf{Y}_d+\mathbf{Y}^{\dagger}_d\mathbf{m}_d^2\mathbf{Y}_d+\mathbf{h}^{\dagger}_d\mathbf{h}_d] \nonumber\\
  &+ 32 g_3^2 \left\{2 |M_3|^2 \mathrm{Tr}[\mathbf{Y}^{\dagger}_d\mathbf{Y}_d] - M_3^\ast \mathrm{Tr}[\mathbf{Y}^{\dagger}_d\mathbf{h}_d]-  M_3 \mathrm{Tr}[\mathbf{h}^{\dagger}_d\mathbf{Y}_d]   \right\} \nonumber\\
& - \frac{4 g_1^2}{5 } \left\{2 (|M_1|^2+|M_K|^2) \mathrm{Tr}[\mathbf{Y}^{\dagger}_d\mathbf{Y}_d] - M_1^\ast \mathrm{Tr}[\mathbf{Y}^{\dagger}_d\mathbf{h}_d]-  M_1 \mathrm{Tr}[\mathbf{h}^{\dagger}_d\mathbf{Y}_d]   \right\}  \nonumber\\ 
 &+  \frac{12g_1^2}{5 }\Big\{ \mathrm{Tr}[(m_{H_{d}}^2 +\mathbf{m}_L^2 )\mathbf{Y}^{\dagger}_e\mathbf{Y}_e+\mathbf{Y}^{\dagger}_e\mathbf{m}_e^2\mathbf{Y}_e+\mathbf{h}^{\dagger}_e\mathbf{h}_e] +2 (|M_1|^2+|M_K|^2) \mathrm{Tr}[\mathbf{Y}^{\dagger}_e\mathbf{Y}_e] \nonumber\\
 &\hspace{6 em}- M_1^\ast \mathrm{Tr}[\mathbf{Y}^{\dagger}_e\mathbf{h}_e]-  M_1 \mathrm{Tr}[\mathbf{h}^{\dagger}_e\mathbf{Y}_e]    \Big\}-\frac{6g_1^2 \mathcal{S}^{\prime}}{5 } +33g_2^4|M_2|^2 
 \nonumber\\
&+\frac{18g_2^2g_1^2}{5 } (|M_2|^2+|M_1|^2+|M_K|^2 +\mathrm{Re}[M_1M_2^\ast]) +\frac{621g_1^4|M_1|^2}{25}+\frac{414g_1^4|M_K|^2}{25}\nonumber\\
&  + 3g_2^2 \sigma_2 +\frac{3g_1^2 \sigma_1}{5 },
\end{align}

\begin{align}
  \beta_{m^{2}_{Q}}^{(1)}   =&     (2m_{H_u}^2+\mathbf{m}_Q^2)\mathbf{Y}_u^{\dagger}\mathbf{Y}_u+(2m_{H_d}^2+\mathbf{m}_Q^2)\mathbf{Y}_d^{\dagger}\mathbf{Y}_d+(\mathbf{Y}_u^{\dagger}\mathbf{Y}_u+\mathbf{Y}_d^{\dagger}\mathbf{Y}_d)\mathbf{m}_Q^2 +2\mathbf{Y}_u^{\dagger}\mathbf{m}_u^2 \mathbf{Y}_u \nonumber\\
&+2\mathbf{Y}_d^{\dagger}\mathbf{m}_e^2\mathbf{Y}_d+2\mathbf{h}_u^{\dagger}\mathbf{h}_u +2 \mathbf{h}_d^{\dagger}\mathbf{h}_d - \frac{32}{3}g_3^2|M_3|^2- 6g_2^2|M_2|^2 -\frac{2g_1^2}{15 }(|M_1|^2+|M_K|^2)\nonumber\\
  &  +\frac{g_1^2\mathcal{S} }{5 } , \\
  \beta_{m^{2}_{Q}}^{(2)}   =&  -(8m_{H_{u}}^2 +2\mathbf{m}_Q^2 )\mathbf{Y}^{\dagger}_u\mathbf{Y}_u\mathbf{Y}^{\dagger}_u\mathbf{Y}_u -4\mathbf{Y}_u^{\dagger}\mathbf{m}_u^2\mathbf{Y}_u\mathbf{Y}_u^{\dagger}\mathbf{Y}_u-4\mathbf{Y}_u^{\dagger}\mathbf{Y}_u\mathbf{m}_Q^2\mathbf{Y}_u^{\dagger}\mathbf{Y}_u  \nonumber\\
&-4\mathbf{Y}_u^{\dagger}\mathbf{Y}_u\mathbf{Y}_u^{\dagger}\mathbf{m}_u^2\mathbf{Y}_u-2\mathbf{Y}_u^{\dagger}\mathbf{Y}_u\mathbf{Y}_u^{\dagger}\mathbf{Y}_u\mathbf{m}_Q^2- (8m_{H_{d}}^2 +2\mathbf{m}_Q^2 )\mathbf{Y}^{\dagger}_d\mathbf{Y}_d\mathbf{Y}^{\dagger}_d\mathbf{Y}_d \nonumber\\
  & -4 \bYd{d}\sqm{d}\bY{d}\bYd{d}\bY{d} -4 \bYd{d}\bY{d}\sqm{d}\bYd{d}\bY{d} -4 \bYd d \bY d \bYd d \sqm d \bY d -2 \bYd d \bY d \bYd d  \bY d \sqm Q \nonumber \\
  &-\left\{ (\sqm Q + 4 \sqmHu)\bYd u \bY u + 2 \bYd u \sqm u \bY u +\bYd u \bY u \sqm Q   \right\}\mr{Tr}[3 \bYd u \bY u]  \nonumber \\
  &-\left\{ (\sqm Q + 4 \sqmHd)\bYd d \bY d + 2 \bYd d \sqm d \bY d +\bYd d \bY d \sqm Q   \right\}\mr{Tr}[3 \bYd d \bY d+\bYd e \bY e]  \nonumber \\
  &-6 \bYd u \bY u \mr{Tr}[\sqm Q \bYd u\bY u + \bYd u \sqm u \bY u] \nonumber\\ 
  &-\bYd d \bY d \mr{Tr}[6 \sqm Q \bYd d \bY d +6 \bYd d \sqm d \bY d +2 \sqm L \bYd e \bY e + 2 \bYd e \sqm e \bY e] \nonumber\\
  &-4 \Big\{ \bYd u \bY u \bhd u \bh u + \bhd u \bh u \bYd u \bY u +\bYd u \bh u \bhd u \bY u + \bhd u \bY u \bYd u \bY u  \Big\} \nonumber\\
&-4 \left\{ \bYd d \bY d \bhd d \bh d + \bhd d \bh d \bYd d \bY d +\bYd d \bh d \bhd d \bY d + \bhd d \bY d \bYd d \bY d   \right\} -\bhd u \bh u \mr{Tr}[6 \bYd u \bY u] \nonumber\\
& - \bYd u \bY u \mr{Tr}[6 \bhd u \bh u] - \bhd u \bY u \mr{Tr}[6\bYd u \bh u]- \bYd u \bh u \mr{Tr}[6 \bhd u \bY u] -\bhd d \bh d \mr{Tr}[6\bYd d \bY d +2 \bYd e \bY e]  \nonumber\\
    &-\bYd d \bY d \mr{Tr}[6\bhd d \bh d +2 \bhd e \bh e] -\bhd d \bY d \mr{Tr}[6\bYd d \bh d +2 \bYd e \bh e]-\bYd d \bh d \mr{Tr}[6\bhd d \bY d +2 \bhd e \bY e] \nonumber\\
  &+  \frac{2g_1^2}{5 }\Big\{( 4\sqmHu + 2\sqm Q)\bYd u \bY u + 4 \bYd u \sqm u \bY u +2 \bYd u \bY u \sqm Q +4 \bhd u \bh u -4 M_1 \bhd u \bY u   \nonumber\\
  & \hspace{5.5 em} - 4 M^\ast_1 \bYd u \bh u +8 (|M_1|^2+|M_K|^2) \bYd u \bY u+(\sqm Q + 2 \sqmHd)\bYd d \bY d+2 \bYd d \sqm d \bY d \nonumber\\
& \hspace{5.5 em}  +\bYd d \bY d \sqm Q +2\bhd d \bh d - 2 M_1 \bhd d \bY d - 2 M^\ast_1\bYd d \bh d + 4 (|M_1|^2+|M_K|^2) \bYd d \bY d \Big\}\nonumber\\ 
 &+\frac{2g_1^2\mathcal{S}^\prime }{5 }- \frac{128}{3}g_3^4|M_3|^2+ 32 g_3^2g_2^2(|M_3|^2 +|M_2|^2 +\mr{Re}[M_2M_3^\ast])  \nonumber\\
 & +\frac{32g_3^2g_1^2}{45 }(|M_3|^2 +|M_1|^2+|M_K|^2 +\mr{Re}[M_1M_3^\ast])+ 33g_2^4 |M_2|^2   \nonumber\\
& +\frac{2g_2^2g_1^2}{5 } (|M_2|^2+|M_1|^2+|M_K|^2 +\mathrm{Re}[M_1M_2^\ast]) +\frac{199g_1^4|M_1|^2}{75}+\frac{398g_1^4|M_K|^2}{225} \nonumber\\
& +\frac{16}{3}g_3^2\sigma_3+3g_2^2 \sigma_2 +\frac{g_1^2\sigma_1}{15},
\end{align}

\begin{align}
  \beta_{m^{2}_{L}}^{(1)}   =&     (2\sqmHd+\sqm L)\bYd e \bY e  +2\bYd e\sqm e \bY e +\bYd e \bY e \sqm L + 2\bhd e\bh e  -6 g_2^2|M_2|^2 \nonumber\\
  &-\frac{6g_1^2}{5 }(|M_1|^2+|M_K|^2) -\frac{ 3g_1^2 \mathcal{S}}{5 } ,  \\
  \beta_{m^{2}_{L}}^{(2)}   =&  -(8\sqmHu +2\sqm L )\bYd e \bY e \bYd e \bY e  -4\bYd e \sqm e  \bY e \bYd e \bY e -4 \bYd e \bY e \sqm L \bYd e \bY e   \nonumber\\
  &-4 \bYd e \bY e \bYd e \sqm e \bY e -2 \bYd e \bY e \bYd e \bY e \sqm L \nonumber \\
& -\left\{ (4 \sqmHd+\sqm L )\bYd e \bY e +2\bYd e \sqm e \bY e +\bYd e \bY e \sqm L \right\} \mr{Tr}[3\bYd d \bY d +\bYd e \bY e] \nonumber\\ 
& -\bYd e \bY e \mr{Tr}[6\sqm Q \bYd d \bY d +6 \bYd d \sqm d \bY d +2\sqm L \bYd e \bY e +2\bYd e \sqm e\bY e] \nonumber\\
  &-4 \Big\{ \bYd e \bY e \bhd e \bh e + \bhd e \bh e \bYd e \bY e +\bYd e \bh e \bhd e \bY e + \bhd e \bY e \bYd e \bY e  \Big\} -\bhd e \bh e \mr{Tr}[6 \bYd d \bY d +2 \bYd e \bY e]  \nonumber\\
    & - \bYd e \bY e \mr{Tr}[6 \bhd d \bh d +2 \bhd e \bh e ] -\bhd e \bY e \mr{Tr}[6 \bYd d \bh d +2 \bYd e \bh e]-\bYd e \bh e \mr{Tr}[6 \bhd d \bY d +2 \bhd e \bY e] \nonumber\\
  &+  \frac{6g_1^2}{5 }\Big\{( 2\sqmHu + \sqm L)\bYd e \bY e + 2 \bYd e \sqm e \bY e + \bYd e \bY e \sqm L +2 \bhd e \bh e -2 M_1 \bhd e \bY e   \nonumber\\
  & \hspace{5 em} - 2 M^\ast_1 \bYd e \bh e +4 (|M_1|^2+|M_K|^2) \bYd e \bY e \Big\}  -\frac{6g_1^2\mathcal{S}^\prime}{5 } +33g_2^4|M_2|^2  \nonumber\\
&+\frac{18g_2^2g_1^2}{5 } (|M_2|^2+|M_1|^2+|M_K|^2 +\mathrm{Re}[M_1M_2^\ast]) +\frac{621g_1^4|M_1|^2}{25}+\frac{414g_1^4|M_K|^2}{25}\nonumber\\&  +3g_2^2 \sigma_2 +\frac{3g_1^2\sigma_1}{5} ,
\end{align}

\begin{align}
  \beta_{m^{2}_{u}}^{(1)}   =&     (4\sqmHu+2\sqm u )\bY u \bYd u  +4\bY u\sqm Q \bYd u +2\bY u \bYd u \sqm u + 4\bh u\bhd u  -\frac{32}{3} g_3^2|M_3|^2 \nonumber\\ &-\frac{32g_1^2}{15 }(|M_1|^2+|M_K|^2) -\frac{4 g_1^2 \mathcal{S}}{5 },  \\
  \beta_{m^{2}_{u}}^{(2)}   =&  -(8\sqmHu +2\sqm u )\bY u \bYd u \bY u \bYd u  -4\bY u \sqm Q  \bYd u \bY u \bYd u -4 \bY u \bYd u \sqm u \bY u \bYd u   \nonumber\\
  &-4 \bY u \bYd u \bY u \sqm Q \bYd u -2 \bY u \bYd u \bY u \bYd u \sqm u-( 4 \sqmHu +4 \sqmHd+2\sqm u) \bY u \bYd d \bY d \bYd u  \nonumber \\ & -4\bY u \sqm Q  \bYd d \bY d \bYd u -4 \bY u \bYd d \sqm d \bY d \bYd u -4 \bY u \bYd d \bY d \sqm Q \bYd u -2 \bY u \bYd d \bY d \bYd u \sqm u
 \nonumber\\
& -\left\{ (4 \sqmHu+\sqm u )\bY u \bYd u +2\bY u \sqm Q \bYd u +\bY u \bYd u \sqm u \right\} \mr{Tr}[6\bYd u \bY u ] \nonumber\\ 
& -12\bY u \bYd u \mr{Tr}[\sqm Q \bYd u \bY u +\bYd u \sqm u \bY u] \nonumber\\ 
  &-4 \Big\{ \bh u \bhd u \bY u \bYd u + \bY u \bYd u \bh u \bhd u +\bh u \bYd u \bY u \bhd u + \bY u \bhd u \bh u \bYd u  \Big\}\nonumber\\ 
  &-4 \Big\{ \bh u \bhd d \bY d \bYd u + \bY u \bYd d \bh d \bhd u +\bh u \bYd d \bY d \bhd u + \bY u \bhd d \bh d \bYd u  \Big\} \nonumber\\
    & -12 \left\{ \bh u \bhd u \mr{Tr}[\bYd u \bY u] +\bY u \bYd u \mr{Tr}[\bhd u \bh u] + \bh u \bYd u \mr{Tr}[\bhd u \bY u] + \bY u \bhd u \mr{Tr}[\bYd u \bh u] \right\}  \nonumber\\
  &+ \left(6g_2^2 -\frac{2g_1^2}{5 }\right )\Big\{( 2\sqmHu + \sqm u)\bY u \bYd u + 2 \bY u \sqm Q \bYd u + \bY u \bYd u \sqm u +2 \bh u \bhd u 
 \Big\}  \nonumber\\
  & +12g_2^2\left\{2|M_2|^2\bY u \bYd u - M_2^\ast \bh u \bYd u -M_2 \bY u \bhd u\right\}\nonumber\\
  &-\frac{4g_1^2}{5}\left\{2(|M_1|^2+|M_K|^2)\bY u \bYd u - M_1^\ast \bh u \bYd u -M_1 \bY u \bhd u\right\} -\frac{8g_1^2\mathcal{S}^\prime}{5 }   \nonumber\\
&-\frac{128}{3}g_3^4|M_3|^2+\frac{512g_3^2g_1^2}{45} (|M_3|^2+|M_1|^2+|M_K|^2 +\mathrm{Re}[M_1M_3^\ast]) +\frac{3424g_1^4|M_1|^2}{75} \nonumber\\ &+\frac{6848g_1^4|M_K|^2}{225} +\frac{16}{3}g_3^2 \sigma_3 +\frac{16g_1^2\sigma_1}{15 } ,
\end{align}

\begin{align}
  \beta_{m^{2}_{d}}^{(1)}   =&     (4\sqmHu+2\sqm d )\bY d \bYd d  +4\bY d\sqm Q \bYd d +2\bY d \bYd d \sqm d + 4\bh d\bhd d  -\frac{32}{3} g_3^2|M_3|^2\nonumber\\
  &-\frac{8g_1^2}{15 }(|M_1|^2+|M_K|^2) +\frac{2 g_1^2 \mathcal{S}}{5 },   \\
  \beta_{m^{2}_{d}}^{(2)}   =&  -(8\sqmHd +2\sqm d )\bY d \bYd d \bY d \bYd d  -4\bY d \sqm Q  \bYd d \bY d \bYd d -4 \bY d \bYd d \sqm d \bY d \bYd d   \nonumber\\
  &-4 \bY d \bYd d \bY d \sqm Q \bYd d -2 \bY d \bYd d \bY d \bYd d \sqm d -( 4 \sqmHu +4 \sqmHd+2\sqm d) \bY d \bYd u \bY u \bYd d  \nonumber \\ 
  & -4\bY d \sqm Q  \bYd u  \bY u \bYd d -4 \bY d \bYd u \sqm u \bY u \bYd d -4 \bY d \bYd u \bY u \sqm Q \bYd d -2 \bY d \bYd u \bY u \bYd d \sqm d
 \nonumber\\
& -\left\{ (4 \sqmHd+\sqm d )\bY d \bYd d +2\bY d \sqm Q \bYd d +\bY d \bYd d \sqm d \right\} \mr{Tr}[6\bYd d \bY d + 2 \bYd e \bY e ] \nonumber\\ 
& -4\bY d \bYd d \mr{Tr}[3\sqm Q \bYd d \bY d +3\bYd d \sqm d \bY d+\sqm L \bYd e \bY e +\bYd e \sqm e \bY e] \nonumber\\ 
  &-4 \Big\{ \bh d \bhd d \bY d \bYd d + \bY d \bYd d \bh d \bhd d +\bh d \bYd d \bY d \bhd d + \bY d \bhd d \bh d \bYd d  \Big\}\nonumber\\  &-4 \Big\{ \bh d \bhd u \bY u \bYd d + \bY d \bYd u \bh u \bhd d +\bh d \bYd u \bY u \bhd d + \bY d \bhd u \bh u \bYd d  \Big\}  -4  \bh d \bhd d \mr{Tr}[3\bYd d \bY d + \bYd e \bY e]\nonumber\\
    & -4\bY d \bYd d \mr{Tr}[3\bhd d \bh d + \bhd e \bh e] -4 \bh d \bYd d \mr{Tr}[3\bhd d \bY d +\bhd e \bY e] -4 \bY d \bhd d \mr{Tr}[3\bYd d \bh d +\bYd e \bh e]  \nonumber\\
  &+ \left(6g_2^2 +\frac{2g_1^2}{5 }\right )\Big\{( 2\sqmHd + \sqm d)\bY d \bYd d + 2 \bY d \sqm Q \bYd d + \bY d \bYd d \sqm d +2 \bh d \bhd d 
 \Big\}  \nonumber\\
  & +12g_2^2\left\{2|M_2|^2\bY d \bYd d - M_2^\ast \bh d \bYd d -M_2 \bY d \bhd d\right\}\nonumber\\
  &+\frac{4g_1^2}{5 }\left\{2(|M_1|^2+|M_K|^2)\bY d \bYd d - M_1^\ast \bh d \bYd d -M_1 \bY d \bhd d\right\} +\frac{4g_1^2\mathcal{S}^\prime}{5 }  \nonumber\\
&-\frac{128}{3}g_3^4|M_3|^2 +\frac{128g_3^2g_1^2}{45 } (|M_3|^2+|M_1|^2+|M_K|^2 +\mathrm{Re}[M_1M_3^\ast]) +\frac{808g_1^4|M_1|^2}{75} \nonumber\\&+\frac{1616g_1^4|M_K|^2}{225}+\frac{16}{3}g_3^2 \sigma_3 +\frac{4g_1^2\sigma_1}{15 },
\end{align}

\begin{align}
  \beta_{m^{2}_{e}}^{(1)}   =&     (4\sqmHd+2\sqm e )\bY e \bYd e  +4\bY e\sqm L \bYd e +2\bY e \bYd e \sqm e + 4\bh e\bhd e-\frac{24g_1^2}{5 }(|M_1|^2+|M_K|^2) \nonumber\\
  &+\frac{6 g_1^2 \mathcal{S}}{5 },   \\
  \beta_{m^{2}_{e}}^{(2)}   =&  -(8\sqmHd +2\sqm e )\bY e \bYd e \bY e \bYd e  -4\bY e \sqm L  \bYd e \bY e \bYd e -4 \bY e \bYd e \sqm e \bY e \bYd e   \nonumber\\
  &-4 \bY e \bYd e \bY e \sqm L \bYd e -2 \bY e \bYd e \bY e \bYd e \sqm e  \nonumber \\ 
   &-\left\{ (4 \sqmHd+\sqm e )\bY e \bYd e +2\bY e \sqm L \bYd e +\bY e \bYd e \sqm e \right\} \mr{Tr}[6\bYd d \bY d + 2 \bYd e \bY e ] \nonumber\\
& -4\bY e \bYd e \mr{Tr}[3\sqm Q \bYd d \bY d +3\bYd d \sqm d \bY d+ \sqm L \bYd e \bY e +\bYd e \sqm e \bY e] \nonumber\\ 
  &-4 \Big\{ \bh e \bhd e \bY e \bYd e + \bY e \bYd e \bh e \bhd e +\bh e \bYd e \bY e \bhd e + \bY e \bhd e \bh e \bYd e  \Big\}-4  \bh e \bhd e \mr{Tr}[3\bYd d \bY d+ \bYd e \bY e] \nonumber\\  
  &  -4\bY e \bYd e \mr{Tr}[3\bhd d \bh d + \bhd e \bh e] -4 \bh e \bYd e \mr{Tr}[3\bhd d \bY d +\bhd e \bY e] -4 \bY e \bhd e \mr{Tr}[3\bYd d \bh d +\bYd e \bh e]  \nonumber\\
  &+ \left(6g_2^2 -\frac{6g_1^2}{5 }\right )\Big\{( 2\sqmHd + \sqm e)\bY e \bYd e + 2 \bY e \sqm L \bYd e + \bY e \bYd e \sqm e +2 \bh e \bhd e 
 \Big\}  \nonumber\\
  & +12g_2^2\left\{2|M_2|^2\bY e \bYd e - M_2^\ast \bh e \bYd e -M_2 \bY e \bhd e\right\}\nonumber\\
  &-\frac{12g_1^2}{5 }\left\{2(|M_1|^2+|M_K|^2)\bY e \bYd e - M_1^\ast \bh e \bYd e -M_1 \bY e \bhd e\right\} +\frac{12g_1^2\mathcal{S}^\prime}{5 }  \nonumber\\
&+\frac{2808g_1^4|M_1|^2}{25}+\frac{1872g_1^4|M_K|^2}{25} +\frac{12g_1^2\sigma_1}{5 } ,
\end{align}
The MSSM beta functions in Ref.~\cite{Martin:1993zk} are recovered in the $g_D \rightarrow 0$ and $\epsilon \rightarrow 0$ limit. The absence of any states carrying dark charge below the messenger scale implies that the beta functions for the MSSM parameters 
have no explicit dependence on $\epsilon$ or $g_D$. 
On the other hand, we observe that the beta functions for scalar soft masses are modified by terms involving $M_K^2$. 
We note that the RG trajectory of dark GMSB can differ significantly from that of conventional GMSB. For instance, $M_1$ and $M_K$ can be the dominant contributions to the beta function for the Higgs mass parameters. 
This could, in turn, greatly affect the determination of parameters related to EWSB.

\section{Evolution gauge couplings and kinetic mixing at high scales}
\label{appendixRunning}

\begin{figure}[tb]
    \centering
    \begin{subfigure}{0.48\textwidth}
    \includegraphics[width=\linewidth]{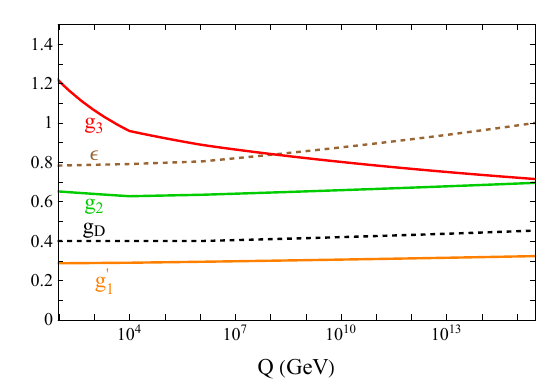}
    \caption{Running of various couplings }\label{subfig:gaugecouplingRUn}
    \end{subfigure}
    \begin{subfigure}{0.48\textwidth}
    \includegraphics[width=\linewidth]{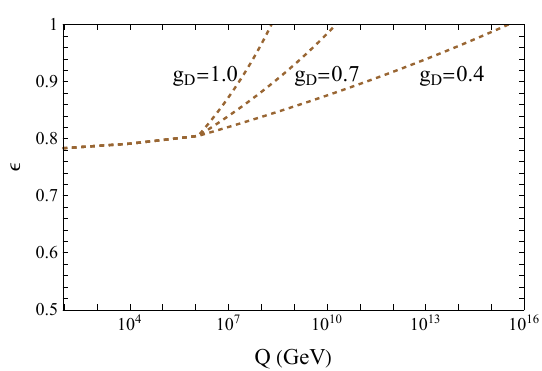}
    \caption{Running of kinetic mixing}\label{subfig:epsilonRun}
    \end{subfigure}
    \caption{
    Examples of the running of gauge couplings and kinetic mixing. We take 
$SU(5)$ complete messenger from Scenario I, and the normalization $g_1^{\prime}\equiv \sqrt{5/3} g_Y^{\prime}$. 
    (a): The parameters are set according to our benchmark point, with $g_Y(m_Z)=0.357$, $g_2(m_Z)=0.652$, $g_3(m_Z)=1.220$ \cite{ParticleDataGroup:2022pth}, $g_D(m_Z)=0.400$, and $\epsilon(m_Z)=0.777$. (b): The evolution of $\epsilon$ for different values of $g_D$ is shown, while the other parameters remain the same as in (a).
   }
    \label{fig:RGrun}
\end{figure}

In this appendix, we discuss the evolution of the gauge coupling constants and kinetic mixing  
in the dark GMSB model above the messenger scale.
We highlight the qualitative impact of a large $\epsilon$ on gauge coupling unification through one-loop RG running.
In the original basis of Eqs.~\eqref{fieldstrength}, \eqref{kineticMix}, the gauge interaction of the matter fields can be expressed as
\begin{equation}
\mathcal{L} \supset \bar{\psi}\gamma^{\mu} (g^{\prime}_Y Y_{\psi} \mathbb{B}_{\mu} + g_D D_{\psi}\mathbb{X}_{\mu})\psi,
\end{equation}
where $\psi$ carries a hypercharge $Y_{\psi}$ and a $U(1)_{D}$ charge $D_{\psi}$. In the canonical basis, obtained through Eq.~(\ref{physical}), where the kinetic terms are diagonal, the interaction terms can be rewritten as
\begin{equation}
\mathcal{L} \supset \bar{\psi}_i\gamma^{\mu} \left[(-g_{\text{eff}} D_{i} + g_{Y} Y_{i}) B_{\mu} + g_D D_{i} X_{\mu}\right]\psi_i,
\end{equation}
where $g_Y = g_Y^{\prime}/ \sqrt{1-\epsilon^2}$, and $g_{\mathrm{eff}} = g_D\epsilon/\sqrt{1-\epsilon^2}$.

The one-loop evolution equations for the gauge couplings are given by \cite{Babu:1996vt}
\begin{equation}
\begin{split}
    &Q\frac{dg_D}{dQ} = \frac{1}{16\pi^2} \mathrm{tr}[g_D^3 D_i^2],\\
    &Q\frac{dg_{Y}}{dQ} = \frac{1}{16\pi^2} \mathrm{tr}[g_Y^3 Y_i^2 + g_Yg_{\mathrm{eff}}^2 D_i^2 - 2g_Y^2g_{\mathrm{eff}}Y_iD_i],\\
    &Q\frac{dg_{\mathrm{eff}}}{dQ} = \frac{1}{16\pi^2} \mathrm{tr}[g_Y^2g_{\mathrm{eff}}Y_i^2 + g_{\mathrm{eff}}^3 D_i^2 + 2g_D^2 g_{\mathrm{eff}}D_i^2  - 2g_D^2 g_Y Y_iD_i - 2g_Yg_{\mathrm{eff}}^2 Y_iD_i ],
\end{split}
\end{equation}
where `$\mathrm{tr}[\;]$' denotes the trace over the chiral supermultiplets. 
The beta functions of the gauge couplings and kinetic mixing in the original basis can be expressed as \cite{Daido:2016kez}
\begin{equation}
\begin{split}
 &\beta_{g_D} = \frac{1}{16\pi^2} \mathrm{tr}[g_D^3 D_i^2],\\
     &\beta_{g_{Y}^{\prime}} = \frac{1}{16\pi^2} \mathrm{tr}[g_Y^{\prime 3} Y_i^2],\\
 &\beta_{\epsilon} = \frac{1}{16\pi^2} \mathrm{tr}[\epsilon(g_{Y}^{\prime2} Y_i^2 + g_D^2 D_i^2) - 2 g_Y g_D Y_i D_i].
 \end{split}
\label{eq:AbelBeta}
\end{equation}
Furthermore, the beta functions for the $SU(2)$ and $SU(3)$ gauge couplings are given as
\begin{equation}
\begin{split}
    &\beta_{g_2} = \frac{g_2^3}{16\pi^2} \left(-6 + \mathrm{Tr}[S_a]\right),\\
    &\beta_{g_3} = \frac{g_3^3}{16\pi^2} \left(-9 + \mathrm{Tr}[S_a]\right),
\end{split}
\label{eq:NonAbelBeta}
\end{equation}
where the trace is taken over all chiral multiples.

\begin{figure}[tb]
    \centering
    \begin{subfigure}{0.48\textwidth}
    \includegraphics[width=\linewidth]{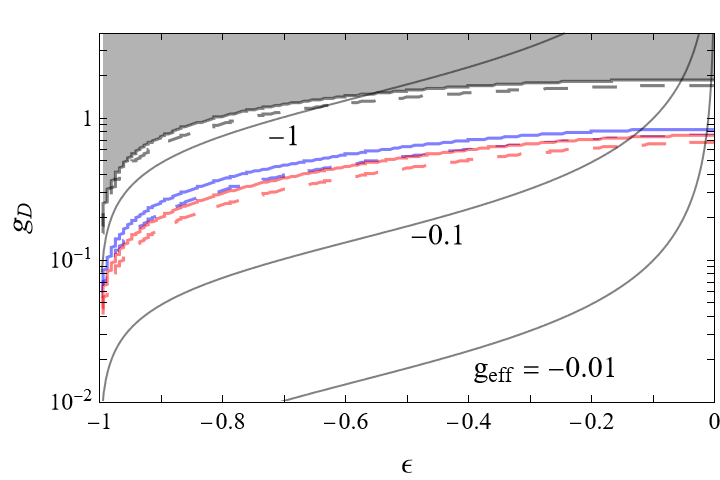}
    \caption{Negative $\epsilon$}\label{subfig:poleNeg}
    \end{subfigure}
    \begin{subfigure}{0.48\textwidth}
    \includegraphics[width=\linewidth]{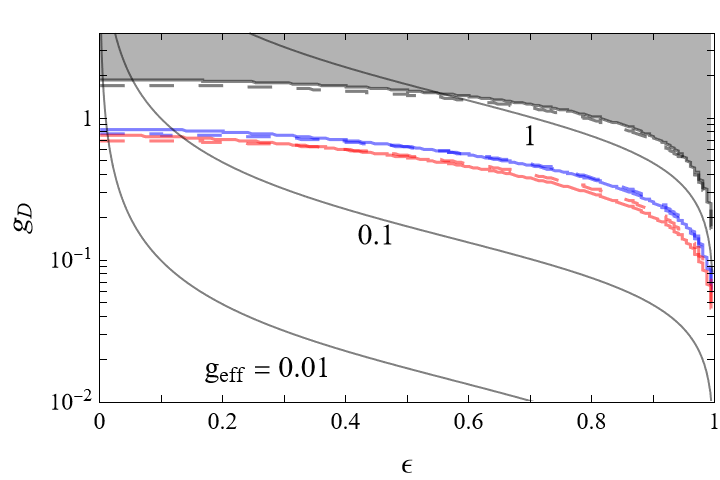}
    \caption{Positive $\epsilon$}\label{subfig:pole}
    \end{subfigure}
    \caption{
    The region of parameter space in which $g_D$ or $\epsilon$ exhibits a Landau pole below $Q = 100 M_{\mathrm{mess}}$ (gray shaded).
    The solid curve represents the bound for Scenario I, while the dashed curve indicates the bound for Scenario II.
    Above the blue and red curves, the pole appears at the GUT and Planck scales, respectively.
    Also shown are isocontours of the effective coupling $g_{\mathrm{eff}} = g_D \epsilon/\sqrt{(1-\epsilon^2)}$. 
    }
\label{fig:poleconst}
\end{figure}

One can calculate the RG evolution of the couplings using the beta functions specified in Eqs.~\eqref{eq:AbelBeta} and \eqref{eq:NonAbelBeta}, provided that appropriate boundary conditions are established. Figure~\ref{fig:RGrun} illustrates an example of the running of gauge couplings in one of our scenarios. We set the initial conditions for the gauge couplings at $m_Z$. There are two characteristic scales for the running: the mass scale of the SM superpartners at $Q \simeq 10$ TeV and the mass scale of the messengers at $Q \simeq 10^6$ GeV. Since only the messengers carry the dark charge, there is no significant running of $\epsilon$ and $g_D$ below the messenger mass scale. However, as shown in Figure~\ref{subfig:epsilonRun}, $\epsilon$ can undergo significant changes above the messenger mass scale. A larger $g_D$ leads to stronger running in both $\epsilon$ and $g_D$.

In dark GMSB, heavy messengers charged under the $U(1)_{D}$ could induce steep running of $g_D$ and $\epsilon$ above the messenger scale, especially in the large $g_D$ and $\epsilon$ limit. Consequently, these quantities might exhibit 
Landau poles not far above the messenger mass scale. Such divergences could indicate the presence of a new physics scale below the pole, which may not be favorable from the perspective of effective field theory. 
In Figure~\ref{fig:poleconst}, we illustrate the 
favored 
parameter space for $\epsilon$ and $g_D$ 
by ensuring  that
in which Landau poles are not generated at least up to $Q \simeq 100 M_{\mathrm{mess}}$ using the one-loop beta functions. This provides a theoretical upper limit on the values of $g_D$ and $\epsilon$.

\begin{figure}[tb]
    \centering    \includegraphics[width=0.7\linewidth]{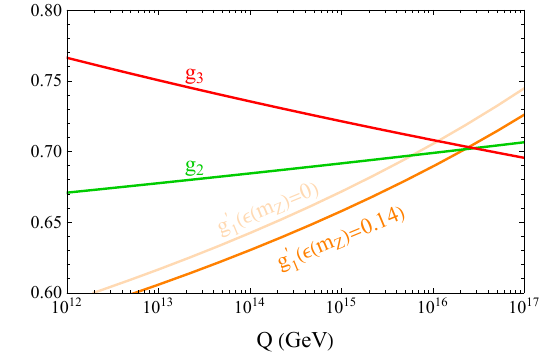}
    \caption{
    Examples of gauge coupling unification. Each curve, colored red, green, and orange, represents the gauge couplings of $SU(3)_C$, $SU(2)_L$, and $U(1)_Y$, respectively. The messenger representation considered is the complete ${\bf 5}$ of $SU(5)$ of Scenario I.
    Gauge coupling unification is achieved with $\epsilon(m_Z) = 0.14$ and occurs at a scale of $Q = 2.7 \times 10^{16}$ GeV.  }
    \label{fig:GaugeUnification}
\end{figure}

\subsection{Gauge coupling unification}\label{App.Gauge coupling unification}

Grand Unified Theory (GUT) aims to unify the gauge interactions at higher energy scales. Although the running of gauge couplings in the SM roughly converges at high energy scales, the convergence is not precise. Interestingly, significant kinetic mixing could potentially lead to the unification of these gauge couplings \cite{Redondo:2008zf,Takahashi:2016iph,Daido:2016kez}. This is due to the rescaling of the physical $U(1)_Y$ gauge coupling, 
$g_Y = g_Y^{\prime}/\sqrt{1-\epsilon^2}$. Notably, 
unification in this scenario is rather insensitive to the size of the dark gauge coupling. 

In contrast to the SM, gauge coupling unification is achieved more naturally in the 
MSSM. Consequently, messengers in GMSB models are typically assumed to form a complete GUT multiplet to facilitate gauge coupling unification 
with similar precision to that obtained in the MSSM. Although this unification is already achieved with improved accuracy compared to the SM, introducing an appropriate value of kinetic mixing, $\epsilon \sim \mathcal{O}(0.1)$, may further improve precise unification \cite{Redondo:2008zf}. This is illustrated in Figure~\ref{fig:GaugeUnification} for the $SU(5)$ complete messenger of Scenario I. Note that the precise value of $\epsilon$ may differ depending on the sparticle mass scale as well as threshold effects.

In general, incomplete messengers will upset gauge coupling unification.
In our scenario II, large kinetic mixing further exacerbates this issue because the hypercharge of the chosen incomplete messenger is relatively small. 
As a result, the $U(1)_Y$  gauge coupling remains smaller than the $SU(2)_L$ and $SU(3)_C$ gauge couplings near the unification scale, even in the absence of kinetic mixing. 
However, there are various ways in which additional physics at an intermediate scale could remedy this. As one example, introducing additional messengers that carry hypercharge while being singlets under $SU(2)_L$ and $SU(3)_C$ (e.g., ({\bf 1},{\bf 1},1)), which could be part of a GUT multiplet, can elevate $g_1$, potentially restoring gauge coupling unification.

\bibliographystyle{JHEP}
\bibliography{ref.bib}

\providecommand{\href}[2]{#2}\begingroup\raggedright\begin{thebibliography}{100}

\bibitem{Martin:1997ns}
S.P.~Martin, \emph{{A Supersymmetry primer}}, \href{https://doi.org/10.1142/9789812839657_0001}{\emph{Adv. Ser. Direct. High Energy Phys.} {\bfseries 18} (1998) 1} [\href{https://arxiv.org/abs/hep-ph/9709356}{{\ttfamily hep-ph/9709356}}].

\bibitem{Dine:1981gu}
M.~Dine and W.~Fischler, \emph{{A Phenomenological Model of Particle Physics Based on Supersymmetry}}, \href{https://doi.org/10.1016/0370-2693(82)91241-2}{\emph{Phys. Lett. B} {\bfseries 110} (1982) 227}.

\bibitem{Nappi:1982hm}
C.R.~Nappi and B.A.~Ovrut, \emph{{Supersymmetric Extension of the SU(3) x SU(2) x U(1) Model}}, \href{https://doi.org/10.1016/0370-2693(82)90418-X}{\emph{Phys. Lett. B} {\bfseries 113} (1982) 175}.

\bibitem{Alvarez-Gaume:1981abe}
L.~Alvarez-Gaume, M.~Claudson and M.B.~Wise, \emph{{Low-Energy Supersymmetry}}, \href{https://doi.org/10.1016/0550-3213(82)90138-9}{\emph{Nucl. Phys. B} {\bfseries 207} (1982) 96}.

\bibitem{Dine:1993yw}
M.~Dine and A.E.~Nelson, \emph{{Dynamical supersymmetry breaking at low-energies}}, \href{https://doi.org/10.1103/PhysRevD.48.1277}{\emph{Phys. Rev. D} {\bfseries 48} (1993) 1277} [\href{https://arxiv.org/abs/hep-ph/9303230}{{\ttfamily hep-ph/9303230}}].

\bibitem{Dine:1994vc}
M.~Dine, A.E.~Nelson and Y.~Shirman, \emph{{Low-energy dynamical supersymmetry breaking simplified}}, \href{https://doi.org/10.1103/PhysRevD.51.1362}{\emph{Phys. Rev. D} {\bfseries 51} (1995) 1362} [\href{https://arxiv.org/abs/hep-ph/9408384}{{\ttfamily hep-ph/9408384}}].

\bibitem{Dine:1995ag}
M.~Dine, A.E.~Nelson, Y.~Nir and Y.~Shirman, \emph{{New tools for low-energy dynamical supersymmetry breaking}}, \href{https://doi.org/10.1103/PhysRevD.53.2658}{\emph{Phys. Rev. D} {\bfseries 53} (1996) 2658} [\href{https://arxiv.org/abs/hep-ph/9507378}{{\ttfamily hep-ph/9507378}}].

\bibitem{Giudice:1998bp}
G.F.~Giudice and R.~Rattazzi, \emph{{Theories with gauge mediated supersymmetry breaking}}, \href{https://doi.org/10.1016/S0370-1573(99)00042-3}{\emph{Phys. Rept.} {\bfseries 322} (1999) 419} [\href{https://arxiv.org/abs/hep-ph/9801271}{{\ttfamily hep-ph/9801271}}].

\bibitem{Meade:2008wd}
P.~Meade, N.~Seiberg and D.~Shih, \emph{{General Gauge Mediation}}, \href{https://doi.org/10.1143/PTPS.177.143}{\emph{Prog. Theor. Phys. Suppl.} {\bfseries 177} (2009) 143} [\href{https://arxiv.org/abs/0801.3278}{{\ttfamily 0801.3278}}].

\bibitem{Dienes:1996zr}
K.R.~Dienes, C.F.~Kolda and J.~March-Russell, \emph{{Kinetic mixing and the supersymmetric gauge hierarchy}}, \href{https://doi.org/10.1016/S0550-3213(97)00173-9}{\emph{Nucl. Phys. B} {\bfseries 492} (1997) 104} [\href{https://arxiv.org/abs/hep-ph/9610479}{{\ttfamily hep-ph/9610479}}].

\bibitem{Suematsu:2006wh}
D.~Suematsu, \emph{{SUSY breaking based on Abelian gaugino kinetic term mixings}}, \href{https://doi.org/10.1088/1126-6708/2006/11/029}{\emph{JHEP} {\bfseries 11} (2006) 029} [\href{https://arxiv.org/abs/hep-ph/0606125}{{\ttfamily hep-ph/0606125}}].

\bibitem{Chun:2008by}
E.J.~Chun and J.-C.~Park, \emph{{Dark matter and sub-GeV hidden U(1) in GMSB models}}, \href{https://doi.org/10.1088/1475-7516/2009/02/026}{\emph{JCAP} {\bfseries 02} (2009) 026} [\href{https://arxiv.org/abs/0812.0308}{{\ttfamily 0812.0308}}].

\bibitem{Arkani-Hamed:2008kxc}
N.~Arkani-Hamed and N.~Weiner, \emph{{LHC Signals for a SuperUnified Theory of Dark Matter}}, \href{https://doi.org/10.1088/1126-6708/2008/12/104}{\emph{JHEP} {\bfseries 12} (2008) 104} [\href{https://arxiv.org/abs/0810.0714}{{\ttfamily 0810.0714}}].

\bibitem{Baumgart:2009tn}
M.~Baumgart, C.~Cheung, J.T.~Ruderman, L.-T.~Wang and I.~Yavin, \emph{{Non-Abelian Dark Sectors and Their Collider Signatures}}, \href{https://doi.org/10.1088/1126-6708/2009/04/014}{\emph{JHEP} {\bfseries 04} (2009) 014} [\href{https://arxiv.org/abs/0901.0283}{{\ttfamily 0901.0283}}].

\bibitem{Cheung:2009qd}
C.~Cheung, J.T.~Ruderman, L.-T.~Wang and I.~Yavin, \emph{{Kinetic Mixing as the Origin of Light Dark Scales}}, \href{https://doi.org/10.1103/PhysRevD.80.035008}{\emph{Phys. Rev. D} {\bfseries 80} (2009) 035008} [\href{https://arxiv.org/abs/0902.3246}{{\ttfamily 0902.3246}}].

\bibitem{Morrissey:2009ur}
D.E.~Morrissey, D.~Poland and K.M.~Zurek, \emph{{Abelian Hidden Sectors at a GeV}}, \href{https://doi.org/10.1088/1126-6708/2009/07/050}{\emph{JHEP} {\bfseries 07} (2009) 050} [\href{https://arxiv.org/abs/0904.2567}{{\ttfamily 0904.2567}}].

\bibitem{Arvanitaki:2009hb}
A.~Arvanitaki, N.~Craig, S.~Dimopoulos, S.~Dubovsky and J.~March-Russell, \emph{{String Photini at the LHC}}, \href{https://doi.org/10.1103/PhysRevD.81.075018}{\emph{Phys. Rev. D} {\bfseries 81} (2010) 075018} [\href{https://arxiv.org/abs/0909.5440}{{\ttfamily 0909.5440}}].

\bibitem{Cohen:2010kn}
T.~Cohen, D.J.~Phalen, A.~Pierce and K.M.~Zurek, \emph{{Asymmetric Dark Matter from a GeV Hidden Sector}}, \href{https://doi.org/10.1103/PhysRevD.82.056001}{\emph{Phys. Rev. D} {\bfseries 82} (2010) 056001} [\href{https://arxiv.org/abs/1005.1655}{{\ttfamily 1005.1655}}].

\bibitem{Kang:2010mh}
Z.~Kang, T.~Li, T.~Liu, C.~Tong and J.M.~Yang, \emph{{Light Dark Matter from the $U(1)_X$ Sector in the NMSSM with Gauge Mediation}}, \href{https://doi.org/10.1088/1475-7516/2011/01/028}{\emph{JCAP} {\bfseries 01} (2011) 028} [\href{https://arxiv.org/abs/1008.5243}{{\ttfamily 1008.5243}}].

\bibitem{Chan:2011aa}
Y.F.~Chan, M.~Low, D.E.~Morrissey and A.P.~Spray, \emph{{LHC Signatures of a Minimal Supersymmetric Hidden Valley}}, \href{https://doi.org/10.1007/JHEP05(2012)155}{\emph{JHEP} {\bfseries 05} (2012) 155} [\href{https://arxiv.org/abs/1112.2705}{{\ttfamily 1112.2705}}].

\bibitem{Baryakhtar:2012rz}
M.~Baryakhtar, N.~Craig and K.~Van~Tilburg, \emph{{Supersymmetry in the Shadow of Photini}}, \href{https://doi.org/10.1007/JHEP07(2012)164}{\emph{JHEP} {\bfseries 07} (2012) 164} [\href{https://arxiv.org/abs/1206.0751}{{\ttfamily 1206.0751}}].

\bibitem{Lee:2017fin}
H.M.~Lee, \emph{{Gauged $U(1)$ clockwork theory}}, \href{https://doi.org/10.1016/j.physletb.2018.01.010}{\emph{Phys. Lett. B} {\bfseries 778} (2018) 79} [\href{https://arxiv.org/abs/1708.03564}{{\ttfamily 1708.03564}}].

\bibitem{Andreas:2011in}
S.~Andreas, M.D.~Goodsell and A.~Ringwald, \emph{{Dark matter and dark forces from a supersymmetric hidden sector}}, \href{https://doi.org/10.1103/PhysRevD.87.025007}{\emph{Phys. Rev. D} {\bfseries 87} (2013) 025007} [\href{https://arxiv.org/abs/1109.2869}{{\ttfamily 1109.2869}}].

\bibitem{Kors:2004ri}
B.~Kors and P.~Nath, \emph{{A Supersymmetric Stueckelberg U(1) extension of the MSSM}}, \href{https://doi.org/10.1088/1126-6708/2004/12/005}{\emph{JHEP} {\bfseries 12} (2004) 005} [\href{https://arxiv.org/abs/hep-ph/0406167}{{\ttfamily hep-ph/0406167}}].

\bibitem{Feldman:2006wd}
D.~Feldman, B.~Kors and P.~Nath, \emph{{Extra-weakly Interacting Dark Matter}}, \href{https://doi.org/10.1103/PhysRevD.75.023503}{\emph{Phys. Rev. D} {\bfseries 75} (2007) 023503} [\href{https://arxiv.org/abs/hep-ph/0610133}{{\ttfamily hep-ph/0610133}}].

\bibitem{Hooper:2008im}
D.~Hooper and K.M.~Zurek, \emph{{A Natural Supersymmetric Model with MeV Dark Matter}}, \href{https://doi.org/10.1103/PhysRevD.77.087302}{\emph{Phys. Rev. D} {\bfseries 77} (2008) 087302} [\href{https://arxiv.org/abs/0801.3686}{{\ttfamily 0801.3686}}].

\bibitem{Ibarra:2008kn}
A.~Ibarra, A.~Ringwald and C.~Weniger, \emph{{Hidden gauginos of an unbroken U(1): Cosmological constraints and phenomenological prospects}}, \href{https://doi.org/10.1088/1475-7516/2009/01/003}{\emph{JCAP} {\bfseries 01} (2009) 003} [\href{https://arxiv.org/abs/0809.3196}{{\ttfamily 0809.3196}}].

\bibitem{Zurek:2008qg}
K.M.~Zurek, \emph{{Multi-Component Dark Matter}}, \href{https://doi.org/10.1103/PhysRevD.79.115002}{\emph{Phys. Rev. D} {\bfseries 79} (2009) 115002} [\href{https://arxiv.org/abs/0811.4429}{{\ttfamily 0811.4429}}].

\bibitem{Katz:2009qq}
A.~Katz and R.~Sundrum, \emph{{Breaking the Dark Force}}, \href{https://doi.org/10.1088/1126-6708/2009/06/003}{\emph{JHEP} {\bfseries 06} (2009) 003} [\href{https://arxiv.org/abs/0902.3271}{{\ttfamily 0902.3271}}].

\bibitem{Feldman:2010wy}
D.~Feldman, Z.~Liu, P.~Nath and G.~Peim, \emph{{Multicomponent Dark Matter in Supersymmetric Hidden Sector Extensions}}, \href{https://doi.org/10.1103/PhysRevD.81.095017}{\emph{Phys. Rev. D} {\bfseries 81} (2010) 095017} [\href{https://arxiv.org/abs/1004.0649}{{\ttfamily 1004.0649}}].

\bibitem{Barnes:2020vsc}
P.~Barnes, Z.~Johnson, A.~Pierce and B.~Shakya, \emph{{Simple Hidden Sector Dark Matter}}, \href{https://doi.org/10.1103/PhysRevD.102.075019}{\emph{Phys. Rev. D} {\bfseries 102} (2020) 075019} [\href{https://arxiv.org/abs/2003.13744}{{\ttfamily 2003.13744}}].

\bibitem{Pierce:2019ozl}
A.~Pierce and B.~Shakya, \emph{{Gaugino Portal Baryogenesis}}, \href{https://doi.org/10.1007/JHEP06(2019)096}{\emph{JHEP} {\bfseries 06} (2019) 096} [\href{https://arxiv.org/abs/1901.05493}{{\ttfamily 1901.05493}}].

\bibitem{Holdom:1985ag}
B.~Holdom, \emph{{Two U(1)'s and Epsilon Charge Shifts}}, \href{https://doi.org/10.1016/0370-2693(86)91377-8}{\emph{Phys. Lett. B} {\bfseries 166} (1986) 196}.

\bibitem{Holdom:1986eq}
B.~Holdom, \emph{{Searching for $\epsilon$ Charges and a New U(1)}}, \href{https://doi.org/10.1016/0370-2693(86)90470-3}{\emph{Phys. Lett. B} {\bfseries 178} (1986) 65}.

\bibitem{Glashow:1961tr}
S.L.~Glashow, \emph{{Partial Symmetries of Weak Interactions}}, \href{https://doi.org/10.1016/0029-5582(61)90469-2}{\emph{Nucl. Phys.} {\bfseries 22} (1961) 579}.

\bibitem{Fabbrichesi:2020wbt}
M.~Fabbrichesi, E.~Gabrielli and G.~Lanfranchi, \emph{{The Dark Photon}},  \href{https://arxiv.org/abs/2005.01515}{{\ttfamily 2005.01515}}.

\bibitem{Burgess:2008ri}
C.P.~Burgess, J.P.~Conlon, L.-Y.~Hung, C.H.~Kom, A.~Maharana and F.~Quevedo, \emph{{Continuous Global Symmetries and Hyperweak Interactions in String Compactifications}}, \href{https://doi.org/10.1088/1126-6708/2008/07/073}{\emph{JHEP} {\bfseries 07} (2008) 073} [\href{https://arxiv.org/abs/0805.4037}{{\ttfamily 0805.4037}}].

\bibitem{Gan:2023jbs}
X.~Gan and Y.-D.~Tsai, \emph{{Cosmic Millicharge Background and Reheating Probes}},  \href{https://arxiv.org/abs/2308.07951}{{\ttfamily 2308.07951}}.

\bibitem{Pan:2018dmu}
J.-X.~Pan, M.~He, X.-G.~He and G.~Li, \emph{{Scrutinizing a massless dark photon: basis independence}}, \href{https://doi.org/10.1016/j.nuclphysb.2020.114968}{\emph{Nucl. Phys. B} {\bfseries 953} (2020) 114968} [\href{https://arxiv.org/abs/1807.11363}{{\ttfamily 1807.11363}}].

\bibitem{Prinz:1998ua}
A.A.~Prinz et~al., \emph{{Search for millicharged particles at SLAC}}, \href{https://doi.org/10.1103/PhysRevLett.81.1175}{\emph{Phys. Rev. Lett.} {\bfseries 81} (1998) 1175} [\href{https://arxiv.org/abs/hep-ex/9804008}{{\ttfamily hep-ex/9804008}}].

\bibitem{Davidson:2000hf}
S.~Davidson, S.~Hannestad and G.~Raffelt, \emph{{Updated bounds on millicharged particles}}, \href{https://doi.org/10.1088/1126-6708/2000/05/003}{\emph{JHEP} {\bfseries 05} (2000) 003} [\href{https://arxiv.org/abs/hep-ph/0001179}{{\ttfamily hep-ph/0001179}}].

\bibitem{PhysRevD.75.032004}
A.~Badertscher, P.~Crivelli, W.~Fetscher, U.~Gendotti, S.N.~Gninenko, V.~Postoev et~al., \emph{Improved limit on invisible decays of positronium}, \href{https://doi.org/10.1103/PhysRevD.75.032004}{\emph{Phys. Rev. D} {\bfseries 75} (2007) 032004}.

\bibitem{Jaeckel:2012yz}
J.~Jaeckel, M.~Jankowiak and M.~Spannowsky, \emph{{LHC probes the hidden sector}}, \href{https://doi.org/10.1016/j.dark.2013.06.001}{\emph{Phys. Dark Univ.} {\bfseries 2} (2013) 111} [\href{https://arxiv.org/abs/1212.3620}{{\ttfamily 1212.3620}}].

\bibitem{Vogel:2013raa}
H.~Vogel and J.~Redondo, \emph{{Dark Radiation constraints on minicharged particles in models with a hidden photon}}, \href{https://doi.org/10.1088/1475-7516/2014/02/029}{\emph{JCAP} {\bfseries 02} (2014) 029} [\href{https://arxiv.org/abs/1311.2600}{{\ttfamily 1311.2600}}].

\bibitem{Chang:2018rso}
J.H.~Chang, R.~Essig and S.D.~McDermott, \emph{{Supernova 1987A Constraints on Sub-GeV Dark Sectors, Millicharged Particles, the QCD Axion, and an Axion-like Particle}}, \href{https://doi.org/10.1007/JHEP09(2018)051}{\emph{JHEP} {\bfseries 09} (2018) 051} [\href{https://arxiv.org/abs/1803.00993}{{\ttfamily 1803.00993}}].

\bibitem{Magill:2018tbb}
G.~Magill, R.~Plestid, M.~Pospelov and Y.-D.~Tsai, \emph{{Millicharged particles in neutrino experiments}}, \href{https://doi.org/10.1103/PhysRevLett.122.071801}{\emph{Phys. Rev. Lett.} {\bfseries 122} (2019) 071801} [\href{https://arxiv.org/abs/1806.03310}{{\ttfamily 1806.03310}}].

\bibitem{Stebbins:2019xjr}
A.~Stebbins and G.~Krnjaic, \emph{{New Limits on Charged Dark Matter from Large-Scale Coherent Magnetic Fields}}, \href{https://doi.org/10.1088/1475-7516/2019/12/003}{\emph{JCAP} {\bfseries 12} (2019) 003} [\href{https://arxiv.org/abs/1908.05275}{{\ttfamily 1908.05275}}].

\bibitem{Fiorillo:2024upk}
D.F.G.~Fiorillo and E.~Vitagliano, \emph{{Self-Interacting Dark Sectors in Supernovae Can Behave as a Relativistic Fluid}}, \href{https://doi.org/10.1103/PhysRevLett.133.251004}{\emph{Phys. Rev. Lett.} {\bfseries 133} (2024) 251004} [\href{https://arxiv.org/abs/2404.07714}{{\ttfamily 2404.07714}}].

\bibitem{Redondo:2008zf}
J.~Redondo, \emph{{The Low energy frontier: Probes with photons}},  in \emph{{43rd Rencontres de Moriond on Electroweak Interactions and Unified Theories}}, pp.~473--478, 5, 2008 [\href{https://arxiv.org/abs/0805.3112}{{\ttfamily 0805.3112}}].

\bibitem{Takahashi:2016iph}
F.~Takahashi, M.~Yamada and N.~Yokozaki, \emph{{Diphoton excess from hidden U(1) gauge symmetry with large kinetic mixing}}, \href{https://doi.org/10.1016/j.physletb.2016.07.013}{\emph{Phys. Lett. B} {\bfseries 760} (2016) 486} [\href{https://arxiv.org/abs/1604.07145}{{\ttfamily 1604.07145}}].

\bibitem{Daido:2016kez}
R.~Daido, F.~Takahashi and N.~Yokozaki, \emph{{Gauge Coupling Unification with Hidden Photon, and Minicharged Dark Matter}}, \href{https://doi.org/10.1016/j.physletb.2017.01.085}{\emph{Phys. Lett. B} {\bfseries 768} (2017) 30} [\href{https://arxiv.org/abs/1610.00631}{{\ttfamily 1610.00631}}].

\bibitem{Dimopoulos:1996gy}
S.~Dimopoulos, G.F.~Giudice and A.~Pomarol, \emph{{Dark matter in theories of gauge mediated supersymmetry breaking}}, \href{https://doi.org/10.1016/S0370-2693(96)01241-5}{\emph{Phys. Lett. B} {\bfseries 389} (1996) 37} [\href{https://arxiv.org/abs/hep-ph/9607225}{{\ttfamily hep-ph/9607225}}].

\bibitem{Allanach:2001kg}
B.C.~Allanach, \emph{{SOFTSUSY: a program for calculating supersymmetric spectra}}, \href{https://doi.org/10.1016/S0010-4655(01)00460-X}{\emph{Comput. Phys. Commun.} {\bfseries 143} (2002) 305} [\href{https://arxiv.org/abs/hep-ph/0104145}{{\ttfamily hep-ph/0104145}}].

\bibitem{ATLAS:2012yve}
{\scshape ATLAS} collaboration, \emph{{Observation of a new particle in the search for the Standard Model Higgs boson with the ATLAS detector at the LHC}}, \href{https://doi.org/10.1016/j.physletb.2012.08.020}{\emph{Phys. Lett. B} {\bfseries 716} (2012) 1} [\href{https://arxiv.org/abs/1207.7214}{{\ttfamily 1207.7214}}].

\bibitem{CMS:2012qbp}
{\scshape CMS} collaboration, \emph{{Observation of a New Boson at a Mass of 125 GeV with the CMS Experiment at the LHC}}, \href{https://doi.org/10.1016/j.physletb.2012.08.021}{\emph{Phys. Lett. B} {\bfseries 716} (2012) 30} [\href{https://arxiv.org/abs/1207.7235}{{\ttfamily 1207.7235}}].

\bibitem{Casas:1994us}
J.A.~Casas, J.R.~Espinosa, M.~Quiros and A.~Riotto, \emph{{The Lightest Higgs boson mass in the minimal supersymmetric standard model}}, \href{https://doi.org/10.1016/0550-3213(94)00508-C}{\emph{Nucl. Phys. B} {\bfseries 436} (1995) 3} [\href{https://arxiv.org/abs/hep-ph/9407389}{{\ttfamily hep-ph/9407389}}].

\bibitem{Carena:1995bx}
M.~Carena, J.R.~Espinosa, M.~Quiros and C.E.M.~Wagner, \emph{{Analytical expressions for radiatively corrected Higgs masses and couplings in the MSSM}}, \href{https://doi.org/10.1016/0370-2693(95)00694-G}{\emph{Phys. Lett. B} {\bfseries 355} (1995) 209} [\href{https://arxiv.org/abs/hep-ph/9504316}{{\ttfamily hep-ph/9504316}}].

\bibitem{Haber:1996fp}
H.E.~Haber, R.~Hempfling and A.H.~Hoang, \emph{{Approximating the radiatively corrected Higgs mass in the minimal supersymmetric model}}, \href{https://doi.org/10.1007/s002880050498}{\emph{Z. Phys. C} {\bfseries 75} (1997) 539} [\href{https://arxiv.org/abs/hep-ph/9609331}{{\ttfamily hep-ph/9609331}}].

\bibitem{Draper:2011aa}
P.~Draper, P.~Meade, M.~Reece and D.~Shih, \emph{{Implications of a 125 GeV Higgs for the MSSM and Low-Scale SUSY Breaking}}, \href{https://doi.org/10.1103/PhysRevD.85.095007}{\emph{Phys. Rev. D} {\bfseries 85} (2012) 095007} [\href{https://arxiv.org/abs/1112.3068}{{\ttfamily 1112.3068}}].

\bibitem{Arbey:2011ab}
A.~Arbey, M.~Battaglia, A.~Djouadi, F.~Mahmoudi and J.~Quevillon, \emph{{Implications of a 125 GeV Higgs for supersymmetric models}}, \href{https://doi.org/10.1016/j.physletb.2012.01.053}{\emph{Phys. Lett. B} {\bfseries 708} (2012) 162} [\href{https://arxiv.org/abs/1112.3028}{{\ttfamily 1112.3028}}].

\bibitem{Ibe:2007km}
M.~Ibe and R.~Kitano, \emph{{Sweet Spot Supersymmetry}}, \href{https://doi.org/10.1088/1126-6708/2007/08/016}{\emph{JHEP} {\bfseries 08} (2007) 016} [\href{https://arxiv.org/abs/0705.3686}{{\ttfamily 0705.3686}}].

\bibitem{Barger:2006dh}
V.~Barger, P.~Langacker, H.-S.~Lee and G.~Shaughnessy, \emph{{Higgs Sector in Extensions of the MSSM}}, \href{https://doi.org/10.1103/PhysRevD.73.115010}{\emph{Phys. Rev. D} {\bfseries 73} (2006) 115010} [\href{https://arxiv.org/abs/hep-ph/0603247}{{\ttfamily hep-ph/0603247}}].

\bibitem{Barger:2005hb}
V.~Barger, P.~Langacker and H.-S.~Lee, \emph{{Lightest neutralino in extensions of the MSSM}}, \href{https://doi.org/10.1016/j.physletb.2005.09.023}{\emph{Phys. Lett. B} {\bfseries 630} (2005) 85} [\href{https://arxiv.org/abs/hep-ph/0508027}{{\ttfamily hep-ph/0508027}}].

\bibitem{Ambrosanio:1996jn}
S.~Ambrosanio, G.L.~Kane, G.D.~Kribs, S.P.~Martin and S.~Mrenna, \emph{{Search for supersymmetry with a light gravitino at the Fermilab Tevatron and CERN LEP colliders}}, \href{https://doi.org/10.1103/PhysRevD.54.5395}{\emph{Phys. Rev. D} {\bfseries 54} (1996) 5395} [\href{https://arxiv.org/abs/hep-ph/9605398}{{\ttfamily hep-ph/9605398}}].

\bibitem{Dimopoulos:1996yq}
S.~Dimopoulos, S.D.~Thomas and J.D.~Wells, \emph{{Sparticle spectroscopy and electroweak symmetry breaking with gauge mediated supersymmetry breaking}}, \href{https://doi.org/10.1016/S0550-3213(97)00030-8}{\emph{Nucl. Phys. B} {\bfseries 488} (1997) 39} [\href{https://arxiv.org/abs/hep-ph/9609434}{{\ttfamily hep-ph/9609434}}].

\bibitem{Griest:1987qv}
K.~Griest and H.E.~Haber, \emph{{Invisible Decays of Higgs Bosons in Supersymmetric Models}}, \href{https://doi.org/10.1103/PhysRevD.37.719}{\emph{Phys. Rev. D} {\bfseries 37} (1988) 719}.

\bibitem{Djouadi:1996mj}
A.~Djouadi, P.~Janot, J.~Kalinowski and P.M.~Zerwas, \emph{{SUSY decays of Higgs particles}}, \href{https://doi.org/10.1016/0370-2693(96)00414-5}{\emph{Phys. Lett. B} {\bfseries 376} (1996) 220} [\href{https://arxiv.org/abs/hep-ph/9603368}{{\ttfamily hep-ph/9603368}}].

\bibitem{Dreiner:2012ex}
H.K.~Dreiner, J.S.~Kim and O.~Lebedev, \emph{{First LHC Constraints on Neutralinos}}, \href{https://doi.org/10.1016/j.physletb.2012.07.058}{\emph{Phys. Lett. B} {\bfseries 715} (2012) 199} [\href{https://arxiv.org/abs/1206.3096}{{\ttfamily 1206.3096}}].

\bibitem{deBlas:2019rxi}
J.~de~Blas et~al., \emph{{Higgs Boson Studies at Future Particle Colliders}}, \href{https://doi.org/10.1007/JHEP01(2020)139}{\emph{JHEP} {\bfseries 01} (2020) 139} [\href{https://arxiv.org/abs/1905.03764}{{\ttfamily 1905.03764}}].

\bibitem{FCC:2018evy}
{\scshape FCC} collaboration, \emph{{FCC-ee: The Lepton Collider}: {Future Circular Collider Conceptual Design Report Volume 2}}, \href{https://doi.org/10.1140/epjst/e2019-900045-4}{\emph{Eur. Phys. J. ST} {\bfseries 228} (2019) 261}.

\bibitem{ILC:2007oiw}
{\scshape ILC} collaboration, \emph{{ILC Reference Design Report Volume 1 - Executive Summary}},  \href{https://arxiv.org/abs/0712.1950}{{\ttfamily 0712.1950}}.

\bibitem{CEPCStudyGroup:2023quu}
{\scshape CEPC Study Group} collaboration, \emph{{CEPC Technical Design Report: Accelerator}}, \href{https://doi.org/10.1007/s41605-024-00463-y}{\emph{Radiat. Detect. Technol. Methods} {\bfseries 8} (2024) 1} [\href{https://arxiv.org/abs/2312.14363}{{\ttfamily 2312.14363}}].

\bibitem{ALEPH:2005ab}
{\scshape ALEPH, DELPHI, L3, OPAL, SLD, LEP Electroweak Working Group, SLD Electroweak Group, SLD Heavy Flavour Group} collaboration, \emph{{Precision electroweak measurements on the $Z$ resonance}}, \href{https://doi.org/10.1016/j.physrep.2005.12.006}{\emph{Phys. Rept.} {\bfseries 427} (2006) 257} [\href{https://arxiv.org/abs/hep-ex/0509008}{{\ttfamily hep-ex/0509008}}].

\bibitem{Pagels:1981ke}
H.~Pagels and J.R.~Primack, \emph{{Supersymmetry, Cosmology and New TeV Physics}}, \href{https://doi.org/10.1103/PhysRevLett.48.223}{\emph{Phys. Rev. Lett.} {\bfseries 48} (1982) 223}.

\bibitem{Weinberg:1982zq}
S.~Weinberg, \emph{{Cosmological Constraints on the Scale of Supersymmetry Breaking}}, \href{https://doi.org/10.1103/PhysRevLett.48.1303}{\emph{Phys. Rev. Lett.} {\bfseries 48} (1982) 1303}.

\bibitem{Viel:2005qj}
M.~Viel, J.~Lesgourgues, M.G.~Haehnelt, S.~Matarrese and A.~Riotto, \emph{{Constraining warm dark matter candidates including sterile neutrinos and light gravitinos with WMAP and the Lyman-alpha forest}}, \href{https://doi.org/10.1103/PhysRevD.71.063534}{\emph{Phys. Rev. D} {\bfseries 71} (2005) 063534} [\href{https://arxiv.org/abs/astro-ph/0501562}{{\ttfamily astro-ph/0501562}}].

\bibitem{Osato:2016ixc}
K.~Osato, T.~Sekiguchi, M.~Shirasaki, A.~Kamada and N.~Yoshida, \emph{{Cosmological Constraint on the Light Gravitino Mass from CMB Lensing and Cosmic Shear}}, \href{https://doi.org/10.1088/1475-7516/2016/06/004}{\emph{JCAP} {\bfseries 06} (2016) 004} [\href{https://arxiv.org/abs/1601.07386}{{\ttfamily 1601.07386}}].

\bibitem{Borgani:1996ag}
S.~Borgani, A.~Masiero and M.~Yamaguchi, \emph{{Light gravitinos as mixed dark matter}}, \href{https://doi.org/10.1016/0370-2693(96)00956-2}{\emph{Phys. Lett. B} {\bfseries 386} (1996) 189} [\href{https://arxiv.org/abs/hep-ph/9605222}{{\ttfamily hep-ph/9605222}}].

\bibitem{Takayama:2000uz}
F.~Takayama and M.~Yamaguchi, \emph{{Gravitino dark matter without R-parity}}, \href{https://doi.org/10.1016/S0370-2693(00)00726-7}{\emph{Phys. Lett. B} {\bfseries 485} (2000) 388} [\href{https://arxiv.org/abs/hep-ph/0005214}{{\ttfamily hep-ph/0005214}}].

\bibitem{Moreau:2001sr}
G.~Moreau and M.~Chemtob, \emph{{R-parity violation and the cosmological gravitino problem}}, \href{https://doi.org/10.1103/PhysRevD.65.024033}{\emph{Phys. Rev. D} {\bfseries 65} (2002) 024033} [\href{https://arxiv.org/abs/hep-ph/0107286}{{\ttfamily hep-ph/0107286}}].

\bibitem{Kamiokande}
{\scshape Kamiokande-II} collaboration, \emph{{Search for fractionally charged particles in Kamiokande-II}}, \href{https://doi.org/10.1103/PhysRevD.43.2843}{\emph{Phys. Rev. D} {\bfseries 43} (1991) 2843}.

\bibitem{Aglietta:1994iv}
M.~Aglietta et~al., \emph{{Search for fractionally charged particles in the Mont Blanc LSD scintillation detector}}, \href{https://doi.org/10.1016/0927-6505(94)90015-9}{\emph{Astropart. Phys.} {\bfseries 2} (1994) 29}.

\bibitem{MACRO:2000bht}
{\scshape MACRO} collaboration, \emph{{A Search for lightly ionizing particles with the MACRO detector}}, \href{https://doi.org/10.1103/PhysRevD.62.052003}{\emph{Phys. Rev. D} {\bfseries 62} (2000) 052003} [\href{https://arxiv.org/abs/hep-ex/0002029}{{\ttfamily hep-ex/0002029}}].

\bibitem{MACRO:2004iiu}
{\scshape MACRO} collaboration, \emph{{Final search for lightly ionizing particles with the MACRO detector}},  \href{https://arxiv.org/abs/hep-ex/0402006}{{\ttfamily hep-ex/0402006}}.

\bibitem{CDMS:2014ane}
{\scshape CDMS} collaboration, \emph{{First Direct Limits on Lightly Ionizing Particles with Electric Charge Less Than $e/6$}}, \href{https://doi.org/10.1103/PhysRevLett.114.111302}{\emph{Phys. Rev. Lett.} {\bfseries 114} (2015) 111302} [\href{https://arxiv.org/abs/1409.3270}{{\ttfamily 1409.3270}}].

\bibitem{Majorana:2018gib}
{\scshape Majorana} collaboration, \emph{{First Limit on the Direct Detection of Lightly Ionizing Particles for Electric Charge as Low as e/1000 with the Majorana Demonstrator}}, \href{https://doi.org/10.1103/PhysRevLett.120.211804}{\emph{Phys. Rev. Lett.} {\bfseries 120} (2018) 211804} [\href{https://arxiv.org/abs/1801.10145}{{\ttfamily 1801.10145}}].

\bibitem{Sbarra:2003ur}
C.~Sbarra, D.~Casadei, L.~Brocco, A.~Contin, G.~Levi and F.~Palmonari, \emph{{Search for fractional charges in cosmic rays with ams}},  \href{https://arxiv.org/abs/astro-ph/0304192}{{\ttfamily astro-ph/0304192}}.

\bibitem{Fuke:2008zza}
H.~Fuke et~al., \emph{{Search for fractionally charged particles in cosmic rays with the BESS spectrometer}}, \href{https://doi.org/10.1016/j.asr.2007.02.042}{\emph{Adv. Space Res.} {\bfseries 41} (2008) 2050}.

\bibitem{Larue:1981jc}
G.S.~Larue, J.D.~Phillips and W.M.~Fairbank, \emph{{Observation of Fractional Charge of $(\dfrac{1}{3})e$ on Matter}}, \href{https://doi.org/10.1103/PhysRevLett.46.967}{\emph{Phys. Rev. Lett.} {\bfseries 46} (1981) 967}.

\bibitem{Marinelli:1983nd}
M.~Marinelli and G.~Morpurgo, \emph{{The Electric Neutrality of Matter: A Summary}}, \href{https://doi.org/10.1016/0370-2693(84)91752-0}{\emph{Phys. Lett. B} {\bfseries 137} (1984) 439}.

\bibitem{Smith:1986ik}
P.F.~Smith, G.J.~Homer, J.D.~Lewin, H.E.~Walford and W.G.~Jones, \emph{{A Search for Fractional Charge Changes on Levitated Niobium Spheres}}, \href{https://doi.org/10.1016/0370-2693(86)91012-9}{\emph{Phys. Lett. B} {\bfseries 171} (1986) 129}.

\bibitem{Smith:1987mj}
P.F.~Smith, G.J.~Homer, J.D.~Lewin, H.E.~Walford and W.G.~Jones, \emph{{Searches for Fractional Electric Charge in Tungsten}}, \href{https://doi.org/10.1016/0370-2693(87)90418-7}{\emph{Phys. Lett. B} {\bfseries 197} (1987) 447}.

\bibitem{Jones:1989cq}
W.G.~Jones, P.F.~Smith, G.J.~Homer, J.D.~Lewin and H.E.~Walford, \emph{{Searches for Fractional Electric Charge in Meteorite Samples}}, \href{https://doi.org/10.1007/BF01506530}{\emph{Z. Phys. C} {\bfseries 43} (1989) 349}.

\bibitem{Homer:1992cz}
G.J.~Homer, P.F.~Smith, J.D.~Lewin, S.J.~Robertson, J.U.D.~Langridge, D.~Evans et~al., \emph{{A Search for fractional electric charge in sea water}}, \href{https://doi.org/10.1007/BF01561292}{\emph{Z. Phys. C} {\bfseries 55} (1992) 549}.

\bibitem{Joyce:1983tj}
D.~Joyce, P.~Abrams, R.~Bland, C.~Hodges, R.T.~Johnson, M.~Lindgren et~al., \emph{{A Search for Fractional Charges in Water}}, \href{https://doi.org/10.1103/PhysRevLett.51.731}{\emph{Phys. Rev. Lett.} {\bfseries 51} (1983) 731}.

\bibitem{Savage:1986vg}
M.L.~Savage et~al., \emph{{A Search for Fractional Charges in Native Mercury}}, \href{https://doi.org/10.1016/0370-2693(86)91305-5}{\emph{Phys. Lett. B} {\bfseries 167} (1986) 481}.

\bibitem{Halyo:1999wq}
V.~Halyo, P.~Kim, E.R.~Lee, I.T.~Lee, D.~Loomba and M.L.~Perl, \emph{{Search for free fractional electric charge elementary particles}}, \href{https://doi.org/10.1103/PhysRevLett.84.2576}{\emph{Phys. Rev. Lett.} {\bfseries 84} (2000) 2576} [\href{https://arxiv.org/abs/hep-ex/9910064}{{\ttfamily hep-ex/9910064}}].

\bibitem{Lee:2002sa}
I.T.~Lee, S.~Fan, V.~Halyo, E.R.~Lee, P.C.~Kim, M.L.~Perl et~al., \emph{{Large bulk matter search for fractional charge particles}}, \href{https://doi.org/10.1103/PhysRevD.66.012002}{\emph{Phys. Rev. D} {\bfseries 66} (2002) 012002} [\href{https://arxiv.org/abs/hep-ex/0204003}{{\ttfamily hep-ex/0204003}}].

\bibitem{Kim:2007zzs}
P.C.~Kim, E.R.~Lee, I.T.~Lee, M.L.~Perl, V.~Halyo and D.~Loomba, \emph{{Search for fractional-charge particles in meteoritic material}}, \href{https://doi.org/10.1103/PhysRevLett.99.161804}{\emph{Phys. Rev. Lett.} {\bfseries 99} (2007) 161804}.

\bibitem{Kudo:2001ie}
A.~Kudo and M.~Yamaguchi, \emph{{Inflation with low reheat temperature and cosmological constraint on stable charged massive particles}}, \href{https://doi.org/10.1016/S0370-2693(01)00938-8}{\emph{Phys. Lett. B} {\bfseries 516} (2001) 151} [\href{https://arxiv.org/abs/hep-ph/0103272}{{\ttfamily hep-ph/0103272}}].

\bibitem{Giudice:2000ex}
G.F.~Giudice, E.W.~Kolb and A.~Riotto, \emph{{Largest temperature of the radiation era and its cosmological implications}}, \href{https://doi.org/10.1103/PhysRevD.64.023508}{\emph{Phys. Rev. D} {\bfseries 64} (2001) 023508} [\href{https://arxiv.org/abs/hep-ph/0005123}{{\ttfamily hep-ph/0005123}}].

\bibitem{ParticleDataGroup:2022pth}
{\scshape Particle Data Group} collaboration, \emph{{Review of Particle Physics}}, \href{https://doi.org/10.1093/ptep/ptac097}{\emph{PTEP} {\bfseries 2022} (2022) 083C01}.

\bibitem{SMITH1979525}
P.~Smith and J.~Bennett, \emph{A search for heavy stable particles}, \href{https://doi.org/https://doi.org/10.1016/0550-3213(79)90006-3}{\emph{Nucl. Phys. B} {\bfseries 149} (1979) 525}.

\bibitem{SMITH1982333}
P.~Smith, J.~Bennett, G.~Homer, J.~Lewin, H.~Walford and W.~Smith, \emph{A search for anomalous hydrogen in enriched d2o, using a time-of-flight spectrometer}, \href{https://doi.org/https://doi.org/10.1016/0550-3213(82)90271-1}{\emph{Nucl. Phys. B} {\bfseries 206} (1982) 333}.

\bibitem{PhysRevD.41.2074}
T.K.~Hemmick, D.~Elmore, T.~Gentile, P.W.~Kubik, S.L.~Olsen, D.~Ciampa et~al., \emph{Search for low-$z$ nuclei containing massive stable particles}, \href{https://doi.org/10.1103/PhysRevD.41.2074}{\emph{Phys. Rev. D} {\bfseries 41} (1990) 2074}.

\bibitem{PhysRevLett.68.1116}
P.~Verkerk, G.~Grynberg, B.~Pichard, M.~Spiro, S.~Zylberajch, M.E.~Goldberg et~al., \emph{Search for superheavy hydrogen in sea water}, \href{https://doi.org/10.1103/PhysRevLett.68.1116}{\emph{Phys. Rev. Lett.} {\bfseries 68} (1992) 1116}.

\bibitem{PhysRevD.47.1231}
T.~Yamagata, Y.~Takamori and H.~Utsunomiya, \emph{Search for anomalously heavy hydrogen in deep sea water at 4000 m}, \href{https://doi.org/10.1103/PhysRevD.47.1231}{\emph{Phys. Rev. D} {\bfseries 47} (1993) 1231}.

\bibitem{Chuzhoy:2008zy}
L.~Chuzhoy and E.W.~Kolb, \emph{{Reopening the window on charged dark matter}}, \href{https://doi.org/10.1088/1475-7516/2009/07/014}{\emph{JCAP} {\bfseries 07} (2009) 014} [\href{https://arxiv.org/abs/0809.0436}{{\ttfamily 0809.0436}}].

\bibitem{Martin:1993zk}
S.P.~Martin and M.T.~Vaughn, \emph{{Two loop renormalization group equations for soft supersymmetry breaking couplings}}, \href{https://doi.org/10.1103/PhysRevD.50.2282}{\emph{Phys. Rev. D} {\bfseries 50} (1994) 2282} [\href{https://arxiv.org/abs/hep-ph/9311340}{{\ttfamily hep-ph/9311340}}].

\bibitem{Fonseca:2011vn}
R.M.~Fonseca, M.~Malinsky, W.~Porod and F.~Staub, \emph{{Running soft parameters in SUSY models with multiple U(1) gauge factors}}, \href{https://doi.org/10.1016/j.nuclphysb.2011.08.017}{\emph{Nucl. Phys. B} {\bfseries 854} (2012) 28} [\href{https://arxiv.org/abs/1107.2670}{{\ttfamily 1107.2670}}].

\bibitem{Babu:1996vt}
K.S.~Babu, C.F.~Kolda and J.~March-Russell, \emph{{Leptophobic U(1) $s$ and the R($b$) - R($c$) crisis}}, \href{https://doi.org/10.1103/PhysRevD.54.4635}{\emph{Phys. Rev. D} {\bfseries 54} (1996) 4635} [\href{https://arxiv.org/abs/hep-ph/9603212}{{\ttfamily hep-ph/9603212}}].

\end{thebibliography}\endgroup

\end{document}